\newif\ifShowKeys
\ShowKeystrue
\ShowKeysfalse

\documentclass[letterpaper,10pt]{article}

\pdfoutput=1
\usepackage[no-natbib-sort]{my-jheppub}

\usepackage{amsmath, amssymb}

\usepackage{xfrac}   

\usepackage{amsthm}

\usepackage{bm,environ,mathrsfs,array,arydshln,booktabs,float,slashed,hyperref}
\usepackage[mathcal]{euscript}
\usepackage{tensor} 	

\usepackage{wick}

\usepackage[nodayofweek]{date time}

\font\sm=cmss10 at 4pt
\usepackage{graphicx,epsfig}
\usepackage{epic}
\usepackage{tikz} 
\usetikzlibrary{decorations.pathreplacing,decorations.markings,snakes}

\tikzset{middlearrow/.style={
        decoration={markings,
            mark= at position 0.5 with {\arrow{#1}} ,
        },
        postaction={decorate}
    }
}

\usepackage{floatflt}

\usepackage{aurical}
\usepackage[T1]{fontenc}

\usepackage{framed}
\definecolor{shadecolor}{RGB}{255, 230, 204}

\allowdisplaybreaks

\newcommand{\be}{\begin{equation}}
\newcommand{\ee}{\end{equation}}
\newcommand{\mc}{\mathcal }

\newcommand{\la}{\label}

\def \bz {\mathsf{z}}
\def \bt {\mathsf{t}}
\def \vp {\varphi}

\newcommand{\EV}[1]{\langle #1 \rangle}
\newcommand{\p}{\partial}

\newcommand{\beqn}{\begin{eqnarray}}
\newcommand{\eeqn}{\end{eqnarray}}
\newcommand{\sft}{{\sf t}}
\newcommand{\sfz}{{\sf z}}

\newcommand{\de}{\Delta}

\newcommand{\chads}{AdS$_{2}$/CFT$_{2}^{\sfrac{1}{2}}$ }

\newcommand{\bC}{{\sf C}}

\makeatletter
\DeclareFontFamily{OMX}{MnSymbolE}{}
\DeclareSymbolFont{MnLargeSymbols}{OMX}{MnSymbolE}{m}{n}
\SetSymbolFont{MnLargeSymbols}{bold}{OMX}{MnSymbolE}{b}{n}
\DeclareFontShape{OMX}{MnSymbolE}{m}{n}{
    <-6>  MnSymbolE5
   <6-7>  MnSymbolE6
   <7-8>  MnSymbolE7
   <8-9>  MnSymbolE8
   <9-10> MnSymbolE9
  <10-12> MnSymbolE10
  <12->   MnSymbolE12
}{}
\DeclareFontShape{OMX}{MnSymbolE}{b}{n}{
    <-6>  MnSymbolE-Bold5
   <6-7>  MnSymbolE-Bold6
   <7-8>  MnSymbolE-Bold7
   <8-9>  MnSymbolE-Bold8
   <9-10> MnSymbolE-Bold9
  <10-12> MnSymbolE-Bold10
  <12->   MnSymbolE-Bold12
}{}

\let\llangle\@undefined
\let\rrangle\@undefined
\DeclareMathDelimiter{\llangle}{\mathopen}%
                     {MnLargeSymbols}{'164}{MnLargeSymbols}{'164}
\DeclareMathDelimiter{\rrangle}{\mathclose}%
                     {MnLargeSymbols}{'171}{MnLargeSymbols}{'171}
\makeatother

\def \RR {{\mathbb R}}

\title{
Non-abelian Toda theory on AdS$_2$\\  and   \chads duality  }

\author[a]{Matteo Beccaria,} 
\author[b]{Hongliang Jiang,}
\author[c]{Arkady A. Tseytlin\footnote{
Also at the   Institute  for Theoretical and Mathematical Physics, Moscow  University
and  Lebedev Institute, Moscow.}} 

\affiliation[a]{Dipartimento di Matematica e Fisica Ennio De Giorgi,\\
Universit\`a del Salento \& INFN, Via Arnesano, 73100 Lecce, 
Italy} 
\affiliation[b]{Albert Einstein Center for Fundamental Physics, Institute for Theoretical Physics, 
University of Bern, \\ Sidlerstrasse 5, 3012 Bern, Switzerland}
\affiliation[c]{Blackett Laboratory, Imperial College, London SW7 2AZ, U.K.}
\emailAdd{matteo.beccaria@le.infn.it,  \ jiang@itp.unibe.ch, \ tseytlin@imperial.ac.uk,}

\abstract{\\
It was recently observed  that boundary correlators of the elementary scalar field  of 
the Liouville theory on AdS$_2$  background are the same (up to a non-trivial proportionality coefficient) 
as the correlators of the  chiral stress tensor of the  Liouville  CFT on the complex plane  restricted to the real line. 
The same relation generalizes  to the conformal abelian Toda theory:  boundary  correlators of Toda scalars on AdS$_2$
are directly related to  the correlation functions of the  chiral $\mc W$-symmetry  generators in the Toda   CFT
and thus are essentially controlled by the underlying infinite-dimensional symmetry.  These  may be viewed  as examples of 
AdS$_2$/CFT$_1$  duality where the CFT$_1$ is the  chiral half  of a 2d CFT;
 we shall refer to  this as 
 AdS$_{2}$/CFT$^{\sfrac{1}{2}}_{2}$.
In this paper we demonstrate that this duality applies also to the non-abelian Toda theory containing a Liouville scalar coupled
to a 2d $\sigma$-model originating from the $SL(2, \mathbb R)/U(1)$ gauged WZW model. 
Here the Liouville scalar is again dual to the chiral  stress tensor $T$ while the other two scalars are dual to the 
parafermionic  operators $V^\pm$ of the non-abelian  Toda CFT.
We explicitly   check the duality 
at the next-to-leading order in the large  central charge   expansion 
by matching the chiral  CFT correlators of $(T, V^+,V^-)$  (computed using a
  free field representation) with   the boundary correlators of the three  Toda scalars
given  by the  tree-level and one-loop Witten  diagrams  in  AdS$_{2}$.
}

\begin{document}


\begin{tabbing}
\hspace*{11.7cm} \=  \kill 
    \> Imperial-TP-AT-2019-03  
\end{tabbing}

\maketitle
\def \kv {\kappa_{_V}}

\def \ov  {\over}
\def \adf {AdS$_2$/CFT$^{1/2}_2$ }
\def \adff {AdS$_2$/CFT$_{2/2}$ }
\def \adfff {AdS$_{2}$/CFT$^{\sfrac{1}{2}}_{2}$}

\def \const {{\rm const}}\def \m {\mu} \def \n {\nu}\def \OO {{\cal O}}\def \b {\beta} \def \ed {\end{document}}
\def \bmxi {\xi} \def \iffa {\iffalse} 
\def \ci {\cite} \def \s  {\sigma}
\def\foot{\footnote}
\newcommand{\rf}[1]{(\ref{#1})}
\def\la{\label}
\def\l{\lambda}
\def \GS  {NAT\ } \def \te {\textstyle} 
\def \ha {{\te{1\ov 2}}}\def \del  {\partial } 
\def \adst  {{AdS$_2$ }}
\def \sm {$\s$-model }
\def \eg {{\em e.g.}}
\def \k {\kappa}
\def \QQ {V}
\def \g {\gamma}
 \def \dil {\varPhi}
 \def \no {\nonumber}
\def \chix {\zeta}
\def \TT {{\rm T}}
\def \dev {\de_{_{V}}} \def \XP {\Phi}


\section{Introduction and summary}



Quantum field theories in rigid AdS$_{2}$   background 
(studied from different perspectives,  {\em e.g.}, in  \cite{DHoker:1983zwg,DHoker:1983msr,Inami:1985di,Callan:1989em,
Zamolodchikov:2001ah,Carmi:2018qzm})   
 were recently  discussed in the context of  AdS$_{5}$/CFT$_{4}$: a
 conformal ``defect''   model  describing correlators
of operators inserted  on a  straight or circular  Wilson line
 \cite{
Drukker:2006xg,Giombi:2017cqn,Beccaria:2018ocq,
Beccaria:2019dws}
is  represented  at   strong coupling  by an effective 2d $\sigma$-model  in  AdS$_2$ background
that follows   from the AdS$_{5}\times S^5$   superstring action expanded near the corresponding minimal surface. 
 To find   strong coupling corrections  to Wilson line 
 correlators  requires  computing loop corrections in 
AdS$_{2}$ \cite{Giombi:2017cqn,Beccaria:2019dws}; this is, in general,  a challenging problem (cf.  \eg  
\ci{Giombi:2017hpr,Bertan:2018afl,Yuan:2018qva,Liu:2018jhs}  and refs. there).

\iffa 
At strong coupling, one starts with the string action in  AdS$_{5}\times S^5$  and  gets a 2d $\sigma$-model
action in  AdS$_2$ background upon expansion around a suitable minimal surface associated with the Wilson line \cite{Giombi:2017cqn}. In this setup, boundary correlators  of elementary 2d excitations are expected to match the strong coupling limit of  
"defect" correlators inserted along the Wilson loop living in the 4d $\mathcal N=4$ super Yang-Mills theory.
Such correlators are covariant under the isometry of AdS$_2$ that is the 
1d conformal group $SO(2,1)\simeq SL(2,\RR)$. In principle, for non-protected observables, 
it is possible to systematically work out subleading strong coupling corrections by computing loop corrections in 
AdS$_{2}$ \cite{Giombi:2017cqn,Beccaria:2019dws}. This task turns out to be the theoretical bottle-neck due to the 
limited number of available tools, despite intense recent efforts in this direction \cite{
\fi

As the    theory   in AdS$_{2}$   originating    \ci{Giombi:2017cqn}   from the AdS$_{5}\times S^5$   superstring action
should  be quantum 
  scale-invariant (having no 2d UV divergences), one may hope to learn some important lessons by 
first  investigating  simpler   examples of   Lagrangian conformal 2d field theories  (like Liouville or Toda) defined on 
  rigid AdS$_2$ background.  
  Having conformal (Weyl) invariance, one  may expect the  bulk correlators 
    in  curved conformally flat space like  AdS$_2$  to be directly related to 
    the correlators in flat space. 
    A  novel feature in non-compact AdS space 
    is that while the elementary  fields (like scalars of Toda theory) are not good conformal fields in flat space,  their  boundary correlators in AdS$_2$ 
   (which are the  observables in AdS)  are well defined  and thus are of interest. 
   
   Somewhat surprisingly, they happen to be   directly related  
   \cite{Ouyang:2019xdd,Beccaria:2019stp,Beccaria:2019ibr}   to the 
   correlators of  chiral (holomorphic)  primary operators   in the 2d  CFT  defined by the same action in flat space. 
   For example, the boundary correlators of the Liouville scalar in \adst  
  have the same form as  the correlators 
   of the holomorphic stress tensor on a plane and thus are essentially controlled by the underlying Virasoro  symmetry.  
   Thus  here   the chiral half of 2d CFT   may be identified with   the effective 
   1d CFT  dual to the conformal field theory  in AdS$_2$.
    We   shall refer to this  relation as  the   \chads duality.\foot{This is thus an example 
     of  $\text{AdS}_{2}/\text{CFT}_{1}$ where the  boundary 1d conformal symmetry is 
enhanced from $SL(2, \mathbb R)$ to the chiral  half  of the  2d Virasoro symmetry. 
This   happens  when  a Weyl-covariant theory is put on a  rigid AdS$_{2}$ background.
While superficially similar,  this  is to be distinguished from what happens 
when one  considers quantum gravity in AdS$_{2}$ \cite{Strominger:1998yg}
where the bulk diffeomorphism symmetry implies  an enlarged asymptotic global symmetry
corresponding to 1d  reparametrizations  or Virasoro symmetry  in  the boundary theory 
 \cite{Hotta:1998iq,Cadoni:1999ja}   (which may be  spontaneously broken 
 \cite{Almheiri:2014cka,Jensen:2016pah,Maldacena:2016upp,Engelsoy:2016xyb}).
      }
    
 Examples of models exhibiting \chads  duality are  conformal 
abelian Toda field theories with exponential potentials  associated with a finite 
Lie algebra $\mathfrak{g}$.
This  correspondence   was  first noticed  at the level of the classical AdS$_2$ theory 
  (or large $c$  CFT)    in \cite{Ouyang:2019xdd} 
for  the $\mathfrak{g}=A_{1}$ Liouville theory and  also  in the particular rank-2 examples of $\mathfrak{g}=A_{2}$  and  $B_{2}$. 
The case of   $\mathfrak{g}=A_{n}$  was  discussed  in  \cite{Beccaria:2019ibr}. 
The   generalization  to \adst loop level (or subleading $1/c$ corrections) was presented 
for the Liouville theory  in \cite{Beccaria:2019stp} where  also the exact   expression  for the map between 
the   \adst  scalar and  CFT stress tensor  correlators was  found.
Below we shall    extend the Liouville theory 
  loop-level results of \cite{Beccaria:2019stp} to the case of 
 the  $A_{2}$ Toda theory.

One  reason  why the  duality between  the elementary scalars in \adst ($ds^2 = { d\bz^2 + d \bt^2 \ov \bz^2}$) and chiral  CFT operators on the plane   is possible 
is that expanded near  the minimum of its potential 
  the  \adst   action describes massive  ($m^2= \Delta(\Delta -1)$) scalar fields $\vp_\Delta$ 
that   should correspond to the dimension $\Delta=2, 3, ...$ operators $V_\Delta$
  at the boundary  (with $V_2\equiv T$). 
  Another is that the 
Weyl symmetry of  the \adst theory suggests   enhancement  of the global  1d conformal $SL(2, \mathbb R)$ symmetry to the  Virasoro symmetry.
 Then  the boundary correlators of $\vp_\Delta$ (with $\bz \to 0$ asymptotics $ \bz^\Delta \Phi_\Delta$)
that we shall  denote as  
$\llangle \Phi_{\Delta_1} (\bt_{1})... \Phi_{\Delta_k} (\bt_{k})   \rrangle$
may be related (on symmetry grounds) to the chiral CFT correlators  
$\langle  V_{\Delta_1} (z_{1})... V_{\Delta_k} (z_{k})   \rangle$
restricted to the boundary of half-plane ($z= \bt + i y \to \bt$). 
A non-trivial question  is a mechanism that  determines 
 the proportionality coefficients    in this duality  relation 
(i.e.,    symbolically,  $\k_\Delta$    in $\Phi_{\Delta}\to \k_\Delta  V_\Delta$).
Using the Weyl invariance 
of the \adst  theory one may  map  it to the flat-space theory  on the upper half-plane 
 and then try to  relate the half-plane 
 boundary asymptotics of the scalar  fields to the  CFT operators using their    free-field (quantum Miura) representation. 
The  free fields are, in general,   related to the elementary bulk scalars 
  by a  non-linear differential (quantum) B\"acklund transformation,
   but this relation should  effectively simplify in  the boundary limit. Details of this remain to be understood.

\iffa 
A non-trivial aspect of the correspondence regards its precise dictionary. When the fields in a conformally covariant bulk correlator 
are moved to the \adst boundary, the associated limit inherits conformal invariance in some sense. However, it is by no means
obvious how to exploit that in practice. The reason is that the AdS local fields do not appear to be  natural objects from the point
of view of the  bulk CFT. For instance, chiral and anti-chiral symmetry generators admit simple free field quantum Miura representations, but the 
involved free fields are related to the AdS ones by non-linear differential (quantum) B\"acklund transformations.
This implies that trying to exploit extended Virasoro symmetry to evaluate AdS local correlators seems hopeless. Nevertheless, 
boundary AdS correlators will turn out to have a simple and direct connection with the generators of the chiral half of the 2d CFT,
guided by the usual mass parameter/scaling dimension correspondence that works also in higher dimensions.
All the dictionary  will reduce to the determination of coupling dependent coefficients relating boundary AdS fields and 
chiral algebra generators, see Fig.~(\ref{fig:intro}).
\begin{figure}[htb]
\centering
\begin{tikzpicture}[line width=1 pt, scale=0.8, rotate=0, baseline=0]
\draw[dotted] (0,0) circle (2);
\coordinate (P1) at (160:1);    \draw[fill=black] (P1) circle (0.06); \node[above,yshift=0.1cm] at (P1) {$\varphi_{\de=2}$};
\coordinate (P2) at (40:1);   \draw[fill=black] (P2) circle (0.06); \node[above,yshift=0.1cm] at (P2) {$\varphi_{\de'}$};
\coordinate (P3) at (-40:1.5);   \draw[fill=black] (P3) circle (0.06); \node[above,yshift=0.1cm] at (P3) {$\varphi_{\de''}$};

\draw[thin,->] (160:1.2)--(160:1.9);
\draw[thin,->] (40:1.2)--(40:1.9);
\draw[thin,->] (-40:1.65)--(-40:1.9);

\node at (0,0) {$\text{AdS}_{2}$};
\node[scale=1.5] at (0:4) {$\Rightarrow$};
\begin{scope}[shift={(8,0)}]
   \draw[dotted] (0,0) circle (2);
   \node at (0,0) {$\text{AdS}_{2}$};
   \coordinate (P1) at (160:2);    \draw[fill=black] (P1) circle (0.06); 
   \node[xshift=-1.3cm,yshift=0.5cm] at (P1) {$\Phi_{\de=2}(\bt) = \kappa_{2}\,T(\bt)$};
   \coordinate (P2) at (40:2);  
   \draw[fill=black] (P2) circle (0.06); \node[right,xshift=0.3cm] at (P2) {$\Phi_{\de'}(\bt') = \kappa_{\de'}\,V_{\de'}(\bt')$};
   \coordinate (P3) at (-40:2);   
   \draw[fill=black] (P3) circle (0.06); \node[right,xshift=0.3cm] at (P3) {$\Phi_{\de''}(\bt'') = 
   \kappa_{\de''}\,V_{\de''}(\bt'')$};
   \node[xshift=-1cm] at (230:2) {$(\text{CFT}_{2})^{\sfrac{1}{2}}$};
\end{scope}
\end{tikzpicture}
\caption{{\bf Pictorial representation of  \chads correspondence.} On the left, we show some local fields $\varphi_{\de}$ in the bulk of 
\adst .  The parameter $\de$ enter the mass term in the usual way. For instance, for a scalar $m^{2}=\de(\de-1)$. In a Weyl invariant theory, their
correlator is in principle governed by a boundary CFT$_{2}$, after mapping to the flat upper half plane, although they are not good conformal 
fields -- natural objects being vertex operators or free fields related by B\"acklund transformations. 
On the right, we have moved the fields to the boundary of \adst and denoted 
$\varphi_{\de}\sim \bz^{\de}\Phi(\bt)$ where $\bz$ is the bulk \adst coordinate and $\bt$ parametrizes $\partial \text{AdS}_{2}$. 
$\Phi_{\de}$ can be fully identified with holomorphic chiral generators $V_{\de}$
with conformal dimension $\de$ and restricted to the 1d boundary. 
One of the $\de=2$ generators is to be identified with the holomorphic stress tensor $T$. 
The AdS boundary correlators are captured by chiral correlators of the $V_{\de}$.
In this $\text{(CFT)}_{2}^{\sfrac{1}{2}}$, the dictionary $\Phi_{\de}\to V_{\de}$ amounts to a coupling dependent 
coefficient $\kappa_{\de}$.
}
\label{fig:intro}
\end{figure}
    To check the \chads   correspondence   one may   compute the 
    boundary correlators  in perturbation theory (expanding in coupling  or  large central charge $c$  and 
      using Witten diagrams in AdS$_2$)  and compare to the correlators of the corresponding 
      chiral primary operators 
      in CFT$_2$   with their arguments  restricted to the boundary of  half-plane. 
  \fi  

\


The main  focus of the present   paper will be on  the  demonstration of  the \chads  duality in 
the case of  the non-abelian conformal  Toda  theory of \ci{Gervais:1992bs}. 
Here the  derivative   part  of the Lagrangian is no longer free but is 
described  by an effective  \sm  originating from the $SL(2,\RR)/U(1)$ gauged 
WZW model \ci{Bardacki:1990wj,Witten:1991yr}. This theory     comes   closer to the string-theory related \adst  model
 in \ci{Giombi:2017cqn} (containing  derivative interactions)
and     turns out to be much more non-trivial  than the abelian Toda one.  
 

\iffa 
Apart from this, our analysis may be useful to collect further information on \chads itself. In principle, 
placing a Weyl-covariant CFT in a conformally flat space does not seem to promise anything new. Nevertheless, 
AdS is special being non-compact and it provides boundary correlators as natural observables, in the standard
AdS/CFT spirit. This opens the way to new results to be obtained in well-studied models. From this point of view, 
specific features of  Liouville/Toda-type theories, like fields with non-trivial conformal transformations, 
suggests that theyre may be an interesting story  to be uncovered. In general, one should understand 
what is truly necessary for \chads to apply. For instance, the considered conformal theories in AdS$_{2}$ have a chiral
decomposition. while string with RR  background does not. Also, it would be important to understand whether there is a higher dimensional  analogue by considering, for instance $\mc N=4$ SYM on AdS$_{4}$. It is not clear whether some relation to a
3d CFT is to be found, {\em cf.} \cite{Beccaria:2016tqy}. That some complication is to be expected may 
be clear from the fact that in higher dimensional non-gravitational theories, a universal bulk field dual to the stress tensor 
is not present in general. Should one try to investigate the 4d case by starting from some kind of higher dimensional 
Liouville theory ? 
\fi 

Below we shall  first review (in  section~\ref{sec:intro:liouv}) 
the   \chads duality    in the case of  the Liouville  theory   and then  summarize   the analogous  statements 
  for the abelian Toda theory in  section~\ref{sec:intro:abel}. 
A  summary of  our   results  for  the non-abelian Toda theory    will 
be given in section~\ref{sec:intro:GS}.

Section \ref{sec:abel}  will  contain details of  new one-loop \adst   boundary correlator 
 computations  in the $A_2$ abelian Toda  model 
providing solid  check of the  \chads duality  in this case. 
Section \ref{sec:non}  will be devoted to  a systematic discussion of the duality in the  non-abelian Toda theory. 
In Appendix \ref{app:kappa3} we will present
a heuristic proposal for the all-order proportionality coefficient  between  the boundary correlators of the 
second  scalar of the $A_{2}$ abelian Toda theory in \adst 
and  correlators  of the dual spin 3 chiral generator in the CFT.
 In Appendix \ref{app:fourier} we will  collect useful relations  
that are used to compute some  \adst  integrals. 
Appendices \ref{app:3p} and \ref{app:c2pm} will contain details  of the  one-loop  \adst 
calculations of the coefficients in the  
three-point boundary correlators   that are required  to check the \chads  duality 
in the $A_{2}$ abelian  and non-abelian  Toda theories.

\subsection{Liouville theory}
\la{sec:intro:liouv}

The action of the Liouville theory defined  on  a 2d space  with curvature $R$  is 
 \cite{Polyakov:1981rd,
 Nakayama:2004vk}
\be
\la{1.1}
\mc S = \frac{1}{4\pi}\int d^{2}x\,\sqrt{g}\, \big[(\partial_a\varphi)^{2}
+\mu^2\,e^{2\,b\,\varphi}+Q\,R\,\varphi\big]\ ,\qquad\qquad  Q=
{b}^{-1}+b\  .
\ee
For the above value of $Q$  this model is Weyl-covariant with
 central charge 
 \be \la{1.2}  c=1+6\,Q^{2}=  b^{-2} + 13 + b^2  \ . \ee
 Considering the  unit-radius  Euclidean AdS$_{2}$   background  with $x=(\bt, \bz)$  and the Poincar\'e plane metric
 \be  \la{1.3} 
 ds^{2} = \frac{1}{\bz^{2}}(d\bt^{2}+d\bz^{2}) \ , \ \ \ \ \ \ \qquad     R=-2 \ , \ee 
 the Liouville  field $\vp$  can be expanded near its constant vacuum  expectation value, 
 \be  \vp = \vp_{0}+\chix \ , \qquad \qquad \vp_0= { 1 \ov 2 b} \log { Q\ov b\mu^2} \ ,  \la{1.4} \ee
  and  then the  fluctuation $\chix$ has   classical  mass $m^2= 2$. 
  Perturbation theory in the \adst bulk was  studied previously in 
 \cite{DHoker:1983msr, Zamolodchikov:2001ah,Menotti:2004uq}.  One   may  also compute \cite{Beccaria:2019stp}
 the boundary correlators of $\chix$ (relevant from the usual AdS/CFT point of view) 
 by    assuming the 
  Dirichlet  boundary conditions  for $\chix$ at the boundary line 
  $\bz=0$, i.e.  
   \be \la{1.5} 
   \chix(\bt, \bz)\big|_{\bz\to 0} =\bz^2 \Phi (\bt) + ...\ . \ee
    The 1d field $\Phi(\bt)$ is  associated  to a  boundary conformal 
  operator  
    with the scaling dimension $\Delta=2$ 
  (here $m^{2}=\Delta(\Delta-1)=2$). 
  Then the boundary correlators are defined as
\be 
\la{1.6}
\llangle \prod_{i=1}^{N}\Phi(\bt_{i})\rrangle \equiv
    \lim_{\bz_i\to 0} \, \langle\prod_{i=1}^{N}\bz_{i}^{-2}\,\chix(\bt_{i}, \bz_{i})\rangle.
    \ee
As was noticed in \cite{Ouyang:2019xdd}, the tree level ($b \to 0$) 
Witten diagrams computing  (\ref{1.6}) in perturbation theory  
match the leading large $c$  limit of the 
 correlators of the chiral  part of stress tensor $T(z)$  of the flat-space CFT (the generator of the Virasoro algebra 
 with  the central charge $c$ given  by  \rf{1.2})
  \be
\la{1.7}
\llangle \prod_{i=1}^{N}\Phi(\bt_{i})\rrangle = \kappa^{N}\, \langle \prod_{i=1}^{N}T(z_{i})
\rangle\Big|_{z_{i}\to \bt_{i}} \ , 
\ee
where $\k=\kappa(b) $ is the  constant in the   identification $\Phi(\bt) \to  \kappa \, T(\bt)$.  
Here the r.h.s. may be viewed as a 
 chiral stress tensor correlator restricted to the real-line  boundary  of the half-plane 
$z_i=\bt_i + i y_i \to \bt_i$. 
The relation (\ref{1.7}) was   demonstrated also to 
hold   \cite{Beccaria:2019stp}  at the one-loop in \adst (i.e. subleading order in large $c$ expansion).
Furthermore, it was argued in 
\cite{Beccaria:2019stp} using boundary CFT considerations 
 that the  
all-order expression for $\kappa(b)$ is  given by  
\be
\la{1.8}
\kappa  = -\frac{4Q}{c} = - \frac{ 4\,b  (1 + b^2)}{(3+2\,b^{2})(2+3\,b^{2})} = 
-\frac{2}{3}\,b+\frac{7}{9}\,b^{3}+\cdots.
\ee
Using that 
\be  \la{180}
\langle T(z_1) T(z_2) \rangle = {c \ov 2z_{12}^4} \ , \qquad \qquad 
\langle T(z_1) T(z_2) T(z_3)  \rangle = {c \ov z_{12}^2 z_{23}^2z_{31}^2} \ , \ee 
 the duality  relation \rf{1.7}  means that the  coefficients 
 in the perturbative expansion of the two-point and three-point  boundary correlators 
(with the structure controlled by the conformal symmetry)\foot{Indices of coefficients indicate dimensions of  operators involved.}
\be\la{1.9}
\llangle \Phi(\bt_{1})\,\Phi(\bt_{2})\,\rrangle  = \frac{C_{22}}{\bt_{12}^{4}},\qquad \qquad \qquad 
\llangle \Phi(\bt_{1})\,\Phi(\bt_{2})\,\Phi(\bt_{3})\,\rrangle  = \frac{C_{222}}{\bt_{12}^{2}\,\bt_{13}^{2}\,
\bt_{23}^{2}},
\ee
should be given by 
\begin{align}
C_{22} &= \kappa^{2}\,\frac{c}{2} = \frac{8\,(1+b^{2})^{2}}{(3+2b^{2})(2+4b^{2})} = 
\frac{4}{3}-\frac{2}{9}\,b^{2}+\frac{13}{27}\,b^{4}+\cdots, 
\notag \\
C_{222} &= \kappa^{3}\,c = -\frac{64\,b\,(1+b^{2})^{3}}{(3+2b^{2})^{2}(2+3b^{2})^{2}} = 
-\frac{16}{9}\,b+\frac{64}{27}\,b^{3}-\frac{100}{27}\,b^{5}+\cdots \ . \la{1.10}
\end{align}
These   expansions were   checked   directly  (for the leading  tree and one-loop terms) in \cite{Beccaria:2019stp}. 
The four-point  boundary correlators   may be decomposed into  the  disconnected and connected parts 
\begin{align}\la{1.11}
\llangle \Phi(\bt_{1})\,\cdots\,\Phi(\bt_{4})\rrangle  = 
\llangle \Phi(\bt_{1})\,\cdots\,\Phi(\bt_{4})\rrangle_{\rm disc}  +
\llangle \Phi(\bt_{1})\,\cdots\,\Phi(\bt_{4})\rrangle_{\rm conn}\ . 
\end{align}
Since the four-point function for 
the chiral part of the stress
tensor  $T$   is controlled by the Virasoro symmetry
 \be
 \la{1333}
 \langle T(z_{1}) \cdots  T(z_{4})\rangle = \frac{c^{2}}{4}\,\Big(\frac{1}{z_{12}^{4}\,z_{34}^{4}}
+\frac{1}{z_{13}^{4}\,z_{24}^{4}}+\frac{1}{z_{14}^{4}\,z_{23}^{4}}\Big)
+c\,\Big(
\frac{1}{z_{12}^{2}\,z_{23}^{2}\,z_{34}^{2}\,z_{14}^{2}}
+\frac{1}{z_{13}^{2}\,z_{24}^{2}\,z_{14}^{2}\,z_{23}^{2}}
+\frac{1}{z_{12}^{2}\,z_{24}^{2}\,z_{34}^{2}\,z_{13}^{2}}
\Big)
\ee
the relation (\ref{1.7}) implies that
 \begin{align}
 \la{1.12}
\llangle \Phi(\bt_{1})\cdots \Phi(\bt_{4})\rrangle_{\rm disc} &= (C_{22})^2\,\Big(
\frac{1}{\bt_{12}^{2}\,\bt_{34}^{2}}
+\frac{1}{\bt_{13}^{2}\,\bt_{24}^{2}}
+\frac{1}{\bt_{14}^{2}\,\bt_{23}^{2}}\Big) \ ,  \notag \\
\llangle \Phi(\bt_{1})\cdots \Phi(\bt_{4})\rrangle_{\rm conn} &= C_{2222}\,\Big(
\frac{1}{\bt_{12}^{2}\,\bt_{23}^{2}\,\bt_{34}^{2}\,\bt_{14}^{2}}
+\frac{1}{\bt_{13}^{2}\,\bt_{24}^{2}\,\bt_{14}^{2}\,\bt_{23}^{2}}
+\frac{1}{\bt_{12}^{2}\,\bt_{24}^{2}\,\bt_{34}^{2}\,\bt_{13}^{2}}
\Big)\ , 
\end{align}
where according to \rf{1.8}
\be\la{1.13}
C_{2222} = \kappa^{4}\,c = \frac{256\,b^{2}\,(1+b^{2})^{4}}{(3+2b^{2})^{3}(2+3b^{2})^{3}} = 
\frac{32}{27}\,b^{2}-\frac{80}{27}\,b^{4}+\cdots\ .
\ee
This  was  also  verified  in \cite{Beccaria:2019stp} (using  numerical computation for  one-loop integrals
appearing in the computation of the four-point boundary correlator \rf{1.12}). 

To conclude, the relation \rf{1.7}   found in the 
 Liouville  model  in \adst  provides  the  simplest  example  of  the \chads  duality. 
 The fact that the bulk theory is conformal  implies that the structure of the  boundary correlators 
 is  essentially fixed by the  Virasoro symmetry.\foot{In particular, 
 the four-point  boundary correlators  of operators that have protected dimension 
   here do not contain    logarithms of 1d cross ratio, etc.} 
 While   this duality  may be viewed as  being essentially  kinematical, 
 the non-trivial   expression for $\k$ in \rf{1.8} receiving corrections   from  all  orders in the small  $b$ expansion 
 provides an important constraint on how the  higher-loop  Witten \adst diagrams are to be evaluated 
 in order to maintain the  underlying infinite-dimensional Virasoro symmetry.

\subsection{Abelian Toda theory}
\la{sec:intro:abel}

The generalization to  abelian Toda theory  for $A_{2}$ and $B_{2}$ algebras was discussed 
in \cite{Ouyang:2019xdd} and for $A_{n}$  algebras  in \cite{Beccaria:2019ibr}. 
A novel feature is that in addition to the Liouville   field  here the Lagrangian  contains  other scalar fields 
that  are massive in  \adst  and are dual to the chiral  generators of the $\mc W_n$  symmetry of the
 Toda theory.\foot{Toda field theories associated with 
 a  finite (non-affine) Lie algebra $\frak g$ of  rank $n$ 
(see, \eg, \cite{Leznov:1979td,
Bilal:1988jg,Hollowood:1989ep}) 
may be formulated in curved space,  see,  for instance,   
 \cite{Fateev:2007ab}.}
 
In section \ref{sec:abel} of  this paper  we will  extend  the discussion of the  corresponding 
\chads  duality to  the one-loop level on the example  of the $A_{2}$ theory
with main results summarized below. 

The  action of the $A_{2}$    abelian Toda theory  in curved  2d background contains in addition to 
the Liouville field $\vp$ another scalar $\psi$, i.e. 
is given by  the following generalization of \rf{1.1} 
\begin{align}
\la{1.14}
&\mc S = \frac{1}{4\pi}\int d^{2}x\,\sqrt{g}\, \big[(\partial_a\varphi )^{2} + (\partial_a\psi)^{2}
+\mu^2\,e^{2b\varphi}\cosh(2\sqrt{3}\, b\, \psi)+Q\,R\,\varphi \big]\ , \\
&\qquad \qquad  Q={b}^{-1}
+4\,b \ ,     \qquad \qquad  c=2+6\,Q^{2} \ . \la{1140} 
\end{align}
This  theory  is  Weyl-covariant at the quantum   level: the required value of $Q$ in \rf{1140}
 can be determined, \eg, by viewing 
  \rf{1.14}  as a string model in a linear dilaton and tachyon backgrounds   and solving the
   corresponding tachyon $\b$-function equation \ci{Tseytlin:1990mz,Grisaru:1990gf}.
  
 Considering the \adst background \rf{1.3}   and expanding near the  constant vacuum   value for $\varphi$  as in the Liouville theory
 (with $\vp_0$   given again by \rf{1.4}) 
   one can develop the perturbation theory in small $b$, i.e.  in powers of  the fluctuation fields $ \chix \equiv \vp - \vp_0$
  and $\psi$. 
 These happen  to have 
 masses    $m^2_\chix=2$  and $m^2_\psi= 6$  corresponding (according to   $m^{2}=\de(\de-1)$) to the 
 dual operator dimensions $\Delta_\zeta=2$ and $\Delta_\psi=3$. 
 Let us    label  the   boundary   fields  as $ \Phi $ and $\Phi _3$ (cf. \rf{1.5})\foot{The index of $\Phi_3$  indicates the dimension of  the associated operator (we omit index $2$ on the Liouville  field $\Phi$).}
  \be \la{1.15} 
   \chix(\bt, \bz)\big|_{\bz\to 0} =\bz^2 \Phi (\bt) + ...\ , \qquad \qquad \psi(\bt, \bz)\big|_{\bz\to 0} =\bz^3 \Phi_3 (\bt) + ...\  .
    \ee
 In this case the dual   CFT is isomorphic to 
  the chiral sector of the $\mc W_{3}$ extension of the Virasoro algebra  \cite{Zamolodchikov:1985wn,Bouwknegt:1992wg}, 
  with $\Phi$  dual to the  dimension   2  stress tensor $T$ and $\Phi_3$  to the  dimension (or spin)    3  generator $\QQ_3$
  of the $\mc W_{3}$  symmetry  algebra.
Defining the boundary correlators as in \rf{1.6} 
\be
\la{1.16}
\llangle \prod_{i=1}^{N}\Phi(\bt_{i})\,\prod_{j=1}^{M}\Phi_{3}(\bt'_{j}) \rrangle
    \equiv  \lim_{\bz_{i},\bz'_{j}\to 0} \langle\, 
    \prod_{i=1}^{N} \bz_{i}^{-2}\chix(\bt_{i}, \bz_{i})\,\prod_{j=1}^{M}\bz_{j}'^{-3} 
    \psi(\bt'_{j}, \bz'_{j})\, \rangle,
\ee
 the expected correspondence is expressed by  a  generalization of     (\ref{1.7}), {\em i.e.}
\begin{align}
\la{1.17}
\llangle \prod_{i=1}^{N}\Phi(\bt_{i})\,\prod_{j=1}^{M}\Phi_{3}(\bt'_{j}) \rrangle
= \kappa^{N}\,\kappa_{3}^{M}\,
\langle \prod_{i=1}^{N}T(z_{i})\,\prod_{j=1}^{M}\QQ_3(z'_{j}) \rangle\Big|_{z_i,z'_j\to \bt_i,\bt'_j} \ .
\end{align}
Here $\kappa$  has the same form  as    in the Liouville theory 
 (cf. \rf{1.8}) and    the coefficient $\kappa_3$  in the duality between 
  $\Phi_3$ and $\QQ_3$   turns out to be   a non-trivial function of $b$ 
\be
\la{1.18}
\kappa = -\frac{4\,Q}{c} = -\frac{2}{3}\,b+\frac{26}{9}\,b^{3}+\cdots,\qquad 
\qquad \kappa_{3} = \frac{24\,Q^{2}}{c\,\sqrt{5c+22}} = \frac{2\sqrt{2}}{\sqrt{15}}\,\big(b-\frac{73}{15}\,b^{3}+
\cdots\big).
\ee
We  shall  verify   (\ref{1.17}) and (\ref{1.18})  by the one-loop \adst   computations  in section \ref{sec:abel}
and present an  argument for the  above  expression for $\k_3$ in Appendix \ref{app:kappa3}. 

The CFT   2- and 3-point functions are 
constrained by the conformal invariance to have the form (we adopt the standard normalization for the spin 3 
generator $\QQ_3$)\footnote{\la{foot:3p}In general, given  a primary field $V_{\de}$ with the dimension $\de$ 
and the two-point function $\langle V_{\de}(z_{1})\,V_{\de}(z_{2})\rangle = {C\ov z_{12}^{2\de}}$ the conformal 
symmetry Ward identity
implies that 
 $\langle T(z_{1})V_{\de}(z_{2})V_{\de}(z_{3})\rangle = {C\,\de\ov z_{12}^{2}\,z_{13}^{2}\,z_{23}^{2\de-2}}$.
 This  is consistent with  the  values of the coefficients in \rf{1.19} 
 (note that  in the  $\langle TTT\rangle$  case  the central term in the  OPE of $T(z)T(0)$  does not contribute
to the three-point function as $\langle T \rangle=0$).
}
%
\begin{align}
\langle T(z_1) T(z_2)\rangle &= \frac{c}{2\,z_{12}^{4}}, & \langle T(z_{1}) T(z_{2})T(z_{3})\rangle &=
 \frac{c}{z_{12}^{2}z_{13}^{2}z_{23}^{2}},
\notag \\
\langle \QQ_3(z_1) \QQ_3(z_2)\rangle &=\frac{c}{3\,z_{12}^{6}}, &  
\langle T(z_{1}) \QQ_3(z_{2}) \QQ_3(z_{3}) \rangle &= \frac{c}{z_{12}^{2}z_{13}^{2}z_{23}^{4}}\ . \la{1.19} 
\end{align}
The corresponding boundary  correlators   can be parametrized as  in \rf{1.9} 
\begin{align}
\llangle \Phi(\bt_1) \Phi(\bt_2)\rrangle &= \frac{C_{22}}{\bt_{12}^{4}}, & 
\llangle \Phi(\bt_{1}) \Phi(\bt_{2}) \Phi(\bt_{3}) \rrangle &= \frac{C_{222}}
{\bt_{12}^{2}\bt_{13}^{2}\bt_{23}^{2}},\notag \\
\llangle \Phi_{3}(\bt_1) \Phi_{3}(\bt_2)\rrangle &=\frac{C_{33}}{\bt_{12}^{6}}, &  
\llangle \Phi(\bt_{1}) \Phi_{3}(\bt_{2}) \Phi_{3}(\bt_{3}) \rrangle &= \frac{C_{233}}
{\bt_{12}^{2}\bt_{13}^{2}\bt_{23}^{4}} \ .  \la{1.20}
\end{align}
We shall explicitly check the relations implied   by \rf{1.17},\rf{1.18} 
\begin{align}
\la{1.21}
C_{22} &= \frac{c}{2}\,\kappa^{2} = \frac{4}{3}-\frac{4}{9}\,b^{2}+\cdots,  &
C_{33} &= \frac{c}{3}\,\kappa_{3}^{2} = \frac{16}{15}-\frac{112}{75}\,b^{2}+\cdots,  \notag \\
C_{233} &= c\,\kappa\,\kappa_{3}^{2} = -\frac{32}{15}\,b+\frac{2752}{225}\,b^{3}+\cdots, &
C_{222} &= c\,\kappa^{3} =  -\frac{16}{9}\,b+\frac{224}{27}\,b^{3}+\cdots,
\end{align}
by the one-loop  computations in \adst   in section~\ref{sec:abel}. 

The non-vanishing four-point  correlators  $\llangle \Phi\Phi\Phi\Phi\rrangle$, $\llangle\Phi\Phi\Phi_{3}\Phi_{3}\rrangle$,
and $\llangle\Phi_{3}\Phi_{3}\Phi_{3}\Phi_{3}\rrangle$ were  considered  at the tree level in \cite{Ouyang:2019xdd}
confirming the expected \chads relations. While  no conceptual difficulties are expected at the one-loop level
where the computation   is similar to the one in the Liouville theory  in \cite{Beccaria:2019stp}
here  will  not discuss it  because of technical limitations.\footnote{\la{foot:num} The  coefficients $C_{233}$ and $C_{222}$ 
in the three-point functions in \rf{1.20} 
are expressed in terms of   finite \adst  integrals. 
Most of them may be computed analytically, but a few have to be 
evaluated numerically because the available analytic tools turn out not to be  sufficient. 
In the case of the coefficients in the four-point correlators 
 the number of diagrams to be evaluated numerically is larger and the expected numerical accuracy 
 is  too low to be  reliable.}

Similar results  should hold also  for  higher rank abelian Toda models  where  there are more scalar fields 
with different masses corresponding to the higher-spin generators of the underlying 
$\mc W$-algebra  symmetry. 
 One implication of the  \chads  duality    \rf{1.17}  is that  it can then be used to  compute the  higher-loop \adst 
  boundary correlators for  the Toda theory    using 
 purely  $\mc W$-algebraic methods.

\subsection{Non-abelian  Toda theory}
\la{sec:intro:GS}

In the abelian Toda theory  the  chiral  CFT operators (conserved  currents) have protected dimensions 
so their correlators are constrained by $\mc W$  symmetry modulo overall normalizations  and thus the 
 corresponding \chads  duality  is a direct generalization of the Liouville theory case. 
 The story becomes   more intricate in the   case of the conformal non-abelian Toda (NAT)  theory \cite{Gervais:1992bs}
 (see also \ci{Bilal:1993rg,Bilal:1995ei}) that we shall discuss in detail  in section~\ref{sec:non} below.
 Here we will summarize the main results.

Keeping only the leading order (one-loop) terms in the corresponding  Weyl-invariant 
 target space  metric and dilaton (cf. \rf{3.8})   the NAT action is the following generalization  of the 
 Liouville  \rf{1.1} or abelian Toda  \rf{1.14}   actions
\begin{align}
\la{1.23}
\mc S &= \frac{1}{4\pi}\int d^{2}x\,\sqrt{g}\Big[
(\partial\varphi)^{2}+(\partial r)^{2}+\tanh^{2}(b r)\,(\partial y)^{2}\notag \\
&\qquad\qquad\qquad\ \ \ \  +\mu^{2}\, e^{2b\varphi}\cosh(2 b r)\,+R\,\Big(Q\varphi-\log\cosh(b r)\Big)\Big]\ , 
\qquad Q = {b}^{-1} +3b.
\end{align}
Here in addition to the Liouville field $\vp$ with linear dilaton term we have 
  the  $(r,y)$  sector with the kinetic term    originating  from the  $SL(2,\RR)/U(1)$ gauged WZW model and the dilaton term 
 $ R \log\cosh(b r)$ required for   the  Weyl invariance 
of this $\s$-model  \cite{Witten:1991yr}. The expression for  $Q$ is fixed  by the condition of 
  quantum Weyl invariance of the  potential term, i.e. by satisfaction of the tachyon equation \rf{3.10}
  \cite{Tseytlin:1990mz,Jack:1993de}. 
  The (exact) relations between  the constant $b$,  the WZW  level $k$ and the total central charge  are    (cf. \rf{3.9},\rf{3.17}) 
  \be
\la{1.24}
b = \frac{1}{\sqrt{k-2}}\ , \qquad \qquad c= 3 + 6 Q^2 + 6 b^2 = 6b^{-2} + 39 + b^2 \ , 
\ee
so that  the small $b$ expansion is the same as the  large $k$ or  large $c$ expansion. 

Expanding \rf{1.23}  around the minimum of the potential  for $\vp$ on the AdS$_{2}$  background  as in \rf{1.4},  
one finds the action for 3 massive fluctuation fields $\chix$ and $\xi_1,\xi_2$   (related to $r,y$  as in \rf{3.13} with $a=b$).
They have the same mass $m^2=2$ (to leading order in  $b$), i.e. should be dual to  
the  boundary operators  with the classical dimension 2. 
We shall denote  the corresponding 
 boundary fields as      $\Phi$   and $\Phi^{\pm}$ (cf. \rf{1.5},\rf{1.15}) 
 \be \la{1.25} 
   \chix(\bt, \bz)\big|_{\bz\to 0} =\bz^2 \Phi (\bt) + ...\ , \qquad  \xi^\pm (\bt, \bz)\big|_{\bz\to 0} =\bz^{\Delta_{_{V}}}  \Phi^\pm (\bt)
   + ...\ ,  \qquad \xi^\pm \equiv  {1\ov \sqrt 2} ( \xi_1 \pm  i \xi_2)
  \  ,  \ee
  where $\Delta_{_{V}}=2+ ...$ (anticipating possible anomalous dimension). 
At the  classical level, the \GS model  in flat space 
has three conserved  holomorphic currents with dimension 2:   the stress tensor $T_{\rm cl}$ and
a $U(1)$ doublet of ``parafermions'' $V^{\pm}_{\rm cl}$  \cite{Bilal:1993rg}  (generalizing the 
classical parafermions of the gWZW model 
\ci{Bardacki:1990wj}).\foot{Expressed in terms  of the  fields in the Lagrangian \rf{1.23} (the left-moving) 
 $V^{\pm}_{\rm cl}$ are   non-local:
$V^\pm_{\rm cl}=
V^\pm_{++} \equiv {1\over \sqrt{2}} \left( 2\del_+  \vp -\del_+ \right)
\left[e^{\pm i\nu}\left(\del_+  r\pm i \tanh br\, \del_+ y\right)\right]$, 
where $\nu$ is defined by $\del_- \nu=\cosh^{-2}br\, \del_-  y, \quad  \del_+ \nu=(1+\tanh^2br)\del_+ y$
and $ \del_- V^\pm_{\rm cl} =0$. 
Together  with $T$,  $V^\pm$ generate a kind of a non-local  generalization of $W$-algebra \ci{Bilal:1995ag}.
}
A natural suggestion is that the  corresponding  quantum operators 
$T$ and $V^{\pm}$  should be related, respectively,  to  the \adst  boundary   fields
 $\Phi$ and $\Phi^{\pm}$. 
To check the  \chads   duality in this case we  will   compute  the \adst boundary correlators 
using Witten diagrams and compare them to the   chiral  CFT correlators of $T$ and $V^{\pm}$. 

The CFT correlators 
   can be found  using  an explicit  free field  representation for  $T$ and  parafermions  
   $V^{\pm}$ that we shall present in section \ref{sec:3.2}.
A novel feature compared  to the abelian Toda theory is  that the dimension of the   primaries $V^{\pm}$  
is not protected, i.e. they 
have a non-zero anomalous  dimension 
\be
\la{1.26}
\Delta_{_{V}} = 2 + \gamma_{_{V}} \ ,  \qquad \qquad  \gamma_{_{V}}={ 1 \ov k} = \frac{b^{2}}{1+2b^{2}}\ .
\ee
Defining the boundary correlators as in \rf{1.6},\rf{1.16}\footnote{Here 
 we use  a short-cut notation:   
the  $\pm$ products are,  of course,   independent. 
 }
\be
\la{1.27}
\llangle \prod_{i=1}^{N}\Phi(\bt_{i})\,
\prod_{j^{\pm}=1}^{M^{\pm}}\Phi^{\pm}(\bt^{\pm}_{j}) \rrangle
    \equiv  \lim_{\bz_{i},\bz^{\pm}_{j}\to 0} \langle
    \prod_{i=1}^{N} \bz_{i}^{-2}\chix(\bt_{i}, \bz_{i})\,\prod_{j^{\pm}=1}^{M^{\pm}}
    (\bz^{\pm}_{j})^{-\de_{_{V}}} 
    \xi^{\pm}(\bt^{\pm}_{j}, \bz^{\pm}_{j})\rangle \ , 
\ee
 the  statement of the \chads  duality should   be (cf. \rf{1.7},\rf{1.17})
\begin{align}
\la{1.28}
\llangle \prod_{i=1}^{N}\Phi(\bt_{i})\,
\prod_{j^{\pm}=1}^{M^{\pm}}\Phi^{\pm}(\bt^{\pm}_{j}) \rrangle
 =  \kappa^{N}\,\kv^{M^{+}+M^{-}}\,
\langle \prod_{i=1}^{N} T(z_{i})\,\prod_{j^{\pm}=1}^{M^{\pm}}
V^{\pm}(z^{\pm}_{j}) \rangle\Big|_{z_i, z_j^{\pm}\to 
\bt_i, \bt^{\pm}_j},
\end{align}
where  $\kappa$ and $\kv$  are the  coefficients in the correspondence 
$\Phi \to \k T, \  \Phi^\pm \to \kv V^\pm$.

As  the dimension 2 of the stress tensor is protected, its  normalization is  universal:
$\langle T(z_1) T(z_2) \rangle = {c \ov 2z_{12}^4} $. As a result, as in the Liouville \rf{1.8} or the abelian 
Toda theory \rf{1.18},  the coefficient $\k$ 
 has  the scheme-independent  expression in terms of  $Q$ and $c$
\be\la{1.29}
\k = -\frac{4Q}{c} = -\frac{4\,b\,(1+3\,b^{2})}{3\,(1+4\,b^{2})(2+5\,b^{2})} = 
-\frac{2}{3}\,b+\frac{7}{3}\,b^{3}+\cdots\  .
\ee
At the same time,  as $V^\pm$   has the non-zero anomalous dimension \rf{1.26}, 
  the    coefficient $C^{+-}$  in  
the corresponding  2-point  function 
\be  \la{1.31}
\langle V^{+}(z_1) V^{-}(z_2)\rangle =  {C^{+-} \ov   z_{12}^{2\de_{_{V}}} } \ ,  \ee
is scheme-dependent (it, in general, 
 contains a factor  of $\Lambda^{-2\de_{_{V}}}$ where $\Lambda $ is a  renormalization  scale). 
We shall  assume  a scheme  in which $C^{+-}$  is given  by 
\be\la{131}    C^{+-}  =  {c \ov  \de_{_{V}}}    =      \frac{3k(k+2)}{k-2}\ ,  \ee
where  $k$ is the WZW level related to $b$ by (\ref{1.24}).
In this scheme our  proposed exact expression for $\kv$ is (cf. \rf{1.26} 
\be \la{1.30}
\kv  
                      ={ 2b\ov 3 (1+4b^2) }\,2^{\g_{_{V}}} = {2\ov 3}  b   -{2\ov 3}\big(4- 
                     \log 2\big)\, b^3 + \cdots \ .
 \ee
Representing  the \adst  boundary two-point  correlators as 
\be
\la{1.32}
\llangle \Phi(\bt_{1})\,\Phi(\bt_{2})\rrangle = \frac{C_{22}}{\bt_{12}^{4}},\qquad \qquad \qquad 
\llangle \Phi^{+}(\bt_{1})\,\Phi^{-}(\bt_{2})\rrangle = \frac{C_{+-}}{(\bt_{12}^{2})^{\de_{_{V}}}},
\ee
the correspondence    \rf{1.28} implies the  following relations 
\begin{align}
C_{22} &= \k^{2}\,\frac{c}{2} = \frac{8\,(1+3b^{2})^{2}}{3\,(1+4b^{2})(2+5b^{2})} = 
\frac{4}{3}-\frac{2}{3}\,b^{2}+\cdots\ , \la{1.33} \\
C_{+-} &= \kv^{2}\,  C^{+-} = \kv^{2}\,\frac{3\,(1+2b^{2})(1+4b^{2})}{b^{2}} = 
\frac{4}{3}+\frac{8}{3}\,(\log 2-1)\,b^{2}+\cdots\ . \la{1.34}
\end{align}
The  conformal invariance (Virasoro algebra)   fixes  the three-point functions 
 of the  chiral primary operators  to have the following form  
 (after using  the Ward identity in  footnote \ref{foot:3p} and (\ref{1.31}),(\ref{131}))
\be\la{1.35}
\EV{T(z_{1})T(z_{2})T(z_{3})} =  \frac{c}{z_{12}^{2}\, z_{13}^{2}\, z_{23}^{2}}\ ,\qquad \qquad \qquad 
\EV{T(z_{1}) V^{+}(z_2) V^{-}(z_3)} =
\frac{c}{z_{12}^2 \, z_{13}^{2}\, z_{23}^{2\de_{_{V}}-2} }\ .
\ee
With the \adst boundary three-point correlators written as
\be
\la{1.36}
\llangle \Phi(\bt_{1})\Phi(\bt_{2})\Phi(\bt_{3})\rrangle =  
\frac{C_{222}}{\bt_{12}^{2}\, \bt_{13}^{2}\, \bt_{23}^{2}}\ ,\qquad \qquad \qquad 
\llangle \Phi(\bt_{1}) \Phi^{+}(\bt_2) \Phi^{-}(\bt_3) \rrangle =
\frac{C_{2+-}}{\bt_{12}^2\,  \bt_{13}^{2}\, \bt_{23}^{2\de_{_{V}}-2} }\ ,
\ee
the duality   \rf{1.28}  then  rests on the validity of the following relations  
\begin{align}
\la{1.37}
C_{222} &= \k^{3}\,c = -\frac{64\,b\,(1+3b)^{3}}{9\,(1+4b^{2})^{2}(2+5b^{2})^{2}} = 
-\frac{16}{9}\,b+\frac{64}{9}\,b^{3}+\cdots, \notag \\
C_{2+-} &= \k\,\kv^{2}\,c = -C_{+-}\,\frac{4b\,(1+3b^{2})}{3\,(1+2b^{2})(1+4b^{2})}
= -\frac{16}{9}\,b+\frac{16}{9}\big(5 - 2\,\log 2\big)\,b^{3}+\cdots.
\end{align}
We shall verify  these predictions  in section~\ref{sec:non} by the  tree and one-loop Witten diagram
 calculations of \rf{1.36}   starting with the \adst action \rf{1.23}  in  expanded in powers of $b$. 
 This   provides strong evidence for the  consistency of the duality \rf{1.28}.

We will  also 
compute (at tree level)
  the non-vanishing four-point correlators: 
 $\llangle\Phi(\bt_{1})\Phi(\bt_{2})\Phi(\bt_{3})\Phi(\bt_{4})\rrangle$ (which has the same form as in \rf{1.11},\rf{1.12}, cf. \rf{1333})
  and also 
\be
\la{1.38}
\llangle\Phi(\bt_{1})\Phi(\bt_{2})\Phi^{+}(\bt_{3})\Phi^{-}(\bt_{4})\rrangle, \qquad \qquad 
\llangle\Phi^{+}(\bt_{1})\Phi^{-}(\bt_{2})\Phi^{+}(\bt_{3})\Phi^{-}(\bt_{4})\rrangle \ . 
 \ee
The  expressions for the latter are  only partially constrained by the  conformal invariance. 
Nevertheless, we  will  demonstrate  that  their 
dependence on the conformally invariant cross ratio, 
 the structure of  kinematical singularities
and the conformal block expansions are in full agreement with the  \chads relations \rf{1.28}. 
It
would be interesting to  evaluate   the four-point functions in (\ref{1.38}) also at the 
one-loop level, but like in the abelian Toda theory,  this would require  developing  more efficient
computational tools (see footnote \ref{foot:num}).

\section{$A_{2}$ abelian Toda theory}
\la{sec:abel}

 Here we  will   discuss  loop corrections
 in the $A_{2}$  Toda theory in \adst  with  the action (\ref{1.14})  with the aim to 
  test the duality  relation (\ref{1.17}). 
 The exact expressions   for the coefficients in (\ref{1.18}) can be argued for   by adapting the conformal Ward identity method
 used in the Liouville  theory case in  \cite{Beccaria:2019stp} as explained  in Appendix \ref{app:kappa3}.

\subsection{Perturbation theory}

As discussed in  detail in \cite{Beccaria:2019stp}, the perturbation theory of  an abelian 
Toda theory in \adst  may be set up in two natural ways that are equivalent at the level of the expressions for the 
boundary correlators. 
In the first approach,  proposed in the context of the Liouville theory in   \cite{Zamolodchikov:2001ah} (and thus 
named {\sf ZZ} in \cite{Beccaria:2019stp}),
one starts with the Toda action in 
flat space and expands around a non-trivial solution corresponding to an effective \adst geometry. 
 In the second 
({\sf AdS}) approach
the theory is put on \adst  background  from the  very beginning and 
the perturbation theory is  developed around the  constant  minimum of the effective 
potential that includes the  curvature coupling term. 
 Technically, the difference between the two schemes   happens to be  in the 
treatment of the short distance singularity of the Toda field propagator or the value of  $g_{\de}(z,z)$.
 In the {\sf ZZ}
scheme, this is a non-trivial quantity, while in the 
{\sf AdS} scheme  it is set to zero. Equivalence  between the two approaches is possible
thanks to special identities that ensure a 
compensation between the tadpole contributions (present in the {\sf ZZ} and absent in the {\sf AdS} case)
and the different effective  couplings in the two schemes.\footnote{In the Liouville theory, the basic identity
is in eq. (A.9) of  \cite{Beccaria:2019stp}. Completely similar results can be proved in the  $A_{2}$  Toda theory; 
 in general, this should be a consequence of the equations of motion.}

Here  we shall adopt the {\sf AdS} scheme 
which is simpler as  one can  simply  ignore all  tadpole contributions.
Thus we start with  the  action (\ref{1.14})
 on the AdS$_{2}$ background \rf{1.3} and  
 expand  $\vp$ near its vacuum value as in  \rf{1.4} 
\be
\la{2.1}
\mc S = \frac{1}{2\pi} \int d\bt\,d\bz\ \Big[ \frac12 (\p_a\chix)^2  + \frac12 (\p_a\psi)^2
 +  \frac{Q}{2 b\, {\sf z}^2 }\, {e^{2b\chix}
 \Big(  \cosh (2\sqrt 3 b \psi)  -2b \chix -1}\Big) 
    \Big]\ .
\ee
Here  the interaction terms in (\ref{2.1}) are multiplied by the common factor $\frac{Q}{b} = 
\frac{1}{b^{2}}(1+4b^{2})$ (see \rf{1140}). 
The  values of the masses   $m^2_\chix=2$, $m^2_\psi=6$  correspond to the 
dimensions of the boundary operators being  $\Delta_\chix=2$ and $\Delta_\psi=3$. 

In general,  the \adst propagator for a   free scalar field $\varphi_{\de}(x)= \vp_\de ( \bt,\bz)$   with mass 
$m^{2}=\de(\de-1)$  and kinetic term normalized as in  (\ref{2.1})
 is 
$\langle\varphi_{\de}(x)\varphi_{\de}(x')\rangle_{\rm free} = 2\,\pi\,G_{\de}(x, x')$ where 
%
 \be
 \la{2.2}
 G_\Delta = \frac{\mathcal C_\Delta}{(4u)^\Delta}\,   {}_2F_1(\Delta, \Delta, 2\Delta, -\frac{4}{u}), 
 \qquad
 \mathcal C_\Delta =\frac{\Gamma(\Delta)}{2\sqrt \pi \Gamma{(\Delta+1/2)}}, 
 \qquad
  u(x, x')=\frac{ ({\sf z} -{\sf z'})^2 + ({\sf t} -{\sf t'})^2  }{4 {\sf z} {\sf z'}}.
 \ee
 In particular, for the free propagators   for the 
  fields  $\chix$ and $\psi$  are\foot{In the following, we shall often omit the arguments in $u\equiv u(x,x')$ and $\eta\equiv \eta(x,x')$.} 
 \begin{align}
g(x, x') &\equiv \EV{\chix(x)\chix(x')}_{\rm free} =
 \begin{tikzpicture}[line width=1 pt, scale=0.6, baseline=-0.1cm]
\draw (-1.2,0)--(1.2,0);
\draw[fill=black] (-1.2,0) circle (0.1); 
\draw[fill=black] (1.2,0) circle (0.1); 
\end{tikzpicture}
= - \frac12 \Big( \frac{1+\eta}{1-\eta} \log\eta+2 \Big),  \qquad \quad \eta\equiv  \eta(x,x')=\frac{u(x,x')}{1+u(x,x')}, \notag  \\
h(x, x') &\equiv \EV{\psi(x)\psi(x')}_{\rm free} =
 \begin{tikzpicture}[line width=1 pt, scale=0.6, baseline=-0.1cm]
\draw[dashed] (-1.2,0)--(1.2,0);
\draw[fill=black] (-1.2,0) circle (0.1); 
\draw[fill=black] (1.2,0) circle (0.1); 
\end{tikzpicture}
= - \frac {3(1-\eta^2 ) +(\eta^2+4\eta+1) \log\eta  } {2(\eta-1)^2} \ . \la{23}
 \end{align}
 As was mentioned above, in the {\sf AdS} approach we shall  use  a particular \adst covariant UV regularization in which 
 \be\la{24}
 g(x,x) = 0, \qquad \qquad  h(x,x)=0 \ , 
 \ee
 and  thus  may ignore  all  tadpole contributions. 


The explicit computations of the AdS$_{2}$ loop  integrals will be often performed by changing the coordinates 
from the  Poincar\'e half plane $x=(\bt, \bz)$ to the unit disk $|z|<1$ as follows 
\be
\la{2.6}
w\equiv  \sft+i \sfz = - i\frac{z+1}{z-1}, \qquad \qquad   z = \frac{w-i}{w+i}, \qquad \qquad 
ds^2 =-\frac{4\,dw\, d\bar w}{(w-\bar w)^2} =\frac{4\,dz d\bar z}{(1-z\bar z)^2} \ .
\ee
Then the  action (\ref{2.1})  becomes  ($z=x_1+i x_2$, $d^2 z=d x_1 dx_2$)
 \be
 \la{2.7}
 \mc S  =\frac{1}{2\pi} \int d^2 z \Big[ \frac12 (\p_a\chix)^2  + \frac12 (\p_a\psi)^2
 +\frac{2Q} { b\, (1-z\bar z)^2}  {e^{2b\chix}\Big( \cosh (2\sqrt 3 b \psi)  -2b\chix-1}\Big)     \Big].
 \ee
The mass and interaction terms  in \rf{2.7}  have the following explicit  expansion in $b$
 \begin{align}
 \la{2.8}
  &\int {\sf d}^2 z \,  \frac{Q}{b}\, \Big(e^{2b\chix} \cosh (2\sqrt 3 b \psi)  -2b \chix -1\Big) =  \int {\sf d}^2 z \,  (1+4b^{2})\, \Big( 2\chix^2 + 6\psi^2\notag \\
&\qquad \qquad +12 b\, \chix \psi^2 +\frac{4}{3}b\, \chix^3+12b^2\, \chix^2\psi^2+6b^2\, \psi^4 +\frac23 b^2\, \chix^4+\cdots \Big),
\quad \qquad {\sf d} ^2 z\equiv \frac{d^2 z}{\pi(1- z \bar z)}\ . \end{align}

\subsection{Two-point functions}
 
 Let us  now discuss  quantum corrections 
 to  the two-point functions of the $\chix$ and $\psi$ fields, first in the bulk, and then 
 in the boundary limit  defined as in  \rf{1.15},\rf{1.16}.
 We shall use the disk coordinates  \rf{2.6},\rf{2.7} and apply 
 the results  of  Appendix \ref{app:fourier}.
 
 \subsubsection*{Bulk two-point function of $\chix$  and its boundary limit $\llangle \Phi\Phi\rrangle$}
 
 The non-vanishing one-loop contributions  to the  two-point function of the $\chix$ field  are  given by\foot{\la{foot:conn}Since we assume  the vanishing of  simple tadpoles \rf{24}  only connected  one-loop diagrams will contribute (for a 
 general discussion of tadpole  contributions  see   \cite{Beccaria:2019stp}). }
\allowdisplaybreaks[0]
\begin{align}
\EV{\chix(z_1) \chix(z_2)} 
&=
\begin{tikzpicture}[line width=1 pt, scale=0.6, baseline=-0.1cm]
\draw (-1.2,0)--(1.2,0);
\draw[fill=black] (-1.2,0) circle (0.1); 
\draw[fill=black] (1.2,0) circle (0.1); 
\end{tikzpicture}
+
\begin{tikzpicture}[line width=1 pt, scale=0.6, baseline=-0.1cm]
\draw (-1.8,0)--(-0.8,0); \draw (0,0) circle(0.8);
\draw (1.8,0)--(0.8,0); 
\draw[fill=black] (-1.8,0) circle (0.1); 
\draw[fill=black] (1.8,0) circle (0.1); 
\end{tikzpicture}
+  
\begin{tikzpicture}[line width=1 pt, scale=0.6, baseline=-0.1cm]
\draw (-1.8,0)--(-0.8,0);
 \draw[dashed] (0,0) circle(0.8);
\draw (1.8,0)--(0.8,0); 
\draw[fill=black] (-1.8,0) circle (0.1); 
\draw[fill=black] (1.8,0) circle (0.1); 
\end{tikzpicture}
+   
\begin{tikzpicture}[line width=1 pt, scale=0.6]
\draw (-1.5,0)--(1.5,0); 
\begin{scope}[shift={(0,0)},scale=0.6]
	   \draw[fill=white,thin] (0,0) circle (0.3);
            \draw[thin] (45:0.3)--(225:0.3);
            \draw[thin] (-45:0.3)--(135:0.3);
	\end{scope}
\draw[fill=black] (-1.5,0) circle (0.1); 
\draw[fill=black] (1.5,0) circle (0.1); 
\end{tikzpicture} 
+\cdots ,\la{28}  \\
& \quad \quad g(z_1,z_2)
\quad \quad\ \ \   \Sigma_{\chix\chix}(z_1,z_2) 
\qquad\  \ \Sigma_{\psi\psi}(z_1,z_2)
\qquad \ \ \Sigma_{\rm ins.}^{(\chix)}(z_{1},z_{2})\notag
\end{align}
\allowdisplaybreaks[1]
where 
\begin{align}
\Sigma_{\chix\chix} =\frac{(-8b)^2}{2} D_{\chix\chix}(z_1,z_2),\quad \quad 
\Sigma_{\psi\psi} =\frac{(-24b)^2}{2} D_{\psi\psi}(z_1,z_2),\quad \quad
\Sigma_{\rm ins.}^{(\chix)} = -16\,b^{2}\,\widehat\Sigma_\zeta(z_{1},z_{2}).
 \la{299}
\end{align}
Here, $D_{\chix\chix}$ is the bubble diagram with virtual $\chix$ fields, $D_{\psi\psi}$
is the similar diagram with virtual $\psi$ fields. The additional order $b^2$  contribution
 $\Sigma_{\rm ins.}$ is associated with the insertion of the vertex $b^2 \chix^{2}$  coming from 
 the $1+4b^{2}$ factor in (\ref{2.8}). 
 Let us give the expressions for the 
various contributions.
  The $\chix$ loop correction $D_{\chix\chix}$  is  
  \be\la{29} 
  D_{\chix\chix}(z_1,z_2)
  =z_1
  \begin{tikzpicture}[line width=1 pt, scale=0.6, baseline=-0.1cm]
\draw (-1.8,0)--(-0.8,0); \draw (0,0) circle(0.8);
\draw (1.8,0)--(0.8,0); 
\draw[fill=black] (-1.8,0) circle (0.1); 
\draw[fill=black] (1.8,0) circle (0.1); 
\end{tikzpicture}
\; z_2
=\int {\sf d}^2z' g(z_1,z') B(z',z_2) \ , \ee  
where 
\be  B(z_1,z_2)=\int {\sf d}^2 z'  \big[g(z_1,z') \big]^2 g(z',z_2)
 = \frac18 -\frac{\eta \log^2\eta}{8(1-\eta)^2} \ . \la{31} \ee
As a result,  
\be 
D_{\chix\chix}= \frac{1}{576} \Big[15+\frac{\pi^2(\eta+1)}{\eta-1} \Big]  
- \frac{\eta \log\eta}{48(\eta-1)}
+\frac{\eta^2\log^2\eta}{64(\eta-1)^2}
+\frac{\log(1-\eta)}{48} \Big[1- \frac{(\eta+1) \log\eta}{\eta-1} \Big]
-\frac{(\eta+1)\text{Li}_2(\eta)}{96(\eta-1)}.\la{32}
\ee
Similarly, the $\psi$ loop contribution $D_{\psi\psi}$   reads
  \be\la{30}
D_{\psi\psi}(z_1,z_2)
  =z_1
  \begin{tikzpicture}[line width=1 pt, scale=0.6, baseline=-0.1cm]
\draw (-1.8,0)--(-0.8,0); \draw[dashed] (0,0) circle(0.8);
\draw (1.8,0)--(0.8,0); 
\draw[fill=black] (-1.8,0) circle (0.1); 
\draw[fill=black] (1.8,0) circle (0.1); 
\end{tikzpicture}
\; z_2
=\int {\sf d}^2z' g(z_1,z') C(z',z_2), \ee 
where 
\be 
C(z_1,z_2)=\int {\sf d}^2 z'  [h(z_1,z') ]^2 g(z',z_2)= \frac{\eta^2-6\eta+1}{16(\eta-1)^2}
-\frac{\eta \big[2-2\eta ^2+(\eta ^2+1) \log \eta  \big] \log \eta }{8 (\eta -1)^4} \ . \la{33}
\ee
 Thus we obtain    
\begin{align}
D_{\psi\psi}(z_1,z_2) =& 
 \frac{\pi^2(\eta^2 - 1)  +12\eta^2 -20\eta+12 }{576(\eta-1)^2} 
- \frac{\eta (\eta^2-\eta+2) \log\eta   }{144(\eta-1)^3}
+\frac{\eta^2 (5\eta^2-10\eta+9)  \log^2 \eta }{576(\eta-1)^4}\no 
\\&
+\frac{\log(1-\eta)}{72} \Big[\frac12- \frac{(\eta+1) \log\eta}{\eta-1} \Big]
-\frac{(\eta+1)\text{Li}_2(\eta)}{96(\eta-1)}.\la{35}
\end{align}
Finally, the insertion diagram in \rf{299} is  determined by   
\begin{align} 
\widehat\Sigma_\zeta &= \int {\sf d}^2 z'   g(z_1,z')   g(z',z_2)
\nonumber\\&=
 -\frac{\eta \log\eta}{6(\eta-1)} 
+ \frac{\log(1-\eta)}{6} \Big[  1+\frac{(\eta+1) \log\eta}{2(\eta-1)} \Big]
+ \frac{(\eta+1) \text{Li}_2( \eta)}{6(\eta-1)} -\frac16-\frac{\pi^2(\eta+1)}{36(\eta-1)}. \la{sigma}
\end{align}
Then \rf{28},\rf{299}   give  the one-loop bulk  two-point function. 
Going back to the 
Poincar\'e plane parametrization (cf. \rf{2.2})  let us define 
 the boundary correlator on the real line  as in   \rf{1.15},\rf{1.16}. 
As a  result,  we get   
\be\la{37}
\llangle \Phi(\sft_1)\Phi(\sft_2)\rrangle =
\lim_{\sfz_1,\sfz_2\rightarrow 0}  \sfz_1^{-2} \, \sfz^{-2}_2 \EV{\chix(\sft_1,\sfz_1)\chix(\sft_2,\sfz_2)} = 
\frac{1}{\sft_{12}^4}\,\Big(\frac43-\frac49 b^2  +\cdots\Big) \ . 
\ee
This   expression is in agreement with  the expected value of $C_{22}$ in \rf{1.20},(\ref{1.21}).

  \subsubsection*{Bulk two-point function of  $\psi$  and its boundary limit $\llangle \Phi_{3}\Phi_{3}\rrangle$}

The  corresponding  two-point function reads
   \begin{align}\la{288}
\EV{\psi(z_1) \psi(z_2)}_{\rm conn}&=
\begin{tikzpicture}[line width=1 pt, scale=0.6, baseline=-0.1cm]
\draw[dashed]  (-1.2,0)--(1.2,0);
\draw[fill=black] (-1.2,0) circle (0.1); 
\draw[fill=black] (1.2,0) circle (0.1); 
\end{tikzpicture}
+
\begin{tikzpicture}[line width=1 pt, scale=0.6, baseline=-0.1cm]
\draw[dashed]  (-1.8,0)--(-0.8,0);  
\draw (0.8,0) arc(0:180:0.8);
\draw[dashed] (-0.8,0) arc( 180:360:0.8);
\draw[dashed]  (1.8,0)--(0.8,0); 
\draw[fill=black] (-1.8,0) circle (0.1); 
\draw[fill=black] (1.8,0) circle (0.1); 
\end{tikzpicture}
+
\begin{tikzpicture}[line width=1 pt, scale=0.6, baseline=-0.1cm]
\draw[dashed]  (-1.5,0)--(1.5,0); 
\begin{scope}[shift={(0,0)},scale=0.6]
	   \draw[fill=white,thin] (0,0) circle (0.3);
            \draw[thin] (45:0.3)--(225:0.3);
            \draw[thin] (-45:0.3)--(135:0.3);
	\end{scope}
\draw[fill=black] (-1.5,0) circle (0.1); 
\draw[fill=black] (1.5,0) circle (0.1); 
\end{tikzpicture} 
+\cdots\\
&
\qquad h(z_1,z_2)
\qquad \quad \Sigma_{\chix\psi}(z_1,z_2)
\qquad \Sigma_{\rm ins.}^{(\psi)}(z_{1},z_{2})\notag
\end{align}
The non-trivial diagram is $\Sigma_{\chix\psi}(z_1,z_2)=(-24b)^2 D_{\chix\psi}(z_1,z_2)$,
while the insertion contribution is 
$\Sigma_{\rm ins.}^{(\psi)}(z_{1},z_{2}) = -48\,b^{2}\,\widehat\Sigma_\psi(z_{1},z_{2})$. 
The bubble is given by 
  \be
  D_{\chix\psi}(z_1,z_2)
  =z_1
\begin{tikzpicture}[line width=1 pt, scale=0.6, baseline=-0.1cm]
\draw[dashed]  (-1.8,0)--(-0.8,0);  
\draw (0.8,0) arc(0:180:0.8);
\draw[dashed] (-0.8,0) arc( 180:360:0.8);
\draw[dashed]  (1.8,0)--(0.8,0); 
\draw[fill=black] (-1.8,0) circle (0.1); 
\draw[fill=black] (1.8,0) circle (0.1); 
\end{tikzpicture}
\; z_2
=\int {\sf d}^2z' h(z_1,z') B_\psi(z',z_2) \ .\la{38} \ee 
Here 
\be 
B_\psi(z_1,z_2)=\int {\sf d}^2 z'   h(z_1,z') g(  z' , z_1)  h(z',z_2)=
  \frac{1+\eta}{16(1-\eta)} - \frac{\eta \log\eta}{8(\eta-1)^2}
  +\frac{\eta(1+\eta) \log^2\eta}{8(\eta-1)^3} \ . \la{39}
  \ee
%
As a result 
\begin{align}
D_{\chix\psi}(z_1,z_2) = &-\frac{15 \left(\eta ^2-1\right)+\pi
   ^2 (\eta  (\eta +4)+1)}{2880
   (\eta -1)^2}-\frac{\eta ^2 (\eta
   +3) \log ^2\eta }{192 (\eta
   -1)^3}+\frac{\eta  (2 \eta +1)
   \log \eta }{160 (\eta
   -1)^2}\notag \\
   &+{ 1\ov 80} \log (1-\eta )
   \Big[ - \frac{\eta +1}{\eta -1 
   }+\frac{(\eta  (\eta +4)+1) \log \eta }{2 (\eta
   -1)^2}\Big]
   +\frac{(\eta  (\eta
   +4)+1) \text{Li}_2(\eta )}{480
   (\eta -1)^2}~.\la{40}
\end{align}
The insertion contribution in \rf{288} is proportional to   
\begin{align}
\widehat\Sigma_\psi &= \int {\sf d}^2 z' \,   h(z_1,z')   h(z',z_2)
\nonumber\\&=
\frac{15 \left(\eta ^2-1\right)+2
   \pi ^2 (\eta  (\eta +4)+1)}{120
   (\eta -1)^2}+\frac{\eta  (3 \eta
   +4) \log (\eta )}{20 (\eta
   -1)^2}
  \nonumber \\ &\quad
   +\log (1-\eta )
   \Big[-\frac{3 (\eta +1)}{20
   (\eta -1)}-\frac{(\eta  (\eta
   +4)+1) \log (\eta )}{20 (\eta
   -1)^2}\Big] -\frac{(\eta  (\eta
   +4)+1) \text{Li}_2(\eta )}{10
   (\eta -1)^2}~.\la{222}
\end{align}
Defining   again  the boundary correlator according to    \rf{1.15},\rf{1.16}
we finish with 
\be\la{42}
\llangle \Phi_{3}(\sft_1)\Phi_{3}(\sft_2)\rrangle  =
\lim_{\sfz_1,\sfz_2\rightarrow 0}   \sfz_1^{-3} \, \sfz^{-3}_2\EV{\psi(\sft_1,\sfz_1)\psi(\sft_2,\sfz_2)}= 
\frac{1}{\sft_{12}^6}\,\Big(\frac{16}{15} -\frac{112}{75}b^2+\cdots\Big) \ ,
\ee
which is in agreement with  the expression for  $C_{33}$ in \rf{1.20},(\ref{1.21}).

\subsection{Three-point functions}

We can  also check the relations \rf{1.28}  for the   three-point functions
\be\la{43}
\llangle \Phi\Phi\Phi\rrangle = \k^{3}\,\langle TTT\rangle, \qquad 
\qquad \llangle \Phi\Phi_{3}\Phi_{3}\rrangle = \k\,\kappa_{3}^{2}\,\langle T \QQ_3\QQ_3\rangle
\ee
 by reproducing the one-loop perturbative expansions of the   coefficients  $C_{222}$ and $C_{233}$ in (\ref{1.21}).
The relevant one-loop diagrams can be  computed  using the  disc parametrization of   AdS$_2$.
The only type of  bulk diagram that cannot be computed analytically is the one 
with a  triangle loop but it can be evaluated numerically with good precision  well below  one  percent level.
As a result, we confirmed  the duality predictions in (\ref{1.21}).
 The  details can be found in Appendix \ref{app:3p}.

\section{Non-abelian Toda theory}
\la{sec:non}

Another non-trivial example of   \chads  duality is  provided by 
  the non-abelian Toda (NAT)  theory of  \ci{Gervais:1992bs} (see also \ci{Bilal:1993rg,Bilal:1995ei}). 
This theory may be viewed 
as a special case of  a conformal model  with a   3-dimensional  target space   with 
the kinetic term  given by  a scalar  plus a    $\s$-model  originating from  the 
 $SL(2,\RR)/U(1)$ gauged WZW model  \ci{Bardacki:1990wj,Witten:1991yr} 
  supplemented   by  a  potential    analogous to the  abelian Toda  model one. 

We shall discuss  the quantum conformal properties of the NAT model, in particular, 
give  a free field representation for  its 3 conserved currents:  the stress tensor $T$ and 
a pair of ``parafermions''    $V^{\pm}$. This will allow us to compute their  CFT correlators.
One  novel feature  will be  that  in contrast to the $\mc W$ symmetry  generators in  the abelian Toda case 
   here  $V^{\pm}$  will have a non-trivial anomalous dimension \rf{1.26}. 
This will allow us to  compare the CFT correlators 
 with the   boundary correlators of the corresponding  elementary  fields
 in the NAT Lagrangian in  AdS$_{2}$ \rf{1.23}. The  boundary  correlators    
given by  the  Witten diagrams  will  be computed at the tree and one-loop level. 
 As as result, we will  verify the   \chads   correspondence  \rf{1.28}. 

\subsection{Definition of  the model on AdS$_2$ background}

Let us start   with  a   $\sigma$-model like   theory  on a curved 2d background 
\be
\la{3.1}
\mc S = \frac{1}{4\pi}\int d^{2}z\,\sqrt{g}\,  \mc L \ , \qquad 
\mc L=  G_{\mu\nu}(x)\,\partial_{a}x^{\mu}\partial^{a}x^{\nu} +   {\rm T}(x)
+ R\,\dil(x)   \ . 
\ee
Here $\m,\n= 1, ..., D$,  \ $R$ is   2d curvature, $\dil$  is a
``dilaton'' and    $\rm T$ is a ``tachyon''. 
The standard Weyl-invariance conditions  of decoupling of the conformal factor of the 2d   metric are  
\ci{Callan:1985ia,Tseytlin:1988rr} 
\begin{align}
&R_{\mu\nu}+2\,\nabla_{\mu}\nabla_{\nu}\dil+\dots=0   \ , \qquad \qquad 
-\ha\nabla^{2}\dil+(\partial\dil)^{2}
+\dots = {\te {c- D\ov 6}}   \ ,  \la{3.2} \\
&\qquad \qquad -\ha \nabla^{2}{\rm T}+G^{\mu\nu}\partial_{\mu}\dil\partial_{\nu}{\rm T} -  2\,{\rm  T}
+\dots=0\ .  \la{3.3} 
\end{align}
Here we  presented  only the leading one-loop  terms in the corresponding Weyl-anomaly coefficients 
and ignored $\OO({\rm T}^2)$ terms in the $\b$-functions corresponding to non-perturbative divergences 
\cite{Tseytlin:1990mz,Grisaru:1990gf} that will not be  relevant for  the model discussed below (with ${\rm T}$  
given by a sum of exponentials with non-constant products). 

 The  $D=3$ model \rf{3.1}  (with $x^\m=(\varphi,r,y)$)  we will be interested  in will  be 
 \be
\la{3.4}
\mc L = (\partial\varphi)^2 +     (\partial r)^2 +  G(r) (\partial y)^2   +   {\rm T}(\varphi,r)    + 
R\, \dil(\varphi,r)\ ,
\ee
with  $G,\dil,\TT$  given  by   a  particular  solution of the leading-order  equations \rf{3.2},\rf{3.3} 
\begin{align} \la{3.5}
& G(r)= \tanh^2 (a r) \ ,\qquad \qquad 
\dil = Q\,\varphi   -\log\cosh(a r),  \\ 
& {\rm T} = \mu^{2}\,e^{2b\varphi}\, \cosh(2 a r) \ .  \la{3.6}
\end{align}
Here the constants $a, b$  and $Q$ are related by 
\be\la{3.7} 
 1  + b^2 -  b\, Q + 2 a^2=0
 \ ,\ \ {\rm i.e.} \ \ \qquad  Q=  {b}^{-1}  (1 + 2  a^2)     + b \ , \ \ \ \ \ \ \qquad   c =  3+6Q^{2} + 6a^{2} \ . 
\ee
The case of $a=0$ brings us back to the  Liouville theory (plus two extra free fields). 

In general, 
the  leading-order metric coefficient $G$ and the dilaton $\dil$ in \rf{3.5} require  a modification  in order to satisfy 
the   Weyl-invariance equations \rf{3.2}  at the 2-loop level   \ci{Tseytlin:1991ht}. 
Including the 3- and 4-loop   terms in the $\beta$-functions  requires further modifications \ci{Jack:1992mk}
 (whose   form depends,   in general,  on a renormalization scheme \ci{Tseytlin:1993df}). 
Recognizing that    \rf{3.5}    is  the ``classical''   background  for the 
$SL(2,\RR)/U(1)$ gWZW model  corresponding to  a coset  CFT   
allows one to determine  the  exact 
metric and dilaton  \ci{Dijkgraaf:1991ba} 
\begin{align} 
&G(r) ={  \tanh^2 (a r)  \ov 1-  { 2 \ov k} \tanh^2 (a r) } \ ,  \qquad \qquad 
\dil= Q\,\varphi   -\log\cosh(a r) - {\te {1\ov 4}}  \log \big[ 1 - {\te {2 \ov k}}  \tanh^2 (a r) \big]  \ ,\la{3.8} \\
& 
\qquad  c=  3+6Q^{2} + {6a^2} = 6Q^2 + { 3k \ov k-2} \  , \qquad \qquad \ \  a = {1 \ov \sqrt{ k-2} }   \ .    \la{3.9} 
\end{align}
Here $k$ is the level of the  gWZW  theory   so that the total central charge  is the sum of the  one for the linear dilaton $\vp$-theory 
and the $SL(2,\RR)/U(1)$  coset theory:   $c=(1 +  6Q^2)  + ({ 3k \ov k-2} -1)$. 
The exact background \rf{3.8} was first found  \ci{Dijkgraaf:1991ba}  by identifying the ``point-particle'' probe equation
($L_0 {\rm T}=2\,{\rm T}$)  in the coset CFT with the  leading-order  tachyon equation \rf{3.3}.\foot{The same   background \rf{3.8}    can be found  also from   the quantum effective action  for the gWZW model 
 \ci{Tseytlin:1992ri,Bars:1993zf,Tseytlin:1993my}. 
} 
This corresponds to the choice of a  ``CFT scheme'' 
 in which  the tachyon equation   is not modified  by quantum (or $\alpha'$)  corrections \ci{Tseytlin:1993df}, i.e. 
 its  exact form  is the leading-order one in \rf{3.3} or
 \be
\la{3.10}
- \frac{1}{\sqrt{G} e^{-2 \dil}}  \partial_{\mu}( \sqrt{G} e^{-2 \dil}\, G^{\mu\nu} 
\partial_{\nu} ) {\rm T} -   4\,{\rm T} =0 \ . 
\ee
As the  potential $\rm T$ in  \rf{3.4}  is assumed  not to depend on the  isometric 
$y$ coordinate  and since  for the exact   background \rf{3.8} one gets no corrections to the combination 
 $ \sqrt{G} e^{-2 \dil} = e^{-2 Q \varphi} \sinh (ar) \cosh (a r)$  
we conclude that the corresponding exact solution for $\rm T$ is still given   by \rf{3.6} 
with $b$ related to $a$ and $Q$  by \rf{3.7}. 

Below     we   will only   consider    the leading   order  terms  in   loop expansion, 
i.e.  will  ignore  finite quantum  counterterms   coming from $1/k$   expansion of \rf{3.8} 
(which are, in general, 
required for the preservation of  the conformal invariance and also integrability  \ci{Hoare:2010fb}
  beyond the  1-loop order). We  shall  thus  start with the   following Lagrangian \rf{3.4} on a unit-radius  AdS$_2$  background ($R=-2$) 
\be\la{3.11}
\mc L = (\partial \varphi)^2 + (\partial r)^2 + \tanh^2 (a r) (\partial y)^2
          +   \mu^{2}  e^{2 b \varphi}\cosh (2a r)\,      - 2 Q \varphi  + 2\log  \cosh(a r)\ .
\ee
Here the  last two terms come from the dilaton coupling in \rf{3.4}.
 The coupling constants $a$ and $b$ are a priori  independent 
with $Q$    expressed in terms of them   by \rf{3.7}.
As in the  Liouville or abelian Toda  model   the presence  of  an extra ``dilaton'' term in the action  on the 
AdS$_2$  background implies the existence of a constant extremum  of the resulting potential (see  \rf{1.4}):
$\varphi_0=  \frac{1}{2b}\log\frac{Q}{b\mu^{2}}, \  r=0$. 
   The  perturbation theory near this vacuum is  then described by 
($\varphi=\varphi_0 +\chix$)
\be
\la{3.12}\te 
\mc L = (\partial \chix)^2 + (\partial r)^2 + \tanh^2 (a r) (\partial y)^2
+{Q\ov b }\,\big[ e^{2b\chix}\cosh(2 a r)-2b \chix\big]+2\log\cosh(a r)\ . 
\ee
Introducing the  new  coordinates $(r, y) \to (\xi_1, \xi_2)$ (with $y$-isometry becoming   rotation in  the $\xi_i$ plane)
\be\la{3.13} 
\xi_{1}={a}^{-1} \sinh(a r)\,\cos(a y),\qquad\qquad 
\xi_{2}={a}^{-1} \sinh(a r)\,\sin(a y) \ , \qquad  \bmxi^2 \equiv \xi_1^2 + \xi_2^2 \ , 
\ee
 the Lagrangian \rf{3.12} takes the  following $SO(2)$  invariant form 
\be
\la{3.14}
\mc L = (\partial \chix)^2 + \frac{(\partial \xi_{1})^2 + (\partial \xi_{2})^2}{1+a^{2}\,\xi^{2}}
+\tfrac{Q}{b}\,\big[e^{2b\chix}(1+2a^{2}\,\xi^{2})-2b\chix\big]+\log(1+a^{2}\xi^{2}) \ . 
\ee
Expanding to quadratic order  in the fluctuation  fields $(\zeta,\xi_1,\xi_2)$ we find that their masses  
 are 
\begin{align}\la{3.15} 
m^{2}_{\chix} = 2\,b\,Q = 2\,(1+2a^{2}+b^{2})\ , \qquad \qquad 
m^{2}_{\xi_{1,2} }= \frac{a^{2}}{b}\,(b+2Q) = \frac{2a^{2}}{b^{2}}(1+2a^{2}+\tfrac{3}{2}b^{2})\ .
\end{align}
We  observe that in the special   case  when 
\be   a= b\ ,  \la{3.16}  \ee
all   masses  have the same  value  $m^2=2$ at leading order in expansion in the  coupling  $b$. 
With \rf{3.16}    we get from \rf{3.7}  (see also \rf{1.23},\rf{1.24}) 
\be\la{3.17} 
   Q=  b^{-1}      + 3 b \ , \ \ \ \ \ \ \qquad   c =  3+6Q^{2} + 6b^{2}= 6 b^{-2} + 39 + b^2  \ . 
\ee
The   case of \rf{3.11}  with \rf{3.16}  corresponds to the NAT  model of \ci{Gervais:1992bs}
with the exact conformal background given by \rf{3.6},\rf{3.8} and $b= {1\ov \sqrt{k-2}}$.\foot{Let us
 mention that the misleading  claim 
in  \ci{Bilal:1995ei} that the  leading-order model of \ci{Gervais:1992bs}
defined by the metric  and tachyon in \rf{3.5},\rf{3.6} is not conformally invariant was  due to ignoring the contribution of the dilaton coupling in \rf{3.5}
(see also \ci{Jack:1993de}).}

This is  the   model we shall study  below. 
It  can be also  obtained by a Hamiltonian reduction \cite{ORaifeartaigh:1990ved,Balog:1990mu,
Feher:1992yx} 
 of the  $B_{2}=SO(5)$  WZW model over a nilpotent subgroup  \ci{Gervais:1992bs}. It 
 should be possible to derive it 
 directly from a gauged WZW model  obtained  by    ``null'' gauging  of a solvable subgroup 
 (similarly to how that was done  for  the  abelian Toda models in \cite{Klimcik:1994wp}). This 
    should  give  a  $\s$-model with a 5d target space,    containing, in addition to the first  three terms  in \rf{3.11},   
     also   $ \Delta  \mc L =[e^{2 b(\varphi+r)} +  e^{2 b(\varphi-r)} ]^{-1} \del_+ u \del_- v $. Solving  for the two 
 extra  fields  $u$ and $v$  as in \cite{Klimcik:1994wp} will then  reproduce  the   $e^{2b \varphi} \cosh ( 2 b r)$  potential  in \rf{3.11}.

\iffa 
Non-abelian Toda theories can be obtained by Hamiltonian reduction of the WZW model
\cite{ORaifeartaigh:1990ved,Feher:1992ed,Feher:1992yx} 
or by gauging a WZW model and integrating out the gauge fields, 
see for instance \cite{Balog:1990mu,Sevrin:1993si}.
From the point of view of its relation with WZW models, the \GS model is obtained 
after null gauging \cite{Klimcik:1994wp} of the $B_{2}=SO(5)$ 
WZW model and integrating out the gauge fields. The null gauging procedure reduces in some cases to the 
Hamiltonian reduction of WZNW theories related to Toda models \cite{Alekseev:1988ce,Bershadsky:1989mf} 
but is more general as discussed in full details in \cite{Klimcik:1994wp}.
\fi

Our starting point in the  computation of the AdS$_2$   boundary correlators 
will thus be the   Lagrangian \rf{3.14} with $a=b$  expanded to quartic order in the  fluctuation fields 
 or  to $b^2$ order in the coupling\foot{Here  we redefine $\mc L$ by $\ha$  to restore  the standard normalization of the fields, i.e. 
the action is now  given by $\mc S = \frac{1}{2\pi}\int d^{2}z\,\sqrt{g}\,  \mc L$.}
\begin{align}
\la{3.18}
\mc L = \tfrac{1}{2} (\partial \chix)^2 + \tfrac{1}{2}(\partial \xi_i)^2 &+ (1+3b^{2})\,\chix^{2}+
(1+\tfrac{7}{2}\,b^{2})\,\xi^{2}\notag \\
&+b(1 + 3 b^2) \big(2\,\chix\,\xi^{2}+\tfrac{2}{3}\,\chix^{3}\big)+ b^2 \big[
2\,\chix^{2}\,\xi^{2}+\tfrac{1}{3}\,\chix^{4}-\tfrac{1}{2}\,\bmxi^{2}\,(\partial\xi_i)^{2}\big]+\OO(b^4)\ . 
\end{align}
Note that the expansion of higher order terms in  ${1\ov k}={b^2\ov 1 + 2 b^2}$    in the exact  background \rf{3.8} 
which  we ignored  in \rf{3.12}  produces  only higher order $\OO(b^4)$   terms in \rf{3.18} that will not contribute to 
the computations performed in this paper.


\subsection{Underlying flat-space  CFT \la{sec:3.2}} 

Following  \cite{Gervais:1992bs}, the classical integrable structure of the \GS
model was  elucidated in \cite{Bilal:1993rg}.  It 
was  shown to admit three  conserved holomorphic  (plus anti-holomorphic)  quantities: 
 the stress tensor $T$ and  the two  currents $V^{\pm}$ with the same classical dimension 2. 
 Using a non-trivial change of variables they  may be written in terms of free fields. 
 After  quantization $T$  and $V^{\pm}$   should  represent  primary operators in the underlying CFT. 
 The  chiral half of this CFT should then  play the role  of the boundary CFT in the 
 \chads  correspondence. 
 

Since our  application of the free-field  construction  to the quantum  \GS model appears to be  new, 
let us  start  with  recalling   first  the free-field   representation 
 for  the $SL(2,\RR)/U(1)$   gWZW model  (see, e.g., \cite{Gerasimov:1990fi,Ford:2000eb,Kruger:2004jm}), 
 i.e.  for the $(r,y)$ subsector of (\ref{3.11}) with no tachyon  coupling.
The classical conserved quantities  in this case are 
\begin{align}
\la{3.19}
T_{_{0,\rm cl}} = -\frac{1}{2}\,(\partial\varphi_{1})^{2}\,-\frac{1}{2}\,(\partial\varphi_{2})^{2}, \qquad \qquad 
U^{\pm}_{_{ cl}} =\frac{1}{\sqrt{2}} \big(\partial\varphi_{1}\pm i\,\partial\varphi_{2}\big)
\,e^{\pm i\,\frac{\sqrt{2}}{\sqrt{k}}\,\varphi_{2}},
\end{align}
where $k$ is the  level. The  quantization    is performed  by assuming that 
 $\varphi_{i}$ ($i=1,2$)  are two free   quantum fields with  the standard 
OPE 
\be
\la{3.20}
\partial \varphi_{i}(z)\,  \partial \varphi_{j}(0) \sim -\frac{1}{z^{2}}\delta_{ij}\ .
\ee
The conserved stress tensor and the  parafermions $U^{\pm}$   
of the quantum gWZW model are    given by the following 
generalizations  of the classical  quantities in (\ref{3.19})
 (see, e.g., \cite{Kruger:2004jm})  
\begin{align}
\la{321}
T_{_{\rm 0}} &= -\frac{1}{2}:(\partial\varphi_{1})^{2}:\,- \frac{1}{2}:(\partial\varphi_{2})^{2}:-\, \frac{1}{\sqrt{2}\,\sqrt{k-2}}\,\partial^{2}
\varphi_{1}\ ,  \\
U^{\pm}
&=\frac{1}{\sqrt{2}} :\big(\sqrt{\tfrac{k-2}{k}}\,\partial\varphi_{1}\pm i\,\partial\varphi_{2}\big)
\,e^{\pm i\,\frac{\sqrt{2}}{\sqrt{k}}\,\varphi_{2}}:\ . \la{3.21}
\end{align}
This  $T_{_0}$ obeys the Virasoro algebra with the central charge of the gWZW theory 
\be\la{3.22} 
c_0 
= \frac{3k}{k-2}-1\ ,
\ee
while  $U^{\pm}$ 
turn out to be  the primary fields    with   dimension  $\Delta_{_{U}} $, {\em i.e.}
\be\la{3.23} 
T_{_{0}}(z)\,U^{\pm} (0) \sim \frac{\Delta_{_{U}}}{z^{2}}\,U^{\pm}(0)
 + \frac{1}{z}\,\partial U^{\pm}(0) + \dots 
 \ , \ \ \ \ \  \ \  \ \ \   \Delta_{_{U}} = 1+\frac{1}{k} \ . 
\ee
 The  operators $U^+ $  and $U^-$    have a    non-trivial 
OPE
\be\la{3.24}
U^{+} (z) \, U^{-} (0) \sim \frac{1}{z^{2\Delta_{_{U}}}}\,\Big[
-1-\frac{2\Delta_{_{U}}}{c_{{0}}}\,T_{_{0}}(0)
+\dots\Big]\ .
\ee
Going back to the  \GS theory, this  free-field  construction may be generalized by adding another 
free   field $\varphi_3$  (again obeying  (\ref{3.20})). 
Then the  analogs  of the classical conserved quantities $T_{_{\rm cl}}$ and $V^\pm_{_{\rm cl}}$ 
 found  in \cite{Bilal:1993rg} are proposed to be (cf. \rf{321},\rf{3.21})
\begin{align}
T &= T_{_0} - \frac{1}{2} : (\partial\varphi_{3})^{2}  : -\,  {1 \ov \sqrt 2} Q \,\partial^{2}\varphi_{3}
= -\frac{1}{2}\sum_{i=1}^{3} : (\partial\varphi_{i})^{2} :-\frac{1}{\sqrt{2}\sqrt{k-2}}\,
\partial^{2}\varphi_{1}-\,  {1 \ov \sqrt 2} Q \,\partial^{2}\varphi_{3}\ , \la{325} \\
V^{\pm} &= \big(\partial\varphi_{3}-p\, \partial\big)  U^\pm =  \  : \big(\partial\varphi_{3}-p\, \partial\big) \,\frac{1}{\sqrt 2}\big(\sqrt{\tfrac{k-2}{k}}\,
\partial\varphi_{1}\pm i\,\partial\varphi_{2}\big)\,
e^{\pm \frac{i\,\sqrt{2}}{\sqrt k}\,\varphi_{2}}:\  ,\la{3.25}
\end{align}
where $Q$  and $p$ are constants to be fixed as functions of the level  $k$.
Note that the  classical dimension of $V^{\pm}(z) $ in \rf{3.25}  differs by 1   from that of  $U^\pm $ in \rf{3.21}.
The  stress tensor $T$ in \rf{325}   obeys  the Virasoro algebra with the central charge (cf. \rf{3.9})
\be\la{3.26} 
c = c_0 + 1 + 6 Q^{2} =  \frac{3k}{k-2}+ 6 Q^{2}\ .
\ee
Requiring that this matches  the NAT central  charge  in \rf{3.17} (where $b= {1\ov \sqrt{ k-2}}$) 
gives the same  value of $Q$ as in \rf{3.17}, i.e. 
\be\la{3.27}
Q = b^{-1} + 3  b=  \frac{k+1}{\sqrt{k-2}}\ . 
\ee
Computing  the OPE  of  $T(z)V^{\pm}(0)$  (generalizing the one in \rf{3.23}) one 
 finds that the leading singularity 
is  $\OO(z^{-3})$ with the residue  $\sim (\sqrt{2}\,k-2\sqrt{k-2}\,p)$, i.e. the necessary  condition for   $V^{\pm}$ to be 
primary  is 
\be\la{3.28} 
p = \frac{k}{\sqrt{2}\,\sqrt{k-2}}\ .
\ee
This turns out  to be also the sufficient  condition, i.e.  assuming \rf{3.28}  
we  find that $V^\pm$  are  primary  with the same 
  dimension $ \Delta_{_{V}} = \Delta_{_{U}}  +1  = 2+\frac{1}{k}$ (cf. \rf{3.23}), i.e. 
\be
\la{3.29}
T(z)\,V^{\pm}(0) \sim \frac{\Delta_{_{V}}}{z^{2}}\,V^{\pm}(0)
 + \frac{1}{z}\,\partial V^{\pm}(0) + \dots \ ,\qquad \qquad \ \Delta_{_{V}}=2+\frac{1}{k}\ .
 \ee
The 
OPE for  $V^+ $   and $V^-$  is  found to be  (cf. \rf{3.24},\rf{1.31}) 
\be
\la{3.30}
V^{+}(z) \, V^{-}(0) \sim \frac{C^{+-}}
{z^{2\Delta_{_{V} }}   }\,\Big[1
+z^{2}\,\frac{2\Delta_{_{V}}}{c}\,T+z^{3}\,\frac{\Delta_{_{V}}}{c}\,\partial T+z^{4}\,\Omega_{4}+\dots\Big],\qquad \qquad
C^{+-}=  \frac{3k(k+2)}{k-2}\ , 
\ee
where $c$ is  the NAT   central charge in   \rf{3.26}.   
 $\Omega_{4}$ is a dimension 4  operator   that  may be decomposed in  
 a dimension 4 descendent of the identity plus a primary
of dimension 4.\footnote{Note that in  the  pure $SL(2,\RR)/U(1)$ gWZW  model (without extra 
 Liouville $\vp$ field  and potential) 
  the 
dimension 3 operator in the corresponding OPE  \rf{3.24}  
 is not simply $\partial T$, but the spin 3 generator of the $\mc W_{3}$ 
 extension  of the Virasoro algebra (see, for instance, \cite{Bakas:1991fs,Kruger:2004jm}).
In the NAT case 
 the first non-trivial correction in \rf{3.30}  appears at order $z^4$  in the 
 square brackets. It would  be  interesting to explore if  this is an indication of an  underlying 
$\mc W_{4}$ symmetry (cf. \ci{Sevrin:1993si}).}

Computing the  OPE of   $V^{+}(z) V^{+}(0)$ 
we find that 
there are  no poles   in $z$  and    first regular term is 
\be
\la{3.31}
V^{+}(z) \,V^{+}(0) \sim z^{{2}/{k}}\, \Omega_{0}(0) + \dots \ , 
\ee
where the  operator $\Omega_{0}$    has somewhat complicated expression (that will not be needed 
below).\foot{ 
It is the normal ordered  product of several monomials in $\varphi_{1}$, $\varphi_{2}$, $\varphi_{3}$
with  total of 4 derivatives 
 times $\exp\left(\frac{2\,i\,\sqrt{2}}{\sqrt{k}}\,\varphi_{2}\right)$.}
One can check that  $\Omega_{0}$   is a primary field with  dimension 
\be
\la{3.32}
\Delta_{_{\Omega_{0}}}  = 4+\frac{4}{k}\ .
\ee


\subsection{Two-point  and three-point functions \la{sec:33}}

To  check  the \chads  duality \rf{1.28}    let us start with  two-point functions.
The coefficient $C_{22}$ in the boundary correlator of $\chix$ fields in 
 (\ref{1.32})  can be computed as in the Liouville theory \cite{Beccaria:2019stp}, 
but  noting that  now we can have also the fields $\xi_{1,2}$ propagating in  the loop.
Taking into account the  coefficients  of the interaction terms in the Lagrangian 
(\ref{3.18}) and the fact that to leading order $\xi_i$ has the  same mass as   $\chix$, 
 the  one-loop result is obtained multiplying by 
3 the one-loop correction in the Liouville theory
\be\la{335}
C_{22} =\frac43 \Big( 1-3\times \frac16 b^2\Big) +...=\frac{4}{3}-\frac{2}{3}\,b^{2}+\cdots\ , 
\ee
which is  indeed   in agreement with (\ref{1.33}). 

The  correlator of the $\xi_{i}$ fields  can  also   be found  by a simple modification
of the previous  calculations (cf. \rf{28},\rf{288}):  
   \begin{align} \la{336}
\EV{\xi_i(z_1) \xi_j(z_2)}_{\rm conn} = \delta_{ij}  \hat  g (z_1,z_2) \ , \qquad   & \hat  g  =
\begin{tikzpicture}[line width=1 pt, scale=0.6, baseline=-0.1cm]
\draw[dashed]  (-1.1,0)--(1.1,0);
\draw[fill=black] (-1.2,0) circle (0.1); 
\draw[fill=black] (1.2,0) circle (0.1); 
\end{tikzpicture}
+
\begin{tikzpicture}[line width=1 pt, scale=0.6, baseline=-0.1cm]
\draw[dashed]  (-1.8,0)--(-0.8,0);  
\draw (0.8,0) arc(0:180:0.8);
\draw[dashed] (-0.8,0) arc( 180:360:0.8);
\draw[dashed]  (1.8,0)--(0.8,0); 
\draw[fill=black] (-1.8,0) circle (0.1); 
\draw[fill=black] (1.8,0) circle (0.1); 
\end{tikzpicture}
+
\begin{tikzpicture}[line width=1 pt, scale=0.6, baseline=-0.1cm]
\draw[dashed]  (-1.5,0)--(1.5,0); 
\begin{scope}[shift={(0,0)},scale=0.6]
	   \draw[fill=white,thin] (0,0) circle (0.3);
            \draw[thin] (45:0.3)--(225:0.3);
            \draw[thin] (-45:0.3)--(135:0.3);
	\end{scope}
\draw[fill=black] (-1.5,0) circle (0.1); 
\draw[fill=black] (1.5,0) circle (0.1); 
\end{tikzpicture} 
+\cdots\\
&=
g (z_1,z_2)
 + (-8b)^2 D_{\chix\chix}(z_1,z_2)
 +\frac72 \cdot (-4)b^{2}\,\widehat\Sigma_\zeta(z_{1},z_{2})+...,  \no 
\end{align}
where $D_{\chix\chix}$ and $\widehat\Sigma_\zeta$  are  given in   \rf{32} and \rf{sigma}.
In the limit when one point is  sent to the boundary,  $\sfz_2\rightarrow 0$, the expression for \rf{336} 
gets simplified to  
\be
\EV{\xi^+(\sfz_1 ,\sft_1) \xi^-(\sfz_2,\sft_2)}_{\rm conn}  = \hat g (\sfz_1 ,\sft_1; \sfz_2,\sft_2)  
= \frac43 \Big[  \frac{ \sfz_1 \sfz_2}{ \sfz_1^2 +   \bt_{12}^2 } +\mc O(\bz_{2}^{2})
\Big]^{2+b^2} \Big[1+ 2b^2( \log 2-1 ) \Big] +\mc O(b^4) ,  \la{337}
\ee
where we  kept only  the leading order term  in  $\bz_{2}\to 0$ and 
defined   $\xi^{\pm} = \frac{1}{\sqrt 2}(\xi_{1}\pm i \xi_{2})$ as in \rf{1.25}.
The above expression is consistent  with the field $\xi$ being dual to an operator with dimension 
\be\la{338}
\Delta_{_V}= 2+\gamma_{_V}\  , \qquad\qquad  \gamma_{_V}=b^2+ \mc O(b^4)~,
\ee  
in agreement with \rf{1.26} obtained from   the CFT. 
Then the 1-loop corrected  bulk-to-boundary propagator turns out to be
\be\la{1loopPropagator}
\hat g
(\sft'; \sft, \sfz)=\lim_{\sfz' \rightarrow 0}  \frac{1}{\sfz'^{\Delta_{V}}}\EV{\xi^+(\sfz ,\sft ) \xi^-(\sfz' ,\sft')}_{\rm conn}  
= \frac43 \Big(  \frac{ \sfz}{ \sfz^2 +    (\sft-\sft')^2 } \Big)^{\Delta_{V}} \Big[1+ 2b^2( \log 2-1 ) \Big]+\mc O(b^{4})~. 
\ee
Setting both legs to the boundary $\sfz,\sfz' \rightarrow 0$,
 we get  the  boundary limit of this  correlator (cf. \rf{1.25},\rf{1.27}) 
 with the coefficient  in (\ref{1.32}) being 
\be
\la{3.35}
C_{+-} = \frac43  \Big[1+ 2b^2(\log2 -1)+\cdots \Big]\ . 
\ee
This is also in agreement with  \rf{1.34} where we substituted  \rf{3.30} and  the 1-loop expression for $\kv $ in   (\ref{1.30}).

The  exact  form of $\kv$ proposed in   (\ref{1.30})
   is motivated by  the free-field representation (\ref{3.25}) of $V^{\pm}$ and the 
OPE (\ref{3.30}).
Following the  argument  used to determine  $\kappa_{3}$ of  the abelian  Toda  theory  \rf{1.18} 
in  Appendix  \ref{app:kappa3}, one may conjecture that 
\be
\la{3.36}
\kv = 2\sqrt{2}\,\,\frac{p}{C^{+-}}\,\,2^{1/k} = \frac{2^{1+1/k}\,\sqrt{k-2}}{3\,(k+2)}\ ,
\ee 
where the factor $2^{1/k}$ takes into account the anomalous dimension \rf{3.29}  of $V^{\pm}$ and the  overall
coefficient is  fixed by matching with  known  tree level  value 
$\kv = \frac{2}{3}\,b+\cdots$.
Written in terms of $b= {1 \ov \sqrt { k-2}}$,   eq. (\ref{3.36})  becomes the same  as    (\ref{1.30}) (see also  \rf{1.26}).

 
Next, let us consider  the three-point  correlator in \rf{1.27}, i.e. 
  {$\llangle\Phi\Phi\Phi\rrangle\sim \EV{TTT}$}.
Again, we can use  the result in the  Liouville theory since the virtual $\xi_{1,2}$ fields behave
as copies of the Liouville field, taking into account the  values of the  coupling coefficients in \rf{3.18}. 
Thus  we find for $C_{222}$   in \rf{1.36} 
\be\la{3.39}
C_{222}= -\frac{16 b}{9}+ 3 \times \frac{64 b^3}{27} +\cdots =
-\frac{16}{9}\,b+\frac{64}{9}\,b^{3}+\cdots \ . 
\ee
This  agrees with the expected expression in   (\ref{1.37}).

 Turning to  the second   three-point function   {$\llangle\Phi \Phi^{+}\Phi^{-}\rrangle\sim \EV{TV^{+}V^{-}}$}, 
using  (\ref{3.35}), the prediction   for its coefficient  in \rf{1.36}  is given in \rf{1.37}, i.e. 
 \be
 \la{3.38}
 C_{2+-}(b) = -\frac{16}{9}\,b+\frac{16}{9} \Big(5- 2\,\log 2\Big)\,b^{3}+\cdots.
 \ee
 We have  confirmed it by a detailed  computation described in  Appendix \ref{app:c2pm}.

\subsection{Four-point functions\la{sec:44}}

The structure of $\langle TTTT\rangle$  is fully  determined by the conformal symmetry \rf{1333}.
The \chads correspondence \rf{1.28} then   predicts that the 
corresponding boundary correlator should be given by 
  (\ref{1.12}),\rf{1.13}.  These predictions can be checked in the same way 
  as in the Liouville theory, taking into account  
 the multiplicity of  virtual exchanges of $\chix$ and $\xi_i$. 
 Indeed,   for the coefficient  of the connected part  in \rf{1.12} 
 we have in  the Liouville  and in  the non-abelian Toda  cases 
 \be \la{339}
 C_{2222} = \k^4 c = \frac{256\,Q^{4}}{c^{3}} = \begin{cases}
 \frac{32}{27}\,b^{2}-\frac{80}{27}\,b^{4}+\frac{496}{81}\,b^{6}+\cdots, & \ \ \text{Liouville} \\
 \frac{32}{27}\,b^{2}-3\times \frac{80}{27}\,b^{4}+\frac{464}{9}\,b^{6}+\cdots, & \ \ \text{NAT}
 \end{cases}
 \ee
 where we used the model-dependent specific values of  $Q$ and $c$ (see \rf{1.1},\rf{1.2},\rf{1.23},\rf{1.24}).
  As expected (cf. \rf{3.35},\rf{3.39}), 
 the one-loop $b^4$ correction in the NAT model is simply three times that of the Liouville theory.
  There is no such  simple relation  at higher orders in $b$.

 \iffa
Similarly, all cases
$C_{222}=-\frac{64Q^{3}}{c^{2}}$ and this gives the expansions
\begin{align}
\la{3.39}
C_{222}^\text{Liouv.} &=  -\frac{16}{9}\,b+\frac{64}{27}\,b^{3}
-\frac{100}{27}\,b^{5}+\cdots, \notag \\
C_{222}^\text{NAT} &= -\frac{16}{9}\,b+3\times \frac{64}{27}\,b^{3}
-\frac{268}{9}\,b^{5}+\cdots \ .   already in \rf{3.39}
\end{align}
and the one-loop correction in the \GS model is three times that of the Liouville theory. 
At higher loop order the pattern is not so trivial and the $\mc O(b^{5})$ coefficients 
in (\ref{3.39})
do not show any simple relation.
\fi

The  expressions for  $\langle TTV^+ V^-\rangle$ and $\langle  V^+ V^- V^+ V^-\rangle$ 
related  to the boundary correlators  in \rf{1.38} are not  fully fixed by the conformal symmetry 
and thus represent a particular interest.
In the first case  the dual  four-point  correlator $\llangle \Phi\Phi\Phi^{+}\Phi^{-}\rrangle$
has a disconnected contribution that is just $\llangle \Phi\Phi\rrangle\llangle\Phi^{+}\Phi^{-}\rrangle$ 
and it then  obeys \chads as a consequence of the relation  between  the two-point functions. 
In the connected part, at the tree  $\mc O(b^{2})$ level, 
the fields $\xi_i$  play again the same role as  the Liouville field $\chix$ and the 
matching  can be easily checked.

The story is more complicated for  the remaining  non-vanishing 
correlator $\llangle \Phi^{+}\Phi^{-}\Phi^{+}\Phi^{-}\rrangle$  which  according to 
\rf{1.28}   should be proportional to  the CFT   correlator 
$\langle V^{+}V^{-}V^{+}V^{-}\rangle$ restricted to the real line. 
The latter is  non-trivial because of the 
non-locality in the  OPE of $VV$  in  (\ref{3.30}) and (\ref{3.31}). 
To compute  this  correlator  one should use the free field
representation (\ref{3.25}) perturbatively  in  small $b$ or large $k$. 


\subsubsection*{\adst boundary correlator  $\llangle \Phi^{+}\Phi^{-}\Phi^{+}\Phi^{-}\rrangle$}

The boundary correlators  in \adst  have,  in general, the same form   as   correlators  in 1d CFT.
If we have four primary operators $\mc O$ in 1d CFT  of  the same  dimension $\Delta$, their correlator is constrained by the 
global  $SO(2,1)$ conformal invariance to   take the following form 
\be
\la{3.40}
\llangle \mc O_{1}(\bt_{1})\,\mc O_{2}(\bt_{2})\,
\mc O_{3}(\bt_{3})\,\mc O_{4}(\bt_{4})\rrangle = \frac{1}{(\bt_{12}\, \bt_{34})^{2\Delta}}
\, G(\chi)\ ,\qquad \qquad \chi = \frac{\bt_{12}\, \bt_{34}}{\bt_{13}\, \bt_{24}} \ . 
\ee
Here $\chi$ is 1d conformally invariant  cross-ratio. 
The function $G(\chi)$ in (\ref{3.40}) admits the $s$-channel expansion   (see, \eg, \cite{Dolan:2011dv})
\be
\la{3.41}
 G(\chi) = \sum_{h}c_{h}\,   {\sf F}_h (\chi)  \ , \qquad\qquad   {\sf F}_h \equiv  \chi^h\  {}_{2}F_{1}(h,h,2h,\chi)\ ,
\ee
where $h$ labels the conformal dimension of the fields  appearing in  the OPE 
$\mc O_{1}\mc O_{2} = [\mc O_{h}]+[\mc O_{h'}]+\cdots$, 
and the coefficients $c_{h}$ may be expressed in terms of the  coefficients in 
the 2-point  and 3-point functions of $\mc O_{1}, \mc O_{2}$ and the exchanged field. 

Let us  consider   the \adst   boundary correlator of   for fields  $\xi_i$ 
using the notation   $\XP_{i}$ for their  boundary values    as in \rf{1.25},\rf{1.27}.
According to (\ref{3.40})  the result should read 
\be\la{344}
\llangle \XP_{i_{1}}(\bt_{1}) \, \XP_{i_{2}}(\bt_{2})\, \XP_{i_{3}}(\bt_{3}) \, \XP_{i_{4}}(\bt_{4}) \rrangle = 
\frac{1}{(\bt_{12}\bt_{34})^{2\dev}} G_{i_{1}i_{2}i_{3}i_{4}}(\chi) \ , \qquad \qquad 
\dev = 2 + b^2 + ... \ ,
\ee
where in general we should have  $\dev =  2 + {b^2\ov 1 + 2 b^2}$ (see \rf{1.26},\rf{3.29}). 
At order $\mc O(b^{2})$ we have trivial disconnected contributions where the two-point function
$\llangle\XP_i \XP_j \rrangle$ includes propagator loop corrections. 
These  contributions
automatically respect the structure in (\ref{3.40}) and also are consistent   with \chads as discussed in section \ref{sec:33}
above. 
Indeed, the disconnected part  can be written 
in the generalized free field form 
\be\la{345}
G^{\rm disc}_{i_{1}i_{2}i_{3}i_{4}}(\chi)   = 
C_{+-}^{2}
\Big[\delta_{i_{1}i_{2}}\delta_{i_{3}i_{4}} + \chi^{2\de_{_{V}}}\,\delta_{i_{1}i_{3}}\delta_{i_{2}i_{4}}
+\Big(\frac{\chi}{1-\chi}\Big)^{2\de_{_{V}}}\,\delta_{i_{1}i_{4}}\delta_{i_{2}i_{3}}\Big] \ , 
\ee
where (cf. \rf{1.34})
\be\la{346}
 C_{+-}= \mc C + \OO(b^2)\ , \qquad  \qquad \mc C = 2\pi\frac{2}{3\pi}=\frac{4}{3} 
 \ee 
is the   coefficient in the two-point  function  of $\XP_i$ in \rf{1.32}. 
The disconnected  $\mc O(b^{2})$ contributions  can be found from the  tree-level  disconnected diagrams where the 
pairs of $\XP_i$ fields are connected by a tree propagator and the $b^{2}$ correction comes from the expansion 
of the prefactor in (\ref{3.40}) depending on the coupling $b$ 
 because of the   non-zero anomalous  dimension in $\dev$. 

The  disconnected contribution can be represented as (cf. \rf{346}) 
\begin{align}
 G^{\rm disc}_{i_{1}i_{2}i_{3}i_{4}}(\chi)  
  &= \mc C^{2}\,(\bt_{12}\bt_{34})^{2(2+b^{2}+\cdots)}\Big[
\frac{\delta_{i_{1}i_{2}}\delta_{i_{3}i_{4}}}{(\bt_{12}\bt_{34})^{2(2+b^{2}+\cdots)}}
+\text{2 crossed terms + self-energy loops}
\Big] \notag \\
&= G_{i_{1}i_{2}i_{3}i_{4}}^{(0)\, \rm disc}(\chi)+\Big[
G_{i_{1}i_{2}i_{3}i_{4}}^{(1)\,\rm disc}(\chi)
+G_{i_{1}i_{2}i_{3}i_{4}}^{(1)\,\rm self}(\chi)\Big]
+\mc O(b^{4}).
\la{3.47}
\end{align}
In particular, 
\begin{align}
G_{i_{1}i_{2}i_{3}i_{4}}^{(0)\, \rm disc}(\chi) &=
\mc C^{2}\Big[\delta_{i_{1}i_{2}}\delta_{i_{3}i_{4}} + \chi^{4}\,\delta_{i_{1}i_{3}}\delta_{i_{2}i_{4}}
+\frac{\chi^{4}}{(1-\chi)^{4}}\delta_{i_{1}i_{4}}\delta_{i_{2}i_{3}}\Big], \notag \\
G_{i_{1}i_{2}i_{3}i_{4}}^{(1)\,\rm disc}(\chi) &= b^{2}\,\mc C^{2}\,\chi^{4}\,\Big[
2\log\chi\,\delta_{i_{1}i_{3}}\delta_{i_{2}i_{4}}+2\frac{\log\chi-\log(1-\chi)}{(1-\chi)^{4}}\delta_{i_{1}i_{4}}
\delta_{i_{2}i_{3}}\Big], \notag \\
G_{i_{1}i_{2}i_{3}i_{4}}^{(1)\,\rm self}(\chi) &=2\,(2\log 2-2)\,b^{2}\,
G_{i_{1}i_{2}i_{3}i_{4}}^{(0)\, \rm disc}(\chi) \ .\la{3.48}
\end{align}

There are also  connected $\mc O(b^{2})$ tree diagrams of  the  two types (see Fig.~\ref{fig:4p}): 
 the  exchange diagrams of the form (a) and  the  contact diagram (b) 
where the internal vertex is the derivative-dependent  $\xi^{2}(\partial\xi)^{2}$ $\sigma$-model  interaction in \rf{3.18},
 i.e. 
the leading-order connected  contribution is 
\be  G_{i_{1}i_{2}i_{3}i_{4}}^{\rm  conn}  = G_{i_{1}i_{2}i_{3}i_{4}}^{(1)\,\rm exch}(\chi)  + G_{i_{1}i_{2}i_{3}i_{4}}^{(1)\,\rm cont}(\chi) + \cdots \ . \la{349} \ee

\begin{figure}[htb]
\centering
\begin{tikzpicture}[line width=1 pt, scale=0.5, rotate=0, baseline=0]
\coordinate (A1) at (140:2);
\coordinate (A2) at (-140:2);
\coordinate (A3) at (40:2);
\coordinate (A4) at (-40:2);
\coordinate (M1) at (-1,0);    \coordinate (M2) at (1,0);

\draw[dotted] (0,0) circle (2);
\draw (A1)--(M1)--(A2); \draw (A3)--(M2)--(A4); \draw (M1)--(M2);

\draw[fill=black] (A1) circle (0.12); \node[left] at (A1) {$\XP_{i_{1}}$};
\draw[fill=black] (A2) circle (0.12); \node[left] at (A2) {$\XP_{i_{2}}$};
\draw[fill=black] (A3) circle (0.12); \node[right, xshift=0.1cm] at (A3) {$\XP_{i_{3}}$};
\draw[fill=black] (A4) circle (0.12); \node[right, xshift=0.1cm] at (A4) {$\XP_{i_{4}}$};
\node[above] at (0,0) {$\chix$};
\node[right] at (2.2,0) {+ crossed diagrams};
\node at (0,-2.7) {(a)};
\end{tikzpicture}\hskip 1.5cm
\begin{tikzpicture}[line width=1 pt, scale=0.5, rotate=0, baseline=0]
\draw[dotted] (0,0) circle (2);
\draw (A1)--(0,0)--(A3); \draw (A2)--(0,0)--(A4); 
\draw[fill=black] (A1) circle (0.12); \node[left] at (A1) {$\XP_{i_{1}}$};
\draw[fill=black] (A2) circle (0.12); \node[left] at (A2) {$\XP_{i_{2}}$};
\draw[fill=black] (A3) circle (0.12); \node[right, xshift=0.1cm] at (A3) {$\XP_{i_{3}}$};
\draw[fill=black] (A4) circle (0.12); \node[right, xshift=0.1cm] at (A4) {$\XP_{i_{4}}$};
\draw[fill=black] (0,0) circle (0.12); 
\node at (0,-2.7) {(b)};
\end{tikzpicture}
\caption{Connected diagrams of order $\mc O(b^{2})$ contributing to 
$\llangle \XP_{i_{1}}(\bt_{1})\cdots \XP_{i_{4}}(\bt_{4}) \rrangle$}
\label{fig:4p}
\end{figure}
%
%
The contribution of the exchange diagrams in Fig.~ \ref{fig:4p} (a) sums up to 
\be\la{3.49}
G_{i_{1}i_{2}i_{3}i_{4}}^{(1)\,\rm exch}(\chi)  = 
\frac{1}{2\pi}\,16\,b^{2}\,\mc C^{4}\,\Big[
\delta_{i_{1}i_{2}}\delta_{i_{3}i_{4}}\frac{D_{1122}}{4\bt_{12}^{2}} + 
\delta_{i_{1}i_{3}}\delta_{i_{2}i_{4}}\frac{D_{1212}}{4\bt_{13}^{2}} + 
\delta_{i_{1}i_{4}}\delta_{i_{2}i_{3}}\frac{D_{1122}}{4\bt_{12}^{2}}\Big]\ ,
\ee
where the $D$-functions are  defined by the \adst integral \cite{DHoker:1999kzh,Dolan:2000ut,Dolan:2003hv}
\be\la{4.50}
D_{\de_{1}\de_{2}\de_{3}\de_{4}}(\bt_{1}, \bt_{2}, \bt_{3}, \bt_{4}) = \int 
\frac{d\bt\,d\bz}{\bz^{2}}\,\prod_{i=1}^{4}
\Big[\frac{\bz}{\bz^{2}+(\bt-\bt_{i})^{2}}\Big]^{\de_{i}}.
\ee
Using the known explicit expressions for  the $D$-functions   in (\ref{3.49}) we obtain 
\begin{align}
 G_{i_{1}i_{2}i_{3}i_{4}}^{(1)\,\rm exch}(\chi)= \frac{b^{2}}{4} \,\mc C^{4}\,\Big\{&
\delta _{i_1 i_2} \delta _{i_3 i_4} \big[ \frac{\chi ^4  \log
\chi }{(\chi -1)^2}-\frac{\chi ^2 }{\chi -1}-(\chi +2) \
\chi  \log (1-\chi )\big] 
\no\\
& +\delta _{i_1 i_4} \delta _{i_2,i_3} \
\big[-\frac{(\chi -3) \chi ^4  \log \chi }{(\chi \
-1)^3}+\frac{\chi ^3 }{(\chi -1)^2}+\chi ^2  \
\log (1-\chi )\big]\notag \\
& +\delta _{i_1 i_3} \delta _{i_2 i_4} \big[ \frac{(2 \chi \
-3) \chi ^4  \log \chi }{(\chi -1)^2}+\frac{\chi ^3 \
}{\chi -1}-(2 \chi +1) \chi ^2  \log (1-\chi )\big]\Big\}. \la{3.51}
\end{align}
The contact diagram in Fig.~\ref{fig:4p} (b) evaluates to 
\begin{align}
 G_{i_{1}i_{2}i_{3}i_{4}}^{(1)\, \rm cont}(\chi) = &\frac{b^{2}}{4}\,\mc C^{4}\,\Big\{ \delta _{i_1 i_4} \delta _{i_2 i_3}
\big[ -\frac{(2 \chi -3) (3 \chi -2) (3 \chi ^2+2 \chi +3) \chi }{4 (\chi 
-1)^2} \la{3.52} \\
&-\frac{1}{2} (3 \chi ^2-4 \chi +3) (3 \chi ^2+4 \chi +3) \log 
(1-\chi ) -\frac{(3 \chi -5) (3 \chi ^2-4 \chi +3) \chi ^4 \log \chi}{2 (1-\chi )^3}\big] \no \\
&  +\delta_{i_1 i_3} \delta _{i_2 i_4} \big[ -\frac{(\chi 
-3) (\chi +2) (8 \chi ^2-8 \chi +3) \chi }{4 (1-\chi )^3} 
 +\frac{1}{2} 
(2 \chi +3) (2 \chi ^2-2 \chi +3) \log (1-\chi ) \no \\
&-\frac{(2 \chi -5) (2 
\chi ^2-2 \chi +3) \chi ^4 \log \chi }{2 (1-\chi )^4}\big]  + \delta_{i_1 i_2} \delta _{i_3 i_4}\big[-\frac{(2 \chi +1) (3 \chi -1) (3 \chi^2-8 \chi +8) \chi ^2}{4 (1-\chi )^3}\no \\
& +\frac{1}{2} (3 \chi +2) (3 \chi 
^2-2 \chi +2) \chi  \log (1-\chi )  -\frac{(3 \chi ^2-10 \chi +10) (3 
\chi ^2-2 \chi +2) \chi ^4 \log \chi }{2 (1-\chi )^4}\big] \Big\}\ . \no 
\end{align}
All the three  of the above  contributions  (disconnected \rf{3.48}, exchange \rf{3.49} and contact \rf{3.52}) 
are separately crossing invariant.
 In particular, 
if we consider $\llangle \Phi^{+}\Phi^{-}\Phi^{+}\Phi^{-}\rrangle$  where $\Phi^\pm = {1\ov \sqrt 2} ( \Phi_1 \pm i \Phi_2)$ 
 correspond to bulk fields  $\xi^\pm$,
 we can check the 
$1\leftrightarrow 3$ crossing invariance relation for each of the  contributions
\be\la{353} 
G_{+-+-}(\chi) = \Big(\frac{\chi}{1-\chi}\Big)^{2(2+b^{2}+\cdots)}\,G_{+-+-}(1-\chi)\ .
\ee
The total  $\mc O(b^{2})$  term in $G=G^{\rm disc} + G^{\rm conn}$ 
turns out to be 
\begin{align}
\la{3.511}
G_{+-+-}(\chi) = C_{+-}^{2}\,\Big[1&+\frac{\chi^4}{(1-\chi)^4}\Big] +b^{2}\Big[-\frac{16\,\chi\,(6-19\chi+19\chi^{2})}{27\,(1-\chi)^{3}} \no \\
&
-\frac{32}{9}\,\Big[1+\frac{\chi^4}{(1-\chi)^4}\Big] \,\log(1-\chi)\Big]+\cdots\ . 
\end{align}
Notice the  cancellation of the $\log\chi$ terms  which  is a non-trivial fact  (cf.  \ci{Giombi:2017cqn,Beccaria:2019dws}). 
This is 
 in agreement with the OPE (\ref{3.30})
where the first exchanged primary has dimension 4 with no anomalous contribution.

Another interesting case is the  four-point function $\llangle \Phi^{+}\Phi^{+}\Phi^{-}\Phi^{-}\rrangle$.
Its coefficient function $G_{+ + - -}$  is related to the  above  one in \rf{3.511}  by the $2\leftrightarrow 3$ crossing transformation 
\be\la{3.55}
G_{+ + - -}(\chi) = \chi^{2(2+b^{2}+\cdots)}\,G_{+-+-}(\chi^{-1})\ .
\ee
From (\ref{3.51}) one finds
\begin{align}
G_{++--}(\chi) = &C_{+-}^{2}\,\Big[\chi^{4}+\frac{\chi^{4}}{(1-\chi)^{4}}\Big]
+b^{2}\,\Big\{
\frac{16 (6 \chi ^2-19 \chi +19) \chi ^4}{27 (1-\chi)^3} \la{3357} \\
& -\frac{32}{9}\,\Big[\chi^{4}+\frac{\chi^{4}}{(1-\chi)^{4}}\Big]
\, [\log (1-\chi )-2\, \log \chi]
\Big\}+\mc O(b^{4}).\no
\end{align}
Expanding in conformal blocks (cf. (\ref{3.41}))
\begin{align} \la{3.57}
G_{++--}(\chi) = \sum_{n=4}^{\infty}(a_{n}+b^{2}d_{n}+\cdots)\ {\sf F}_{h_n}(\chi) \ , \qquad \qquad 
h_n= {n+b^{2}\,\gamma_{n}+\cdots}\ , 
\end{align}
we find non-zero contributions only with  even values of  index $n$. The leading order
coefficients $a_{2n}$ are 
\be
a_{2n} = \frac{\sqrt\pi\,4^{3-2n}\,(2n-1)(2n-3)(n-1)\Gamma(2n+3)}
{81\,\Gamma(2n-\frac{1}{2})}\ , \la{358}
\ee
while the  first  few values of $d_{2n}$ are 
\be\la{359}
d_{4} = \frac{304}{27}, \qquad d_{6} = \frac{5056}{729}, 
\qquad d_{8} = \frac{219320}{61347},  \qquad \dots\ .
\ee
Remarkably, all anomalous dimensions $\gamma_n$ of the exchanged operators  are equal, {\em i.e.} 
\be\la{360}
\gamma_{n}=4 \ . 
\ee 
This  follows from the fact that  the $b^{2}\log\chi$ term in \rf{3357}  is 
proportional to the tree level disconnected contribution. Again,  this feature has a simple explanation on CFT side -- 
from the point of view of the OPE of $V^{+}V^{+}$.
 All operators in this  OPE have a common  exponential 
  factor of the form $\exp(i\frac{2\sqrt{2}}{\sqrt{k}}\vp_2)$
that is just the squared exponential in $V^{+}$ in \rf{3.25} 
and this  is the origin of the common anomalous contribution $\frac{1}{2}
(\frac{2\sqrt{2}}{\sqrt{k}})^{2} = \frac{4}{k}$. In particular, this is true for $n=4$, 
in agreement with  (\ref{3.32}), {\em i.e.}
$h_4=\Delta_{_{\Omega_0}}=4+\frac{4}{k} = 4+4b^{2}+\cdots$. 

\subsubsection*{CFT correlator  $\langle V^{+}V^{-}V^{+}V^{-}\rangle$}

We can now  test  \chads  by  comparing  the above result   for $\llangle \Phi^{+}\Phi^{-}\Phi^{+}\Phi^{-}\rrangle$     in \rf{344},\rf{3357} 
with  the CFT correlator 
$\langle V^{+}V^{-}V^{+}V^{-}\rangle$. The latter can be  found 
 using 
the free field representation (\ref{3.25})  and   expanding in  large $k$.
 This reduces to a purely Wick contraction
calculation since after the  $1/k$ expansion   the exponential  operators in \rf{3.25}   become a sum 
of polynomial  local operators.
 Representing the large $k$ expansion  as 
\be\la{361}
\langle V^{+}V^{-}V^{+}V^{-} \rangle = k^{2}\, \mc G_{0}(\chi) + k\,\mc G_{1}(\chi) + \cdots\ ,
\ee
we find 
\begin{align}
\mc G_{0} &= 9\Big[1+\frac{\chi^{4}}{(1-\chi)^{4}}\Big],\notag \\
\mc G_{1} &= \frac{3\,(24-102\chi+169\chi^{2}-134\chi^{3}+67\chi^{4})}{(1-\chi)^{4}}
- 18\Big[1+\frac{\chi^{4}}{(1-\chi)^{4}}\Big]\log(1-\chi)\ .\la{362} 
\end{align}
Notice that the logarithmic term  in (\ref{362}) arises from the 
contraction of two $\varphi_{2}$ fields  with no derivatives which appear from 
the  large $k$ expansion  of the exponential in (\ref{325}).
 As a result,  the coefficient of this  logarithmic term   is proportional to $\mc G_{0} $.

To satisfy  the  \chads relation \rf{1.28} we should have 
\be
\la{3.577}
C_{+-}^{2}\,{\sf G_{0}}+b^{2}{\sf G_{1}}+\cdots  = \kv^{2}(k^{2}\mc G_{0}+k \mc  G_{1} +\cdots)\ ,
\ee
where from    (\ref{3.51})  
\begin{align} \la{364}
{\sf G}_{0} &= 1+\frac{\chi^{4}}{(1-\chi)^{4}}\ ,\quad \no \\
{\sf G}_{1} &= -\frac{16\,\chi\,(6-19\chi+19\chi^{2})}{27\,(1-\chi)^{3}}
-\frac{32}{9 }\Big[1+\frac{\chi^{4}}{(1-\chi)^{4}}\Big]\,\log(1-\chi)\ .
\end{align}
Expanding (\ref{3.577}) in small $b= {1\ov k} + ...$ using the values of $\kv$  and   $C_{+-}(b)$  from \rf{1.30}  and   (\ref{1.34}) gives 
\be
\mc G_{0} = 9\,{\sf G}_{0},\qquad\qquad\qquad   \mc G_{1} = \tfrac{81}{16}\,{\sf G}_{1}+72\,{\sf G_{0}} \ . \la{365}
\ee
Comparing \rf{362}   with \rf{364}   we find that these relations  are indeed satisfied.

\section*{Acknowledgments}

We would like to thank  S. Giombi, E. Perlmutter  and  V. Rosenhaus  for 
useful  discussions of related questions. 
HJ was supported by Swiss National Science Foundation.
AAT was supported by the STFC grant ST/P000762/1.

\appendix

\section{Expression for $\kappa_{3}$ in   $A_{2}$ Abelian  Toda theory}
\la{app:kappa3}


To  suggest  an  exact expression for $\kappa_{3}(b)$ in (\ref{1.18}) 
we shall   use similar considerations as in the Liouville case in \ci{Beccaria:2019stp}.
We  shall start with a 
free field realization of the  spin 2  and spin 3  $\mc W_{3}$ symmetry  generators  $(T, V_3)$ 
based on two real free bosons $\phi_{1}$,$\phi_2$
with normalization (different by $\sqrt 2$ from the one in \rf{3.20}) 
\be
\phi_{i}(z)\phi_{j}(0) \sim -2\,\delta_{ij}\,\log z\  .\la{A.1} 
\ee
Explicitly (conformal normal ordering is understood in all composite operators) \cite{Fateev:1987zh}
\begin{align}
T &= -\frac{1}{4}\,(\partial\phi_{1})^{2}-\frac{1}{4}\,(\partial\phi_{2})^{2}+i\,\alpha_{0}\,\partial^{2}\phi_{1}, \la{a1}  \\
\QQ_3 &= \frac{\beta}{12\,i}\,\Big[
(\partial\phi_{2})^{3}-3\,(\partial\phi_{1})^{2}\,\partial\phi_{2}+3\,i\,\alpha_{0}\,\partial^{2}\phi_{1}\,\partial\phi_{2}
+9\,i\,\alpha_{0}\,\partial\phi_{1}\partial^{2}\phi_{2}+6\,\alpha_{0}^{2}\,\partial^{3}\phi_{2}\Big] \ , \la{A.2}
\end{align}
where (cf. \rf{1140})
\be\la{a2}
\alpha_{0} = \sqrt{\frac{2-c}{24}}=-\frac{i}{2}\,Q,\qquad\qquad  \beta = -\frac{4}{\sqrt{5c +22}}\ , \qquad \qquad c= 2 + 6 Q^2 \ . 
\ee
Naively, if we could ignore the  contribution of the potential term in the flat-space  Toda action \rf{1.14}  to its stress tensor 
(or  assume that the stress tensor becomes traceless at the quantum level) 
 we could  formally  identify  the fields in \rf{a1}  as  $(\phi_{1}, \phi_{2}) = 2\,(\vp, \psi)$.\foot{In fact, the transformation between 
 the fields in  the  Liouville or Toda action and the free fields
  involves a non-trivial B\"acklund transformation  \ci{Braaten:1982yn}.}

 In the semiclassical limit starting with \rf{1.14} one 
  may formally  eliminate   the  conformal factor of the AdS$_2$  metric by  redefining  the Liouville  field   as  
 $\varphi(t,\bz) \to  \phi(t,\bz) +  Q \log\bz$, i.e. transform  the  action  into the flat-space one.
  This leads to  
 the  Toda  CFT  for the  field $(\phi,\psi)$  defined on a   flat upper half-plane ($w=\bt+i\,\bz, \  \bz>0$)
 with the   stress  tensor $T(w) = -(\partial_w\phi)^{2} - (\partial_w\psi)^{2}  + Q\,\partial^{2}_w\phi$
 (where $\partial_{w} = \frac{1}{2}(\partial_{\bt}-i\,\partial_{\bz})$).
 The field $\phi$ has then the  
 boundary asymptotics $\phi(\bt,\bz)\big|_{\bz\to 0} =\bz^{2}\,\Phi(\bt) - Q\log\bz+\dots$.
 Taking the boundary 
 limit in $T$  ($w\to \bt$) gives 
$
 T(\bt)  =-\tfrac{3}{2}\, Q\,\Phi(t) + \mc O(\bz^{2}).
 $
 This is  precisely  the operator relation   which is required  for the validity of of the expression for $\k$ in \rf{1.8}, i.e.  
  $\kappa(b) = -\frac{2}{3}\,b+\cdots$.
Identifying $(\phi_{1}, \phi_{2}) = 2\,(\vp, \psi)$ we get in the boundary limit  $\bz \to 0$: 
  $\phi_{1} \to  2\,\bz^{2}\Phi-2\,Q\,\log \bz, \  \phi_{2} \to  2\,\bz^{3}\Phi_3$. 
The boundary limit  of  the product $\QQ_3\phi_{2}$ is then 
\be\la{a5} 
\QQ_3(w) \phi_{2}(0) \Big|_{\bz \to 0}  
\ {\to} \ 2 \QQ_3(\bt) \, \bz^{3}\Phi_{3}(0) = 2\,\bz^{3}\,\kappa_{3}\ 
\QQ_3(\bt)\QQ_3(0)\ , 
\ee
where we replaced $\Phi_3 \to \k_3 V_3$  as it should be in the correlation functions  in \rf{1.17}. 
Hence, using that   $\QQ_3(w)\QQ_3(0)\sim {c\ov 3} w^{-6}+\cdots$ we can write $\k_3$ in terms of the 
coefficient $K$ in the leading singularity in  the OPE \rf{a5}
\be
\la{A.5}
\kappa_{3} = \frac{3}{2\,c}\, K \ , \qquad \qquad 
\QQ_3(w)\phi_{2}(0)\Big|_{\bz\to 0} \sim  \bz^{3} \frac{K }{\bt ^{6}}  + \cdots \ . 
\ee
A straightforward computation using the free-field representation  \rf{A.2}  gives  
\be
\la{A.6}V_3 (w)\phi_{2}(0) \sim\frac{2\,i\,\beta\,\alpha_{0}^{2}}{w^{3}}
+\frac{3}{2}\frac{\beta\,\alpha_{0}\,\partial\phi_{1}(0)}
{w^{2}}+ \,\frac{\beta}{w}\Big[-\frac{i}{12}\mc X(0)+\frac{3}{2}\alpha_{0}\partial^{2}\phi_{1}(0)\Big]\ ,
\ee
where $\mc X = 6\,(\partial\phi_{1})^{2}-6\,(\partial\phi_{2})^{2}-6\,i\,\alpha_{0}\,\partial^{2}\phi_{1}$.
Then contributions to (\ref{A.5}) come from the first $\sim w^{-3}$ term 
and  also 
terms originating  from the $-2Q\log \bz$ piece in $\phi_{1}$, i.e.  
\begin{align}\la{a8}
\partial\phi_{1} = \frac{i\,Q}{\bz}+\cdots,\quad\qquad
-\frac{i}{12}\mc X+\frac{3}{2}\alpha_{0}\partial^{2}\phi_{1} = \frac{3\,i\,Q^{2}}{4\bz^{2}}+\cdots.
\end{align}
To  find  the latter one has to take into account the  upper half plane 
 mirror poles in the OPE (see,  for instance,   \cite{Schomerus:2002dc})
and expand near
the boundary $\bz \to 0$ 
according to\footnote{The Dirichlet boundary conditions for  the free fields $\phi_{1}$ and $\phi_{2}$
require a non trivial gluing map for the odd spin chiral generators, {\em i.e.} the non-trivial reflection relation 
$V_{3}(z) = -\overline{V}_{3}(\overline z)$ for the odd spin field $V_{3}$. The same 
happens in the familiar example of the spin 1 current $J(z) = i\,\partial\phi$ when $\phi$ has a Dirichlet boundary condition, 
see, \eg,  Sec. 4.1 of \cite{Recknagel:2013uja}.
}
\begin{align}
&\frac{1}{(\bt -i\,\bz)^{3}}-\frac{1}{(\bt+i\,\bz)^{3}} = 
\cdots -\frac{20\,i\,\bz^{2}}{ \bt^{6}}+\cdots\ ,  \qquad   \frac{1}{\bz}\Big[
\frac{1}{(\bt -i\,\bz)^{2}}+\frac{1}{(\bt+i\,\bz)^{2}}\Big] = \cdots+\frac{10\,\bz^{2}}{\bt^{6}}+\cdots,     \notag \\
&\qquad 
\frac{1}{\bz^{2}}\Big[
\frac{1}{(\bt -i\,\bz)}-\frac{1}{(\bt+i\,\bz)}\Big] = \cdots+\frac{2\,i\,\bz^{2}}{\bt^{6}}+\cdots.
\end{align}
This gives in total the following expression for the coefficient $K$ in the OPE in \rf{A.5}
\begin{align}
K = 2\,i\,\beta\,\alpha_{0}^{2}\times (-20\,i)+\frac{3}{2}\,\beta\,\alpha_{0}
\times i\,Q\times 10\,i+\beta\times \tfrac{3i}{4} Q^{2}\times 2i = - 4Q^{2}\,\beta \ . 
\end{align}
As a result, 
\be \la{a10}
\kappa_{3} = \frac{3}{2\,c}\, K  =  -\frac{6\,Q^{2}\,\beta}{c} = 
\frac{24\,Q^{2}}{c\,\sqrt{5c+22}} \ , \ee
which is the expression given in 
(\ref{1.18}).

\section{Fourier expansion of the $\de=3$ propagator in \adst}
\la{app:fourier}

A useful tool to compute the  \adst integrals in disk parametrization \rf{2.6}
is the Fourier representation of the $\de=2,3$ propagators \rf{23}.
As illustrated in \cite{Menotti:2003km}, see also App. A of \cite{Beccaria:2019stp},
the simplest diagrams that occur in the calculations of this paper are 
multiple \adst integrals that depend on a number of fixed  points and on internal 
points that are integrated in  $SU(1,1)$ covariant way. Using a Fourier representation,
one first integrates the relative angles of involved points and performs radial integration 
as the last step. The coefficients of the Fourier expansion in the 
 $\de=2$  case of $g(x,x')$  are given  in \cite{Menotti:2003km}.
 Here we need the generalization to  the  $\de=3$ case,
{\em i.e.} the propagator $h(x,x')$ in (\ref{23}). Using   the  disk coordinates 
 ($x\equiv z$, $x'\equiv z'$) we have  $\eta$ in \rf{23}   given by 
$\eta = \left|\frac{z-z'}{1-z\overline{z}'}\right|^{2}$.
The  Fourier expansion of  $h(z,z')$ then reads 
\begin{align}
h(z,z') &= \sum_{n=0}^{\infty} h_{n}(|z|^{2}, |z'|^{2})\,\cos(n\,\vartheta(z,z')),\la{b1} \\
h_{n}(x,y) &= \theta(x-y)\,c_{n}(y)\,d_{n}(x)+\theta(y-x)\,c_{n}(x)\,d_{n}(y) \ ,
\end{align}
where $\vartheta(z,z')$ is the angle between the disc points $z$ and $z'$. 
The coefficients $c_{n}(x)$ are given by (for any $n\ge 0$)
\begin{align}
c_{n}(x) &= \frac{x^{n/2}}{(1-x)^{2}}\Big[1-\frac{2\,(n-2)}{n+1}\,x+\frac{(n-1)(n-2)}{(n+2)(n+1)}\,x^{2}\Big].
\end{align}
The coefficients $d_{n}(x)$ take the following special values for $n=0,1,2$
\begin{align}
d_{0}(x) &= \frac{3 (x+1)}{2 (x-1)}-\frac{(x^2+4 x+1) \log x}{2 (x-1)^2}, \no \\
d_{1}(x) &= \frac{-x^2-10 x-1}{(x-1) \sqrt{x}}+\frac{6 \sqrt{x} (x+1) \log x}{(x-1)^2}, \notag \\
d_{2}(x) &= \frac{x^3-7 x^2-7 x+1}{2 x-2 x^2}-\frac{6 x \log x}{(x-1)^2}, \la{b4}
\end{align}
while  for all $n\ge 3$
\begin{align}
d_{n}(x) =& \frac{1}{n(n-1)(n-2)\,x^{n/2}\,(1-x)^{2}}\,\Big[
-(n-2) (n-1) x^{n+2}
+2 (n-2) (n+2) x^{n+1}
\notag \\
&-(n+1) (n+2) x^n +(n+1) (n+2) x^2-2 (n-2) (n+2) x+(n-2) (n-1)
\Big]\ . \la{b5}
\end{align}

\section{Perturbative calculation of $C_{222}$ and $C_{233}$ in  $A_{2}$ abelian Toda theory}
\la{app:3p}


To compute three-point boundary correlators in \rf{43}  it is useful first to recall 
the form of  the tree level    bulk-to-boundary  scalar propagator in \adst  in disc parametrization. 
On the Poincar\'e half plane,   assuming Dirichlet boundary conditions, 
for generic  mass or  $\de$ ($m^2 = \de (\de-1)$) we may define (here $w= (\bt,\bz)$)
\be\la{c1}
g_{ \Delta}   (\sft, w') = \bz^{-\de}\,  g_\Delta (w,w')\Big|_{\sfz\rightarrow 0}\  , 
\ee
where for the $\de=2,3$  fields  $\chix$ and  $\psi$ of  the $A_{2}$ abelian Toda theory (cf. \rf{23}) 
\be\la{c2}
g(\sft,w'  ) \equiv g_2(\sft,w'  ) =  \frac43 \Big[  \frac{\sfz'}{\sfz'{}^2  +(\sft-\sft')^2} \Big]^2, \qquad 
h (\sft,w'  ) \equiv g_3(\sft,w'  )=  \frac{16}{ 15} \Big[  \frac{\sfz'}{\sfz'{}^2  +(\sft-\sft')^2} \Big]^3\ .
\ee
Under the map (\ref{2.6}) 
the disk boundary $|z|=1$ is mapped to the real axis $\sfz=0$ of Poincar\'e plane as 
\be
\la{2.28}
\sft(\theta)=- i\frac{e^{i\theta}+1}{e^{i\theta}-1}=-\cot \frac{\theta}{2} ,\qquad \qquad z=r\, e^{i\theta}\ .
\ee
Then the bulk-to-boundary propagators \rf{c2}   become 
\be
\la{2.29}
g(\theta,z') =\frac43 \frac{\sin^4\frac{\theta}{2} \, (1-|z'|)^2 }{|e^{i\theta}-z'|^4}\ , \qquad \qquad 
h(\theta,z')=  \frac{16}{ 15}  \frac{\sin^6\frac{\theta}{2}\,( 1-|z'|)^3 }{|e^{i\theta}-z'|^6}.
\ee
In general, for three   fields  of dimensions $\de_k$ we will have 
\begin{align} \la{cc1}
&\llangle\Phi_{\de_1}\Phi_{\de_2}\Phi_{\de_3}\rrangle = C_{\de_1 \de_2 \de_3}\,  \mathcal K_{\de_{1} \de_{2} \de_{3}}(\theta_1, \theta_2, \theta_3) \ , \\
&\mathcal K_{\de_{1} \de_{2} \de_{3}}(\theta_1, \theta_2, \theta_3) 
= |\sft(\theta_1) -\sft(\theta_2) |^{\de_{3}-\de_{1}-\de_{2}}\,
|\sft(\theta_2) -\sft(\theta_3) |^{\de_{1}-\de_{2}-\de_{3}}\,
|\sft(\theta_3) -\sft(\theta_1) |^{\de_{2}-\de_{1}-\de_{3}} \ . \no 
\end{align}
Let us now  consider the Witten  diagrams
contributing the two non-vanishing three-point  boundary correlators  $\llangle \Phi\Phi\Phi\rrangle$ and 
$\llangle\Phi\Phi_{3}\Phi_{3}\rrangle$.

 \subsection*{Coefficient $C_{222}$ in $\llangle\Phi\Phi\Phi\rrangle$}  

Let us start with  computing  the  coefficient $C_{222}$ of the 
contributions proportional to  $\mc K \equiv \mc K_{222}$ in \rf{cc1}.
The expression   for  $\llangle\Phi\Phi\Phi\rrangle$ will differ, in general, 
 from the one in the  Liouville theory because  here the second  field $\psi$ may also appear in virtual exchanges. 

\paragraph{Tree level diagram} At leading order in $b$  there is a single diagram 
\be
C^{(0)}_{222}\, b=
 \begin{tikzpicture}[line width=1 pt, scale=0.4, rotate=0,baseline=-0.1cm]
\coordinate (A1) at (90:2);  \coordinate (A2) at (210:2);  \coordinate (A3) at (-30:2);
\coordinate (B1) at (90:1);  \coordinate (B2) at (210:1);  \coordinate (B3) at (-30:1);
\draw[dotted] (0,0) circle (2);
\draw (A1)--(0,0); \draw (A2)--(0,0); \draw (A3)--(0,0);
\draw[fill=black] (A1) circle (0.1); \draw[fill=black] (A2) circle (0.1); \draw[fill=black] (A3) circle (0.1); 
\end{tikzpicture}  
=\frac{1}{\mathcal K} \,(-8b)\, \int {\sf d}^2 w \, g(\sft_1, w) g(\sft_2, w) g(\sft_3, w) 
=\frac{2 }{9}\, (-8 b ) =-\frac{16}{9} b\  . \la{c6}
\ee
This is a special  case of the  general expression
{\small
\be
\la{2.32}
\frac{1}{\mc K}\int {\sf d}^2 w \, g_{ \de_{1}}(\sft_1, w) g_{ \de_{2}}(\sft_2, w) g_{ \de_{3}}(\sft_3, w)
= \frac{\pi}{8}\,\frac{
\Gamma(\frac{\de_{1}+\de_{2}-\de_{3}}{2})
\Gamma(\frac{\de_{1}+\de_{3}-\de_{2}}{2})
\Gamma(\frac{\de_{2}+\de_{3}-\de_{1}}{2})
\Gamma(\frac{\de_{1}+\de_{2}+\de_{3}-1}{2})}
{\Gamma(\frac{1}{2}+\de_{1})\Gamma(\frac{1}{2}+\de_{2})\Gamma(\frac{1}{2}+\de_{3})},
\ee
}
giving $\frac{2}{9}$ for $\de_{1}=\de_{2}=\de_{3}=2$.

\paragraph{Diagrams with dressed propagators} These diagrams  give 
\be
C^{(1)}_{222,1}\,b^{3} = 
\begin{tikzpicture}[line width=1 pt, scale=0.4, rotate=0,baseline=-0.1cm]
\coordinate (A1) at (90:2);  \coordinate (A2) at (210:2);  \coordinate (A3) at (-30:2);
\coordinate (B1) at (90:1);  \coordinate (B2) at (210:1);  \coordinate (B3) at (-30:1);
\draw[dotted] (0,0) circle (2);
\draw (A1)--(0,0); \draw (A2)--(0,0); \draw (A3)--(0,0);
\draw[    fill=white] (0,1) circle (0.5);
\draw[fill=black] (A1) circle (0.1); \draw[fill=black] (A2) circle (0.1); \draw[fill=black] (A3) circle (0.1); 
\end{tikzpicture}
\quad+ \quad
\begin{tikzpicture}[line width=1 pt, scale=0.4, rotate=0,baseline=-0.1cm]
\coordinate (A1) at (90:2);  \coordinate (A2) at (210:2);  \coordinate (A3) at (-30:2);
\coordinate (B1) at (90:1);  \coordinate (B2) at (210:1);  \coordinate (B3) at (-30:1);
\draw[dotted] (0,0) circle (2);
\draw (A1)--(0,0); \draw (A2)--(0,0); \draw (A3)--(0,0);
\draw[dashed , fill=white] (0,1) circle (0.5);
\draw[fill=black] (A1) circle (0.1); \draw[fill=black] (A2) circle (0.1); \draw[fill=black] (A3) circle (0.1); 
\end{tikzpicture}
\quad+ \quad \text{2  permutations} \ . 
\ee
Their contribution is 
\be
C^{(1)}_{222,1} =3 \, (-\frac13)\,C^{(0)}_{222} =-C^{(0)}_{222}.
\ee

\paragraph{Diagrams with  $\chix$ loop} The  contribution of the diagrams
 \be
 C^{(1)}_{222,2} b^3 = 
\begin{tikzpicture}[line width=1 pt, scale=0.4, rotate=0,baseline=0]
\coordinate (A1) at (90:2);  \coordinate (A2) at (230:2);  \coordinate (A3) at (-50:2);
\draw[dotted] (0,0) circle (2);
\draw (A1)--(0,0.7); \draw (0,0) circle(0.7); \draw (A2)--(0,-0.7)--(A3);
\draw[fill=black] (A1) circle (0.1); \draw[fill=black] (A2) circle (0.1); \draw[fill=black] (A3) circle (0.1); 
\node[above] at (A1) {$\theta_{1}$};
\node[left] at (A2) {$\theta_{2}$};
\node[right] at (A3) {$\theta_{3}$};
\end{tikzpicture}
\quad+ \quad \text{2  permutations},
\ee
may be found  by taking  the boundary limit of  (the propagator  on  the r.h.s. is  bulk-to-boundary one)
 \be
 \la{C.11}
\lim_{\sfz_1 \to 0}\frac{1}{\mathsf{z}_{1}^{2}}\ \ 
\begin{tikzpicture}[line width=1 pt, scale=0.4, rotate=0,baseline=-0.1cm]
\draw (-2,0)--(-1,0); \draw (0,0) circle(1); 
\node[left] at (-1.8,0) {$z_{1}$};
\draw[fill=black] (1,0) circle(0.12); \draw[fill=black] (-2,0) circle(0.12);
\node[right] at (1.12,0) {$z_{2}$};
\end{tikzpicture}
= \frac{1}{8}
\begin{tikzpicture}[line width=1 pt, scale=0.4, rotate=0,baseline=-0.1cm]
\draw (-2,0)--(1,0);
\node[left] at (-1.8,0) {$\theta_{1}$};
\draw[fill=black] (1,0) circle(0.12); 
\node[right] at (1.12,0) {$z_{2}$};
\draw[thin] (200:2) arc(200:160:2);\draw[fill=black] (-2,0) circle(0.12);
\end{tikzpicture}\ .
\ee
This gives
\be
C^{(1)}_{222,2} b^3= \frac18 \, (-8b)\,(-16b^2) \, \frac12 \, C^{(0)}_{222}b\,   (- { 1 \ov 8b})  \times 3
=-3\, C^{(0)}_{222}\, b^3.
\ee

\paragraph{Diagrams with  $\psi$ loop} Similarly, the diagrams
\be
C^{(1)} _{222, 3} b^3 = 
\begin{tikzpicture}[line width=1 pt, scale=0.4, rotate=0,baseline=0]
\coordinate (A1) at (90:2);  \coordinate (A2) at (230:2);  \coordinate (A3) at (-50:2);
\draw[dotted] (0,0) circle (2);
\draw   (A1)--(0,0.7); \draw[dashed]   (0,0) circle(0.7); \draw  (A2)--(0,-0.7)--(A3);
\draw[fill=black] (A1) circle (0.1); \draw[fill=black] (A2) circle (0.1); \draw[fill=black] (A3) circle (0.1); 
\node[above] at (A1) {$\theta_{1}$};
\node[left] at (A2) {$\theta_{2}$};
\node[right] at (A3) {$\theta_{3}$};
\end{tikzpicture}\quad+ \quad \text{2  permutations}\ ,
\ee
may be computed from 
 \be
 \la{C.14}
\lim_{\sfz_1 \to 0}\frac{1}{\mathsf{z}_{1}^{2}}\ \ 
\begin{tikzpicture}[line width=1 pt, scale=0.4, rotate=0,baseline=-0.1cm]
\draw (-2,0)--(-1,0); \draw[dashed] (0,0) circle(1); 
\node[left] at (-1.8,0) {$z_{1}$};
\draw[fill=black] (1,0) circle(0.12); \draw[fill=black] (-2,0) circle(0.12);
\node[right] at (1.12,0) {$z_{2}$};
\end{tikzpicture}
= \frac{1}{24}
\begin{tikzpicture}[line width=1 pt, scale=0.4, rotate=0,baseline=-0.1cm]
\draw (-2,0)--(1,0);
\node[left] at (-1.8,0) {$\theta_{1}$};
\draw[fill=black] (1,0) circle(0.12); 
\node[right] at (1.12,0) {$z_{2}$};
\draw[thin] (200:2) arc(200:160:2);\draw[fill=black] (-2,0) circle(0.12);
\end{tikzpicture} \ ,
\ee
and give
\be
C^{(1)} _{222, 3} b^3= \frac{1}{24}  \, \big( -12b \, 2! \big) \, \big( -12b^2  \, 2! \, 2! \big) \, \frac12 \, C^{(0)}_{222} b\, ( - {1\ov 8b})  
\times 3
= - 9\,C^{(0)}_{222}\, b^3.
\ee

 \paragraph{Diagram with vertex insertion} 
 
According to the structure of the  interaction Lagrangian in  (\ref{2.8}), at order $b^3$ we  need to  include 
also a special tree diagram    
\be
C^{(1)}_{222, 4}\,b^{3} = 
\begin{tikzpicture}[line width=1 pt, scale=0.4, rotate=0,baseline=-0.1cm]
\coordinate (A1) at (90:2);  \coordinate (A2) at (210:2);  \coordinate (A3) at (-30:2);
\coordinate (B1) at (90:1);  \coordinate (B2) at (210:1);  \coordinate (B3) at (-30:1);
\draw[dotted] (0,0) circle (2);
\draw (A1)--(0,0); \draw (A2)--(0,0); \draw (A3)--(0,0);
\draw[fill=black] (A1) circle (0.1); \draw[fill=black] (A2) circle (0.1); \draw[fill=black] (A3) circle (0.1); 
\begin{scope}[shift={(0,0)},scale=0.8]
	   \draw[fill=white,thin] (0,0) circle (0.3);
            \draw[thin] (45:0.3)--(225:0.3);
            \draw[thin] (-45:0.3)--(135:0.3);
	\end{scope}
\end{tikzpicture}\ ,
\ee
where the insertion  represents  the $b^3$   vertex  appearing due to the overall   factor  $1+4b^{2}$ in \rf{2.8}.
%
%
This gives (cf. \rf{c6}) 
\be
C^{(1)}_{222, 4}   = 4 \, C^{(0)}_{222}. 
\ee

\paragraph{Triangle diagram with  $\chix$  loop } 

This non-trivial diagram is the same as in the Liouville theory and thus 
can be found  from    \cite{Beccaria:2019stp}:
\be
 C^{(1)}_{222,5}  = b^{-3}\,
\begin{tikzpicture}[line width=1 pt, scale=0.4, rotate=0,baseline=-0.1cm]
\coordinate (A1) at (90:2);  \coordinate (A2) at (210:2);  \coordinate (A3) at (-30:2);
\coordinate (B1) at (90:1);  \coordinate (B2) at (210:1);  \coordinate (B3) at (-30:1);
\draw[dotted] (0,0) circle (2);
\draw (B1)--(B2)--(B3)--(B1);
\draw (A1)--(B1); \draw (A2)--(B2); \draw (A3)--(B3);
\draw[fill=black] (A1) circle (0.1); \draw[fill=black] (A2) circle (0.1); \draw[fill=black] (A3) circle (0.1); 
\end{tikzpicture}
=\frac76 \, C^{(0)}_{222}
=-\frac{56}{27}.
\ee

\paragraph{Triangle diagrams with $\psi$  loop}    

This diagram is given by the following  finite disk integral
\begin{align}
& C^{(1)}_{222,6}\, b^3 = 
\begin{tikzpicture}[line width=1 pt, scale=0.4, rotate=0,baseline=-0.1cm]
\coordinate (A1) at (90:2);  \coordinate (A2) at (210:2);  \coordinate (A3) at (-30:2);
\coordinate (B1) at (90:1);  \coordinate (B2) at (210:1);  \coordinate (B3) at (-30:1);
\draw[dotted] (0,0) circle (2);
\draw[dashed] (B1)--(B2)--(B3)--(B1);
\draw (A1)--(B1); \draw (A2)--(B2); \draw (A3)--(B3);
\draw[fill=black] (A1) circle (0.1); \draw[fill=black] (A2) circle (0.1); \draw[fill=black] (A3) circle (0.1); 
\end{tikzpicture}\notag \\
&= 
  \frac{\big( -24b \big)^3}{\mathcal K}     
 \int {\sf d}^2 w_1 {\sf d}^2 w_2 {\sf d}^2 w_3 \, g(\sft_1, w_1) g(\sft_2, w_2) g(\sft_3, w_3)  h(w_1,w_2) h(w_1,w_3) h(w_2,w_3).
\end{align}
Its  numerical calculation by the same method as  discussed in \cite{Beccaria:2019stp} gives\foot{In principle,
 one could  try to use  an analytic  approach based on 
 an exact integral representation for the triangle diagram.
  One such option is the split-representation discussed, \eg,  in  \cite{Liu:2018jhs}. 
  However, if one is interested in the final number  and 
   not just  in special analytical features (like residues)  such representation does not appear to be  very useful.} 
\be
\la{2.46}
C^{(1)}_{222, 6}  = -5.619  \pm 0.011.
\ee

\paragraph{Total  coefficient of three-point correlator}

The duality  prediction (\ref{1.21}) amounts to 
 \be
\frac{C^{(1)}_{222}}{C^{(0)}_{222}}\equiv\sum_{a=1}^6  \frac{C^{(1)}_{222,a}}{C^{(0)}_{222}} = 
 -\frac{14}{3}.
\ee
Thus, we should have the following  exact value of the contribution $C^{(1)}_{222,6}$ in \rf{2.46}
\be
 C^{(1)}_{222, 6} =C^{(0)}_{222} \Big( -\frac{14}{3} +1-4+3+9 -\frac76 \Big) = 
 \frac{19}{6} C^{(0)}_{222} =- \frac{152}{27}= -5.\overline{629}.
\ee
Comparing this to (\ref{2.46}), a relative error is just   $0.2\%$, 
thus  confirming  the   validity of  (\ref{1.21}).

\subsection*{Coefficient $C_{233}$ in $\llangle\Phi\Phi_{3}\Phi_{3}\rrangle$} 

Let us now check the prediction \rf{1.21} for $C_{233}$. 
Here we shall use the notation  $\mc K\equiv \mc K_{233}$  (see (\ref{cc1})).  

\paragraph{Tree level diagram}  The  leading order  $\OO(b)$ contribution is given by 
 \be
C_{233}^{(0)} b=
 \begin{tikzpicture}[line width=1 pt, scale=0.4, rotate=0,baseline=-0.1cm]
\coordinate (A1) at (90:2);  \coordinate (A2) at (210:2);  \coordinate (A3) at (-30:2);
\coordinate (B1) at (90:1);  \coordinate (B2) at (210:1);  \coordinate (B3) at (-30:1);
\draw[dotted] (0,0) circle (2);
\draw (A1)--(0,0); \draw[dashed]  (A2)--(0,0); \draw[dashed]  (A3)--(0,0);
\draw[fill=black] (A1) circle (0.1); \draw[fill=black] (A2) circle (0.1); \draw[fill=black] (A3) circle (0.1); 
\node[above] at (A1) {$\theta_{1}$};
\node[left] at (A2) {$\theta_{2}$};
\node[right] at (A3) {$\theta_{3}$};
\end{tikzpicture}  
=\frac{1}{\mathcal K} \,(-24b)\, \int {\sf d}^2 w \, g(\sft_1, w) h(\sft_2, w) h(\sft_3, w) 
=\frac{4 }{45}\, (-24 b ) =-\frac{32}{15} b\ .\la{c23}
\ee

\paragraph{Diagrams with dressed propagators}  

The diagrams that are obtained from the tree level diagram by  adding loop to  one of the propagators are 
\be
C_{233, 1}^{(1)}\,b^{3} = 
\begin{tikzpicture}[line width=1 pt, scale=0.4, rotate=0,baseline=-0.1cm]
\coordinate (A1) at (90:2);  \coordinate (A2) at (210:2);  \coordinate (A3) at (-30:2);
\coordinate (B1) at (90:1);  \coordinate (B2) at (210:1);  \coordinate (B3) at (-30:1);
\draw[dotted] (0,0) circle (2);
\draw (A1)--(0,0); \draw[dashed] (A2)--(0,0); \draw[dashed] (A3)--(0,0);
\draw[    fill=white] (0,1) circle (0.5);
\draw[fill=black] (A1) circle (0.1); \draw[fill=black] (A2) circle (0.1); \draw[fill=black] (A3) circle (0.1); 
\end{tikzpicture}
\ \ +\quad
\begin{tikzpicture}[line width=1 pt, scale=0.4, rotate=0,baseline=-0.1cm]
\coordinate (A1) at (90:2);  \coordinate (A2) at (210:2);  \coordinate (A3) at (-30:2);
\coordinate (B1) at (90:1);  \coordinate (B2) at (210:1);  \coordinate (B3) at (-30:1);
\draw[dotted] (0,0) circle (2);
\draw (A1)--(0,0); \draw[dashed] (A2)--(0,0); \draw[dashed] (A3)--(0,0);
\draw[ dashed, fill=white] (0,1) circle (0.5);
\draw[fill=black] (A1) circle (0.1); \draw[fill=black] (A2) circle (0.1); \draw[fill=black] (A3) circle (0.1); 
\end{tikzpicture}
\ \ +\quad
\begin{tikzpicture}[line width=1 pt, scale=0.4, rotate=0,baseline=-0.1cm]
\coordinate (A1) at (90:2);  \coordinate (A2) at (210:2);  \coordinate (A3) at (-30:2);
\coordinate (B1) at (90:1);  \coordinate (B2) at (210:1);  \coordinate (B3) at (-30:1);
\draw[dotted] (0,0) circle (2);
\draw (A1)--(0,0); \draw[dashed] (A2)--(0,0); \draw[dashed] (A3)--(0,0);
\draw[ fill=white] (-30   : 1.4  )arc(  -30:150:0.5);
\draw[dashed, fill=white] (-30   : 1.4  )arc(  330:150:0.5);
\draw[fill=black] (A1) circle (0.1); \draw[fill=black] (A2) circle (0.1); \draw[fill=black] (A3) circle (0.1); 
\end{tikzpicture}
\ \ + \quad
\begin{tikzpicture}[line width=1 pt, scale=0.4, rotate=0,baseline=-0.1cm]
\coordinate (A1) at (90:2);  \coordinate (A2) at (210:2);  \coordinate (A3) at (-30:2);
\coordinate (B1) at (90:1);  \coordinate (B2) at (210:1);  \coordinate (B3) at (-30:1);
\draw[dotted] (0,0) circle (2);
\draw (A1)--(0,0); \draw[dashed] (A2)--(0,0); \draw[dashed] (A3)--(0,0);
\draw[ fill=white] (30   :-1.4  )arc( 210:30:0.5);
\draw[dashed, fill=white] (30   :-1.4  )arc( 210:390:0.5);
\draw[fill=black] (A1) circle (0.1); \draw[fill=black] (A2) circle (0.1); \draw[fill=black] (A3) circle (0.1); 
\end{tikzpicture}\ .
\ee
Their contribution is 
\be
C_{233, 1}^{(1)} =\Big[ (-\frac16 -\frac16)+2 \cdot (-\frac75)\Big]C^{(0)}_{233}   
=  -\frac{47}{15}\,C^{(0)}_{233}.
\ee

\paragraph{Diagram with  $\chix$ loop}  
 Taking into account (\ref{C.11})
 we find 
 \be
C^{(1)}_{233, 2} b^3 = \begin{tikzpicture}[line width=1 pt, scale=0.4, rotate=0,baseline=0]
\coordinate (A1) at (90:2);  \coordinate (A2) at (230:2);  \coordinate (A3) at (-50:2);
\draw[dotted] (0,0) circle (2);
\draw   (A1)--(0,0.7); \draw (0,0) circle(0.7); \draw[dashed]  (A2)--(0,-0.7)--(A3);
\draw[fill=black] (A1) circle (0.1); \draw[fill=black] (A2) circle (0.1); \draw[fill=black] (A3) circle (0.1); 
\node[above] at (A1) {$\theta_{1}$};
\node[left] at (A2) {$\theta_{2}$};
\node[right] at (A3) {$\theta_{3}$};
\end{tikzpicture}\ 
\ee
 \be
=\frac18 \, \Big( -\frac{4b}{3 } \, 3! \Big) \, \Big( -12b^2 \, 2! \, 2! \Big) 
\, \frac12 \, C^{(0)}_{233} b\,  (-{1\ov 24 b} ) 
=- C^{(0)}_{233}\, b^3.
\ee

\paragraph{Diagram with  $\psi$ loop}

In a similar way, the diagram   
\be
C^{(1)}_{233, 3}\, b^3 = 
\begin{tikzpicture}[line width=1 pt, scale=0.4, rotate=0,baseline=0]
\coordinate (A1) at (90:2);  \coordinate (A2) at (230:2);  \coordinate (A3) at (-50:2);
\draw[dotted] (0,0) circle (2);
\draw   (A1)--(0,0.7); \draw[dashed] (0,0) circle(0.7); \draw[dashed]  (A2)--(0,-0.7)--(A3);
\draw[fill=black] (A1) circle (0.1); \draw[fill=black] (A2) circle (0.1); \draw[fill=black] (A3) circle (0.1); 
\node[above] at (A1) {$\theta_{1}$};
\node[left] at (A2) {$\theta_{2}$};
\node[right] at (A3) {$\theta_{3}$};
\end{tikzpicture}\ ,
\ee
may be computed by using (\ref{C.14}) giving 
\be
C^{(1)}_{233, 3}\, b^3= \frac{1}{24}  \, \Big( -12b \, 2! \Big) \, \Big( -6b^2  \, 4! \Big) 
\, \frac12 \, C^{(0)}_{233}\, b\, ( -\frac{1}{24b} ) 
=-3\, C^{(0)}_{233}\, b^3.
\ee
 
 \paragraph{Diagrams with mixed $\chix\psi$ loop}
The two diagrams  
 \be
  C^{(1)}_{233,4}\,b^{3} = 
\begin{tikzpicture}[line width=1 pt, scale=0.4, rotate=0,baseline=0]
\coordinate (A1) at (90:2);  \coordinate (A2) at (230:2);  \coordinate (A3) at (-50:2);
\draw[dotted] (0,0) circle (2);
\draw[dashed]   (A1)--(0,0.7);  \draw[dashed]  (A2)--(0,-0.7); \draw (0,-0.7)--(A3);
\draw  (0,-0.7) arc(-90:90:0.7); 
\draw[dashed]  (0, 0.7) arc(90:270:0.7); 
\draw[fill=black] (A1) circle (0.1); \draw[fill=black] (A2) circle (0.1); \draw[fill=black] (A3) circle (0.1); 
\node[above] at (A1) {$\theta_{3}$};
\node[left] at (A2) {$\theta_{2}$};
\node[right] at (A3) {$\theta_{1}$};
\end{tikzpicture}
\quad +\quad
\begin{tikzpicture}[line width=1 pt, scale=0.4, rotate=0,baseline=0]
\coordinate (A1) at (90:2);  \coordinate (A2) at (230:2);  \coordinate (A3) at (-50:2);
\draw[dotted] (0,0) circle (2);
\draw[dashed]   (A1)--(0,0.7);  \draw[dashed]  (A2)--(0,-0.7); \draw (0,-0.7)--(A3);
\draw  (0,-0.7) arc(-90:90:0.7); 
\draw[dashed]  (0, 0.7) arc(90:270:0.7); 
\draw[fill=black] (A1) circle (0.1); \draw[fill=black] (A2) circle (0.1); \draw[fill=black] (A3) circle (0.1); 
\node[above] at (A1) {$\theta_{2}$};
\node[left] at (A2) {$\theta_{3}$};
\node[right] at (A3) {$\theta_{1}$};
\end{tikzpicture}\ ,
\ee
can be computed  using by  the relation
  \be
\lim_{\sfz_1 \to 0}\frac{1}{\mathsf{z}_{1}^{3}}\ \ 
\begin{tikzpicture}[line width=1 pt, scale=0.4, rotate=0,baseline=-0.1cm]
\draw[dashed] (-2,0)--(-1,0); 
\draw  (1,0) arc(0:180:1); 
\draw[dashed]  (-1,0) arc(180:360:1); 
\node[left] at (-1.8,0) {$z_{1}$};
\draw[fill=black] (1,0) circle(0.12); \draw[fill=black] (-2,0) circle(0.12);
\node[right] at (1.12,0) {$z_{2}$};
\end{tikzpicture}
=  \frac{1}{12}
\begin{tikzpicture}[line width=1 pt, scale=0.4, rotate=0,baseline=-0.1cm]
\draw [dashed](-2,0)--(1,0);
\node[left] at (-1.8,0) {$\theta_{1}$};
\draw[fill=black] (1,0) circle(0.12); 
\node[right] at (1.12,0) {$z_{2}$};
\draw[thin] (200:2) arc(200:160:2);\draw[fill=black] (-2,0) circle(0.12);
\end{tikzpicture}\ ,
\ee
leading to 
\be
 C^{(1)}_{233,4} =  \frac{1}{12} \, \Big( -12b \, 2! \Big) \, \Big( -12b^2 \, 2! \, 2! \Big) \,  
 C{(0)}_{233}  b\, ( -\frac{1}{24b} )   \times 2
= - 8\, C^{(0)}_{233}\, b^3.
\ee
 
\paragraph{Diagram with vertex insertion}

The diagram
\be
C^{(1)}_{233, 5}\, b^3  = 
\begin{tikzpicture}[line width=1 pt, scale=0.4, rotate=0,baseline=-0.1cm]
\coordinate (A1) at (90:2);  \coordinate (A2) at (210:2);  \coordinate (A3) at (-30:2);
\coordinate (B1) at (90:1);  \coordinate (B2) at (210:1);  \coordinate (B3) at (-30:1);
\draw[dotted] (0,0) circle (2);
\draw (A1)--(0,0); \draw[dashed]  (A2)--(0,0); \draw[dashed]  (A3)--(0,0);
\draw[fill=black] (A1) circle (0.1); \draw[fill=black] (A2) circle (0.1); \draw[fill=black] (A3) circle (0.1); 
\begin{scope}[shift={(0,0)},scale=0.8]
	   \draw[fill=white,thin] (0,0) circle (0.3);
            \draw[thin] (45:0.3)--(225:0.3);
            \draw[thin] (-45:0.3)--(135:0.3);
	\end{scope}
\end{tikzpicture}\ ,
\ee
comes  from  the $b^3$  cubic vertex  in the action (\ref{2.8}), giving 
\be
 C^{(1)}_{233, 5} =4\, C^{(0)}_{233}\ .
 \ee

\paragraph{Triangle  loop diagrams}

The  two most complicated  diagrams
 \be
 C^{(1)}_{233,6} b^3 = 
 \begin{tikzpicture}[line width=1 pt, scale=0.4, rotate=0,baseline=-0.1cm]
\coordinate (A1) at (90:2);  \coordinate (A2) at (210:2);  \coordinate (A3) at (-30:2);
\coordinate (B1) at (90:1);  \coordinate (B2) at (210:1);  \coordinate (B3) at (-30:1);
\draw[dotted] (0,0) circle (2);
\draw (B3)--(B1)--(B2) ;
\draw[dashed]  (B2)--(B3);
\draw (A1)--(B1); \draw[dashed]  (A2)--(B2); \draw[dashed]  (A3)--(B3);
\draw[fill=black] (A1) circle (0.1); \draw[fill=black] (A2) circle (0.1); \draw[fill=black] (A3) circle (0.1); 
\end{tikzpicture}
\quad +\quad
\begin{tikzpicture}[line width=1 pt, scale=0.4, rotate=0,baseline=-0.1cm]
\coordinate (A1) at (90:2);  \coordinate (A2) at (210:2);  \coordinate (A3) at (-30:2);
\coordinate (B1) at (90:1);  \coordinate (B2) at (210:1);  \coordinate (B3) at (-30:1);
\draw[dotted] (0,0) circle (2);
\draw[dashed]  (B3)--(B1)--(B2) ;
\draw  (B2)--(B3);
\draw (A1)--(B1); \draw[dashed]  (A2)--(B2); \draw[dashed]  (A3)--(B3);
\draw[fill=black] (A1) circle (0.1); \draw[fill=black] (A2) circle (0.1); \draw[fill=black] (A3) circle (0.1); 
\end{tikzpicture}\ ,
\ee
correspond to 
 \begin{align}
& C^{(1)}_{233,6} b^3 =\notag \\
& \frac{1}{\mathcal K}  \,( -8b ) \, ( -24b)^2 
\, \int {\sf d}^2 w_1 {\sf d}^2 w_2 {\sf d}^2 w_3 \, g(\sft_1, w_1) h(\sft_2, w_2) h(\sft_3, w_3)  g(w_1,w_2) g(w_1,w_3) h(w_2,w_3)
    \notag \\
&+
\frac{1}{\mathcal K}   \, ( -24b )^3\, 
 \int {\sf d}^2 w_1 {\sf d}^2 w_2 {\sf d}^2 w_3 \, g(\sft_1, w_1) h(\sft_2, w_2) h(\sft_3, w_3)
h(w_1,w_2) h(w_1,w_3) g(w_2,w_3)\  .
  \la{2.66}
 \end{align}
These  integrals require a 
numerical evaluation following the same   method  as in  \cite{Beccaria:2019stp} giving 
\be
\la{2.67}
C^{(1)}_{233,6}  =-11.47 \pm  0.03.
\ee

\paragraph{Total coefficient in three-point correlator}

From the expected  result for $C_{233}$ in \rf{1.21} we should have 
\be
\frac{C^{(1)}_{233}}{C^{(0)}_{233}}\equiv\sum_{a=1}^6 \frac{C^{(1)}_{233, a}}{C^{(0)}_{233}} = -\frac{86}{15}.
\ee
This corresponds to the following prediction for the numerical coefficient $C^{(1)}_{233,6}$
\be
C^{(1)}_{233,6}= C^{(0)}_{233}\,( -\frac{86}{15}  +\frac{47}{15}+1+3+8-4)=\frac{27}{ 5} C^{(0)}_{233} 
= -\frac{288}{ 25}=-11.52\dots\ .
\ee 
Compared with the numerical  value in  (\ref{2.67}), we find  good agreement with a relative error 
of about $0.4\%$. This strongly supports the validity of (\ref{1.21}).

\section{Perturbative calculation of $C_{2+-} $ in non-abelian Toda theory}
\la{app:c2pm}

Here  we collect the  details of the calculation of the coefficient $C_{2+-}$
in the boundary correlator in \rf{1.36} 
 leading to the result in \rf{1.37},(\ref{3.38}). The aim is  to reproduce the expansion 
 \be 
 \la{D.1}
 \llangle \XP^{+}(\sft_1) \XP^{-}(\sft_2)  \Phi(\sft_3)\rrangle  =\frac{\bC^{(0)}\,b\,}   
 {(\sft_1-\sft_2)^2(\sft_3-\sft_1)^2(\sft_3-\sft_2)^2}   
 \  \Big[ 1+ b^2 \,\Big(\bC^{(1)} + \bC^{(1)}_{\text{log}}   \log|\sft_{12}| \Big)+\cdots  \Big],
\ee
where 
\be
\la{D.2}
\bC^{(0)} = -\frac{16}{9}, \qquad\qquad  \bC^{(1)}=2\log 2-5, \qquad\qquad   \bC^{(1)}_{\text{log}}= -2 \ .
\ee
We shall  consider in turn various classes of loop diagrams   contributing to \rf{D.1} 
starting with the \adst action \rf{3.18}. The coefficient $\bC^{(1)}_{\rm log}$
will get contributions only from diagrams with dressed $\xi$ propagators while 
$\bC^{(1)}$ will receive several types of contributions.

\subsection*{Diagrams with dressed $\xi$ propagator}

The diagrams with dressed $\xi$ propagators can be obtained from the three-point tree-level  contact diagrams by replacing one of the bulk-to-boundary
$\xi$-propagator by its loop corrected version $g^{\text{1-loop}}(\sft, w)$  in \rf{1loopPropagator}. Explicitly it   can be computed as follows
{\small
\begin{align}
& (-8b) \, \int {\sf d}^2 w \; g^{\text{1-loop}}(\sft_1, w)\ g^{\text{1-loop}} (\sft_2, w)\  g (\sft_3, w) \notag \\
&=
(-8b)\Big[1+ b^2(2\log 2-2)  \Big] ^2\, \frac{1}{4\pi} \, \Big( \frac{4}{3} \Big)^3
\int \frac{d\sft d\sfz}{\sfz^2}   \Big(  \frac{ \sfz }{\sfz ^2 + (\sft-\sft_1)^2} \Big)^{2+b^2} 
\Big(  \frac{ \sfz }{\sfz ^2 + (\sft-\sft_2)^2} \Big)^{2+b^2}  \Big(  \frac{ \sfz }{\sfz ^2 + (\sft-\sft_3)^2} \Big)^2
\notag \\
&=
- \frac{16b}{ 9 } 
\frac{1}{ |\sft_{12}|^{2}  |\sft_{23}|^{2 }  
 |\sft_{31}|^{2 }  }
 \Big[1+  b^2   (2\log 2- \frac{10}{3})- 2 b^2 \log    |\sft_ {12}|   \Big] +\mc O(b^4)\ ,
\end{align}
}
where we used the relation  (\ref{2.32}) written in the form 
\be
\int \frac{d\sft d\sfz}{\sfz^2}  \prod_{i=1}^3  \Big(  \frac{ \sfz }{\sfz ^2 + (\sft-\sft_i)^2} \Big)^{\Delta_i}
= \frac{\sqrt{\pi}}{2}
 \frac{  \Gamma(\frac{\Delta_1+\Delta_2-\Delta_3}{2})  \Gamma(\frac{\Delta_2+\Delta_3-\Delta_1}{2})  \Gamma(\frac{\Delta_1+\Delta_3-\Delta_2}{2})  \Gamma(\frac{\Delta_1+\Delta_2+\Delta_3-1}{2}) } 
 {  \Gamma(\Delta_1)  \Gamma(\Delta_2)  \Gamma(\Delta_3)   
 |\sft_{12}|^{\Delta_1+\Delta_2-\Delta_3}  |\sft_{23}|^{\Delta_2+\Delta_3-\Delta_1}  
 |\sft_{31}|^{\Delta_3+\Delta_1-\Delta_2}    }.
\ee
Comparing with (\ref{D.1}) we see that the overall coefficient $\bC^{(0)}$ and the logarithmic coefficient
$\bC^{(1)}_{\rm log}$  already match (\ref{D.2})  while the  coefficient $\bC^{(1)}$ gets the following   contribution
\be
\la{D.5}
\bC^{(1)}_{1}=2\log 2- \frac{10}{3}.
\ee

\subsection*{Diagrams with non-derivative vertices}

There is a set of { simple} diagrams  which do not 
involve the derivative interaction vertex  $\xi^{2}(\partial\xi)^{2}$ in (\ref{3.18})
and thus 
 do not require new calculations: their contributions  can be obtained
from previous results  in the Liouville or abelian $A_{2}$ Toda theory by adjusting combinatorial coefficients
(below we indicate  this  by using the symbol ``$\times$'').  These are: 
\begin{enumerate}
\item Diagrams with dressed $\zeta$ propagator:
\be
\la{D.6}
\bC^{(1)}_{2} = -\frac{1}{6}\times 3 = -\frac{1}{2}.
\ee
\item Diagram  with extra  cubic  vertex from the factor $\frac{Q}{b} =\frac{1}{b^{2}}\,(1+3\,b^{2})$ in  (\ref{3.14}):
\be
\la{D.7}
\bC^{(1)}_{3} = 3\times 1 = 3.
\ee
\item Triangle diagrams  with two possible fields in the loop: 
\be
\la{D.8}
\bC^{(1)}_{4} = 2\times \frac{7}{6} = \frac{7}{3}.
\ee
\item Diagrams with a cubic and a non-derivative quartic vertex: (i) 
for  only $\chix$ in the loop (one cubic $\chix^{3}$ and one quartic $\chix^{2}\xi^{2}$ vertex) 
\be
\la{D.9}
\bC^{(1)}_{5} = -1 \ , 
\ee
and  with  both  $\chix$ and $\xi$ in the loop (one cubic $\chix \xi^{2}$ and one quartic $\chix^{2}\xi^{2}$ vertices) we get 
\be
\la{D.10}
\bC^{(1)}_{6} = 2\times (-2) = -4.
\ee

\end{enumerate}
 
 \subsection*{Diagrams with  derivative  vertex $\xi^{2}(\partial\xi)^{2}$}
 
 This are the diagrams  involve the  $\sigma$-model  derivative interaction 
 $\xi^{2}(\partial\xi)^{2}$ in (\ref{3.18}). There are  different types of them depending on  
 which leg (bulk-to-bulk or bulk-to-boundary) is acted on by  the two  derivatives.

\paragraph{Type I}

The  first relevant diagram is 
\be
\bC^{(1)}_{7}\,\bC^{(0)} b^3    = 
\begin{tikzpicture}[line width=1 pt, scale=0.8, rotate=0,baseline=0]
\coordinate (A1) at (90:2);  \coordinate (A2) at (230:2);  \coordinate (A3) at (-50:2);
\coordinate (B0) at ( 0:1 ); \coordinate (B1) at ( 180:1.6 );  \coordinate (B2) at (220:1.3);  \coordinate (B3) at (-60:1);
\draw[dotted] (0,0) circle (2);
\draw   (A1)--(0,0.7); \draw[dashed] (0,0) circle(0.7); \draw[dashed]  (A2)--(0,-0.7)--(A3);
\draw[fill=black] (A1) circle (0.1); \draw[fill=black] (A2) circle (0.1); \draw[fill=black] (A3) circle (0.1); 
\node[above] at (A1) {$\theta_{3}$};
\node[left] at (A2) {$\theta_{2}$};
\node[right] at (A3) {$\theta_{1}$};
\node[right] at (B0) {$\xi$};
\node[right] at (B1) {$ \xi$};
\node at (B2) {$\p\xi$};
\node[right] at (B3) {$ \p\xi$};
\node[above] at  (0,-0.7)  {$ w$};
\node[below] at  (0,0.7)  {$ w'$};
\end{tikzpicture}\ 
\ee
To evaluate it we find it convenient to use the identity in Eq.~(4.10) of \cite{Giombi:2017cqn}, {\em i.e.}
 \begin{align}
g^{ab} \p_a   & \widetilde K_{\Delta_1}(\sfz,\sft;\sft_1)\, \p_b   \widetilde K_{\Delta_2}(\sfz,\sft;\sft_2)\notag \\
&=\Delta_1\Delta_2 \Big[ 
  \widetilde K_{\Delta_1}(\sfz,\sft;\sft_1)  \widetilde K_{\Delta_2}(\sfz,\sft;\sft_2)
  -2\sft_{12}^2  \widetilde K_{\Delta_1+1} (\sfz,\sft;\sft_1)  \widetilde K_{\Delta_2+1}  (\sfz,\sft;\sft_2)
 \Big],
 \end{align}
 where $g^{ab}=\sfz^2  \delta^{ab}$ is the \adst metric and 
 $\widetilde{K}_{\de}(\bz, \bt, \bt') = \left(\frac{\bz}{\bz^{2}+(\bt-\bt')^{2}}\right)^{\de}$. This allows us  to 
 transform the expression for the above  diagram  into another one
  with  $\de=3$  external  propagators    and  only  non-derivative interactions. 
 Using that  (here $\mc C_{2}={2\ov 3 \pi}$ as in (\ref{2.2}))
\begin{align}
& \int  {\sf d}^2 w\;  \p^{a} g(w,\sft_1)   \p_a g(w ,\sft_2)g(w,\sft_3)
= (2\pi \mathcal C_2)^3   \int  {\sf d}^2 w\;  \p^{a} \widetilde K_2(w,\sft_1)   \p_a \widetilde K_2(w ,\sft_2)\widetilde K_2(w,\sft_3)\notag \\
&= 
(2\pi \mathcal C_2)^3   \int  {\sf d}^2 w\; 4\Big(   \widetilde K_2(w,\sft_1)      \widetilde K_2(w',\sft_2)  
-2 \sft_{12}^{2} \widetilde K_3(w,\sft_1)      \widetilde K_3(w ,\sft_2)    \Big)   \widetilde K_2(w,\sft_3)\notag \\
&= 
4(2\pi \mathcal C_2)^3  \Big( 
 \frac{3}{32\,  |\sft_{12} |^2 |\sft_{ 23} | ^2 |\sft_{31} |^2} -
 2\sft_{12}^2 \frac{15}{256\,   |\sft_{12   }|^4 |\sft_{ 23} |^2\sft_{31}|^2  } 
  \Big) =
   -\frac29\   \frac{1}{   |\sft_{12} |^2 |\sft_{ 23} |^2 |\sft_{31} |^2} \ , 
\end{align}
 this  diagram evaluates to (cf. \rf{31},(\ref{C.11}))
\begin{align}
\la{D.14}
 & \int  {\sf d}^2 w\;  \p^{a} g(w,\sft_1)   \p_a g(w ,\sft_2)\ \frac{1}{\sfz_3^2} B(w,w_3) \stackrel{\bz_{3}\to 0}{=} 
\frac18  \int  {\sf d}^2 w\;  \p^{a} g(w,\sft_1)   \p_a g(w ,\sft_2)g(w,\sft_3)\notag\\
&\qquad = 
  \frac18(   -\frac29  )  \,   \frac{1}{   |\sft_{12} |^2 |\sft_{ 23} |^2\sft_{31} |^2}= 
  -\frac{1}{36} \,   \frac{1}{   |\sft_{12} |^2 |\sft_{ 23} |^2\sft_{31} |^2} \ . 
\end{align}
As a result, its contribution to  $\bC^{(1)}$ is given by 
\be
\la{D.15}
\bC^{(1)}_{7}\,\bC^{(0)} b^3     =  (-8b) \, (b^2 \times 2\times 2) \, 2 \times\frac12 \,   \big(  -\frac{1}{36}  \big) =- \frac12 \bC^{(0)} b^3 \quad \to\quad \bC^{(1)}_{7} = -\frac{1}{2}\ , 
\ee
where the factor of 2 comes from the two species of the fields in the loop and  $\frac12$ is a symmetry factor.

\paragraph{Type II} If one derivative from $\xi^{2}(\partial\xi)^{2}$ is on a bulk propagator while the other is on a bulk-to-boundary
propagator, we get the diagram
\be
\bC^{(1)}_{8}\,\bC^{(0)} b^3    = 
\begin{tikzpicture}[line width=1 pt, scale=0.8, rotate=0,baseline=0]
\coordinate (A1) at (90:2);  \coordinate (A2) at (230:2);  \coordinate (A3) at (-50:2);
\coordinate (B0) at ( 0:1 ); \coordinate (B1) at ( 180:1.6 );  \coordinate (B2) at (220:1.3);  \coordinate (B3) at (-60:1);
\draw[dotted] (0,0) circle (2);
\draw   (A1)--(0,0.7); \draw[dashed] (0,0) circle(0.7); \draw[dashed]  (A2)--(0,-0.7)--(A3);
\draw[fill=black] (A1) circle (0.1); \draw[fill=black] (A2) circle (0.1); \draw[fill=black] (A3) circle (0.1); 
\node[above] at (A1) {$\theta_{3}$};
\node[left] at (A2) {$\theta_{2}$};
\node[right] at (A3) {$\theta_{1}$};
\node[right] at (B0) {$  \xi$};
\node[right] at (B1) {$\p \xi$};
\node[right] at (B2) {$ \xi$};
\node[right] at (B3) {$ \p\xi$};
\node[above] at  (0,-0.7)  {$ w$};
\node[below] at  (0,0.7)  {$ w'$};
\end{tikzpicture}\ 
\ee
It can be computed in terms of the function defined in (\ref{31})
\be
B (w,w_3) =\int {\sf d}^2 w'\;     \big[g(w,w')\big]^{2}\, g(w',w_3).
\ee
Taking a derivative over  $w^a$ gives 
\be
\la{D.18}
\p_a B (w,w_3) =2\int {\sf d}^2 w'\;   \p_a g(w,w')g(w,w')   g(w',w_3).
\ee
On the other hand, sending  $w_3$ to the boundary, we have 
\be
\lim_{\sfz_3\to 0} \frac{1}{\bz_{3}^{2}}B (w,w_3) =\frac{1}{8} g(w,\sft_3) \ , 
\ee
and thus 
\be
\int {\sf d}^2 w'\;   \p_a g(w,w')g(w,w')   g(w',w_3)
=\frac{1}{16}  \p_a g(w,\sft_3)\ .
\ee
These simple manipulations allow us  to write the contribution of  the above diagram as 
\begin{align}
&  \int {\sf d}^2 w'\; {\sf d}^2 w\;  \p^{a} g(w,\sft_1)   
\p_a g(w,w')g(w,w')   g(w',w_3)g(w,\sft_2)\notag \\
&=
\frac{1}{16}  \int  {\sf d}^2 w\;  \p^{a} g(w,\sft_1)   \p_a g(w',w_3)g(w,\sft_2).
\end{align}
This is the same expression as for the diagram of Type I  (cf. (\ref{D.14}))  and thus we get 
\be
\frac{1}{16} \,   ( -\frac29 )   \,   \frac{1}{   |\sft_{12} |^2 |\sft_{ 23} |^2\sft_{31} |^2}
=  -\frac{1}{72} \,   \frac{1}{   |\sft_{12} |^2 \sft_{ 23} |^2|\sft_{31} |^2} \ . 
\ee
As a result 
\be
\la{D.24}
\bC^{(1)}_{8}\,\bC^{(0)}\ b^3    = (-8b) \, (b^2 \times 2\times 2) \,  \times 2 \, \big(  -\frac{1}{72}  \big) 
=- \frac12 \bC^{(0)}b^3\quad \to \quad \bC^{(1)}_{8} = -\frac{1}{2}\ , 
\ee
where the factor $2$ comes from the two possibilities of assigning the  derivative (on either  leg 1 or 2).
 
\paragraph{ Type III} When both derivatives  from the \sm vertex are on the 
bulk propagators we obtain 
\be
\bC^{(1)}_{9}\,\bC^{(0)} b^3    = 
\begin{tikzpicture}[line width=1 pt, scale=0.8, rotate=0,baseline=0]
\coordinate (A1) at (90:2);  \coordinate (A2) at (230:2);  \coordinate (A3) at (-50:2);
\coordinate (B0) at ( 0:1 ); \coordinate (B1) at ( 180:1.6 );  \coordinate (B2) at (220:1.3);  \coordinate (B3) at (-60:1);
\draw[dotted] (0,0) circle (2);
\draw   (A1)--(0,0.7); \draw[dashed] (0,0) circle(0.7); \draw[dashed]  (A2)--(0,-0.7)--(A3);
\draw[fill=black] (A1) circle (0.1); \draw[fill=black] (A2) circle (0.1); \draw[fill=black] (A3) circle (0.1); 
\node[above] at (A1) {$\theta_{3}$};
\node[left] at (A2) {$\theta_{2}$};
\node[right] at (A3) {$\theta_{1}$};
\node[right] at (B0) {$ \p \xi$};
\node[right] at (B1) {$ \p\xi$};
\node[right] at (B2) {$ \xi$};
\node[right] at (B3) {$ \xi$};
\node[above] at  (0,-0.7)  {$ w$};
\node[below] at  (0,0.7)  {$ w'$};
\end{tikzpicture}\ 
\ee
The diagram requires the calculation of 
\be
B_{\p\p} (w,w_3) =\int {\sf d}^2 w'\;  \p^{a} g(w,w') \p_{a} g(w,w') g(w',w_3).
\ee
Taking another derivative of (\ref{D.18})  
gives ($\Box \equiv  {1\ov \sqrt g} \del^a\del_a$)
\be
\la{D.26}
\Box B (w,w_3) =2\int {\sf d}^2 w'\;    \Box g(w,w')g(w,w')   g(w',w_3)
+ 2\,B_{\p\p}(w,w_{3}).
\ee
The $m^2=2$  scalar  propagator in  \adst  satisfies 
\be
\Box g(w,w') = 2 g(w,w') -\frac{1}{\sqrt{g}}\,\delta^{(2)}(w,w')\ .
\ee
Using this in \rf{D.26} the  $\delta$-function  gives a  contact term  proportional to the 
propagator at coincident points $g(w,w)$ which we set  to zero 
in the  {\sf AdS} scheme (consistent with the \adst symmetry).
Thus 
%
\beqn
B_{\p\p} (w,w_3) &=& \frac12 \Box B (w,w_3) -2 B(w,w_3).
\eeqn
To compute $B_{\p\p}$ as  a   function of the  $\eta$-invariant in (\ref{23})  we observe that 
\be
\Box B=
\eta(\eta-1)^2  \p_\eta \p_\eta B +  (\eta-1)^2 \p_\eta B\ ,
\ee
where we used the relations $\Box \eta= \eta(\eta-1)^2$ and $\p^{a}\eta\p_a \eta  =(\eta-1)^2$,
that follow  from the definition of $\eta$ in (\ref{23}).
Sending $w_3$ to the boundary, we obtain 
\be
\lim_{\bz_{3}\to 0}\frac{1}{\bz_{3}^{2}}B_{\p\p} (w,w_3) 
=  -  \frac{1}{8}    g(w,\sft_3).
\ee
Finally,  the  above  diagram  thus  gives 
\be
-\frac{1}{8} \frac29   \frac{1}{   |\sft_{12} |^2 \sft_{ 23} |^2\sft_{31} |^2}
= -\frac{1}{36 }\frac{1}{   |\sft_{12} |^2 \sft_{ 23} |^2\sft_{31} |^2},
\ee
 which results in the following   contribution to  the coefficient  $\bC^{(1)}$ 
\be
\la{D.33}
\bC^{(1)}_{9}\,\bC^{(0)}\, b^3    = 
(-8b) \, (b^2 \times 2\times 2) \times 2 \times \frac12 \,   \big(  -\frac{1}{36}  \big) 
= - \frac12 \bC^{(0)}\, b^3\quad\to\quad \bC^{(1)}_{9} = -\frac{1}{2}\ , 
\ee
where the factor of 2  accounts for the two possible  fields in the loop and  $\frac12$ is the symmetry factor. 

\subsection*{Total result for $\bC^{(1)}$ }

Summing up all the 9 contributions 
given   above in  \rf{D.5}, \rf{D.6}, \rf{D.7}, \rf{D.8}, \rf{D.9},\rf{D.10}, \rf{D.15}, \rf{D.24},\rf{D.33}
 we get for  the total value of the coefficient $\bC^{(1)}$ in \rf{D.1} 
\be
\bC^{(1)}  = \sum_{i=1}^9 \bC^{(1)}_i= {2\log 2-\frac{10}{3}}
{-\frac{1}{2}}
{\vphantom{\frac{1}{2}}+3}
{+\frac{7}{3}}
{\vphantom{\frac{1}{2}}-1}
{\vphantom{\frac{1}{2}}-4}
{-\frac{1}{2}}
{-\frac{1}{2}}
{-\frac{1}{2}} = 2\,\log 2-5 \ , 
\ee
which is in agreement with (\ref{D.2}).

\bibliography{BT-Biblio}

\providecommand{\href}[2]{#2}\begingroup\raggedright\begin{thebibliography}{10}

\bibitem{DHoker:1983zwg}
E.~D'Hoker and R.~Jackiw, \emph{{Space translation breaking and
  compactification in the Liouville theory}},
  \href{http://dx.doi.org/10.1103/PhysRevLett.50.1719}{\emph{Phys. Rev. Lett.}
  {\bf 50} (1983) 1719--1722}.

\bibitem{DHoker:1983msr}
E.~D'Hoker, D.~Z. Freedman and R.~Jackiw, \emph{{SO(2,1) Invariant Quantization
  of the Liouville Theory}},
  \href{http://dx.doi.org/10.1103/PhysRevD.28.2583}{\emph{Phys. Rev.} {\bf D28}
  (1983) 2583}.

\bibitem{Inami:1985di}
T.~Inami and H.~Ooguri, \emph{{Dynamical breakdown of sypersymmetry in
  two-dimensional Anti de Sitter space}},
  \href{http://dx.doi.org/10.1016/0550-3213(86)90255-5}{\emph{Nucl. Phys.} {\bf
  B273} (1986) 487--500}.

\bibitem{Callan:1989em}
C.~G. Callan, Jr. and F.~Wilczek, \emph{{Iinfrared behaviour at negative
  curvature}},
  \href{http://dx.doi.org/10.1016/0550-3213(90)90451-I}{\emph{Nucl. Phys.} {\bf
  B340} (1990) 366--386}.

\bibitem{Zamolodchikov:2001ah}
A.~B. Zamolodchikov and A.~B. Zamolodchikov, \emph{{Liouville field theory on a
  pseudosphere}},  \href{http://arxiv.org/abs/hep-th/0101152}{{\tt
  hep-th/0101152}}.

\bibitem{Carmi:2018qzm}
D.~Carmi, L.~Di~Pietro and S.~Komatsu, \emph{{A Study of Quantum Field Theories
  in AdS at Finite Coupling}},
  \href{http://dx.doi.org/10.1007/JHEP01(2019)200}{\emph{JHEP} {\bf 01} (2019)
  200}, [\href{http://arxiv.org/abs/1810.04185}{{\tt 1810.04185}}].

\bibitem{Drukker:2006xg}
N.~Drukker and S.~Kawamoto, \emph{{Small deformations of supersymmetric Wilson
  loops and open spin-chains}},
  \href{http://dx.doi.org/10.1088/1126-6708/2006/07/024}{\emph{JHEP} {\bf 07}
  (2006) 024}, [\href{http://arxiv.org/abs/hep-th/0604124}{{\tt
  hep-th/0604124}}].

\bibitem{Giombi:2017cqn}
S.~Giombi, R.~Roiban and A.~A. Tseytlin, \emph{{Half-BPS Wilson loop and
  AdS$_2$/CFT$_1$}},
  \href{http://dx.doi.org/10.1016/j.nuclphysb.2017.07.004}{\emph{Nucl. Phys.}
  {\bf B922} (2017) 499--527}, [\href{http://arxiv.org/abs/1706.00756}{{\tt
  1706.00756}}].

\bibitem{Beccaria:2018ocq}
M.~Beccaria and A.~A. Tseytlin, \emph{{On non-supersymmetric generalizations of
  the Wilson-Maldacena loops in $\mathcal N=4$ SYM}},
  \href{http://dx.doi.org/10.1016/j.nuclphysb.2018.07.019}{\emph{Nucl. Phys.}
  {\bf B934} (2018) 466--497}, [\href{http://arxiv.org/abs/1804.02179}{{\tt
  1804.02179}}].

\bibitem{Beccaria:2019dws}
M.~Beccaria, S.~Giombi and A.~A. Tseytlin, \emph{{Correlators on
  non-supersymmetric Wilson line in $ \mathcal{N}=4 $ SYM and
  AdS$_{2}$/CFT$_{1}$}},
  \href{http://dx.doi.org/10.1007/JHEP05(2019)122}{\emph{JHEP} {\bf 05} (2019)
  122}, [\href{http://arxiv.org/abs/1903.04365}{{\tt 1903.04365}}].

\bibitem{Giombi:2017hpr}
S.~Giombi, C.~Sleight and M.~Taronna, \emph{{Spinning AdS Loop Diagrams: Two
  Point Functions}},
  \href{http://dx.doi.org/10.1007/JHEP06(2018)030}{\emph{JHEP} {\bf 06} (2018)
  030}, [\href{http://arxiv.org/abs/1708.08404}{{\tt 1708.08404}}].

\bibitem{Bertan:2018afl}
I.~Bertan, I.~Sachs and E.~D. Skvortsov, \emph{{Quantum $\phi^4$ Theory in
  AdS${}_4$ and its CFT Dual}},
  \href{http://dx.doi.org/10.1007/JHEP02(2019)099}{\emph{JHEP} {\bf 02} (2019)
  099}, [\href{http://arxiv.org/abs/1810.00907}{{\tt 1810.00907}}].

\bibitem{Yuan:2018qva}
E.~Y. Yuan, \emph{{Simplicity in AdS Perturbative Dynamics}},
  \href{http://arxiv.org/abs/1801.07283}{{\tt 1801.07283}}.

\bibitem{Liu:2018jhs}
J.~Liu, E.~Perlmutter, V.~Rosenhaus and D.~Simmons-Duffin,
  \emph{{$d$-dimensional SYK, AdS Loops, and $6j$ Symbols}},
  \href{http://dx.doi.org/10.1007/JHEP03(2019)052}{\emph{JHEP} {\bf 03} (2019)
  052}, [\href{http://arxiv.org/abs/1808.00612}{{\tt 1808.00612}}].

\bibitem{Ouyang:2019xdd}
H.~Ouyang, \emph{{Holographic four-point functions in Toda field theories in
  AdS$_{2}$}}, \href{http://dx.doi.org/10.1007/JHEP04(2019)159}{\emph{JHEP}
  {\bf 04} (2019) 159}, [\href{http://arxiv.org/abs/1902.10536}{{\tt
  1902.10536}}].

\bibitem{Beccaria:2019stp}
M.~Beccaria and A.~A. Tseytlin, \emph{{On boundary correlators in Liouville
  theory on AdS$_{2}$}},  \href{http://arxiv.org/abs/1904.12753}{{\tt
  1904.12753}}.

\bibitem{Beccaria:2019ibr}
M.~Beccaria and G.~Landolfi, \emph{{Toda theory in AdS$_{2}$ and $\mathcal
  WA_{n}$-algebra structure of boundary correlators}},
  \href{http://arxiv.org/abs/1906.06485}{{\tt 1906.06485}}.

\bibitem{Strominger:1998yg}
A.~Strominger, \emph{{AdS$_{2}$ quantum gravity and string theory}},
  \href{http://dx.doi.org/10.1088/1126-6708/1999/01/007}{\emph{JHEP} {\bf 01}
  (1999) 007}, [\href{http://arxiv.org/abs/hep-th/9809027}{{\tt
  hep-th/9809027}}].

\bibitem{Hotta:1998iq}
M.~Hotta, \emph{{Asymptotic isometry and two-dimensional anti-de Sitter
  gravity}},  \href{http://arxiv.org/abs/gr-qc/9809035}{{\tt gr-qc/9809035}}.

\bibitem{Cadoni:1999ja}
M.~Cadoni and S.~Mignemi, \emph{{Asymptotic symmetries of AdS$_{2}$ and
  conformal group in d = 1}},
  \href{http://dx.doi.org/10.1016/S0550-3213(99)00398-3}{\emph{Nucl. Phys.}
  {\bf B557} (1999) 165--180}, [\href{http://arxiv.org/abs/hep-th/9902040}{{\tt
  hep-th/9902040}}].

\bibitem{Almheiri:2014cka}
A.~Almheiri and J.~Polchinski, \emph{{Models of AdS$_{2}$ backreaction and
  holography}}, \href{http://dx.doi.org/10.1007/JHEP11(2015)014}{\emph{JHEP}
  {\bf 11} (2015) 014}, [\href{http://arxiv.org/abs/1402.6334}{{\tt
  1402.6334}}].

\bibitem{Jensen:2016pah}
K.~Jensen, \emph{{Chaos in AdS$_2$ Holography}},
  \href{http://dx.doi.org/10.1103/PhysRevLett.117.111601}{\emph{Phys. Rev.
  Lett.} {\bf 117} (2016) 111601}, [\href{http://arxiv.org/abs/1605.06098}{{\tt
  1605.06098}}].

\bibitem{Maldacena:2016upp}
J.~Maldacena, D.~Stanford and Z.~Yang, \emph{{Conformal symmetry and its
  breaking in two dimensional Nearly Anti-de-Sitter space}},
  \href{http://dx.doi.org/10.1093/ptep/ptw124}{\emph{PTEP} {\bf 2016} (2016)
  12C104}, [\href{http://arxiv.org/abs/1606.01857}{{\tt 1606.01857}}].

\bibitem{Engelsoy:2016xyb}
J.~Engelsoy, T.~G. Mertens and H.~Verlinde, \emph{{An investigation of
  AdS$_{2}$ backreaction and holography}},
  \href{http://dx.doi.org/10.1007/JHEP07(2016)139}{\emph{JHEP} {\bf 07} (2016)
  139}, [\href{http://arxiv.org/abs/1606.03438}{{\tt 1606.03438}}].

\bibitem{Gervais:1992bs}
J.-L. Gervais and M.~V. Savelev, \emph{{Black holes from nonAbelian Toda
  theories}}, \href{http://dx.doi.org/10.1016/0370-2693(92)91774-4}{\emph{Phys.
  Lett.} {\bf B286} (1992) 271--278},
  [\href{http://arxiv.org/abs/hep-th/9203039}{{\tt hep-th/9203039}}].

\bibitem{Bardacki:1990wj}
K.~Bardakci, M.~J. Crescimanno and E.~Rabinovici, \emph{{Parafermions From
  Coset Models}},
  \href{http://dx.doi.org/10.1016/0550-3213(90)90365-K}{\emph{Nucl. Phys.} {\bf
  B344} (1990) 344--370}.

\bibitem{Witten:1991yr}
E.~Witten, \emph{{On string theory and black holes}},
  \href{http://dx.doi.org/10.1103/PhysRevD.44.314}{\emph{Phys. Rev.} {\bf D44}
  (1991) 314--324}.

\bibitem{Polyakov:1981rd}
A.~M. Polyakov, \emph{{Quantum Geometry of Bosonic Strings}},
  \href{http://dx.doi.org/10.1016/0370-2693(81)90743-7}{\emph{Phys.Lett.} {\bf
  B103} (1981) 207--210}.

\bibitem{Nakayama:2004vk}
Y.~Nakayama, \emph{{Liouville field theory: A Decade after the revolution}},
  \href{http://dx.doi.org/10.1142/S0217751X04019500}{\emph{Int. J. Mod. Phys.}
  {\bf A19} (2004) 2771--2930},
  [\href{http://arxiv.org/abs/hep-th/0402009}{{\tt hep-th/0402009}}].

\bibitem{Menotti:2004uq}
P.~Menotti and E.~Tonni, \emph{{Standard and geometric approaches to quantum
  Liouville theory on the pseudosphere}},
  \href{http://dx.doi.org/10.1016/j.nuclphysb.2004.11.003}{\emph{Nucl. Phys.}
  {\bf B707} (2005) 321--346}, [\href{http://arxiv.org/abs/hep-th/0406014}{{\tt
  hep-th/0406014}}].

\bibitem{Leznov:1979td}
A.~N. Leznov and M.~V. Saveliev, \emph{{Representation of zero curvature for
  the system of nonlinear partial differential equations $\chi_{\alpha, z
  \overline{z}} = \exp(k\chi)_{\alpha}$ and its integrability}},
  \href{http://dx.doi.org/10.1007/BF00401930}{\emph{Lett. Math. Phys.} {\bf 3}
  (1979) 489--494}.

\bibitem{Bilal:1988jg}
A.~Bilal and J.-L. Gervais, \emph{{Systematic Construction of Conformal
  Theories with Higher Spin Virasoro Symmetries}},
  \href{http://dx.doi.org/10.1016/0550-3213(89)90633-0}{\emph{Nucl. Phys.} {\bf
  B318} (1989) 579--630}.

\bibitem{Hollowood:1989ep}
T.~J. Hollowood and P.~Mansfield, \emph{{Quantum Group Structure of Quantum
  Toda Conformal Field Theories. 1.}},
  \href{http://dx.doi.org/10.1016/0550-3213(90)90129-2}{\emph{Nucl. Phys.} {\bf
  B330} (1990) 720--740}.

\bibitem{Fateev:2007ab}
V.~A. Fateev and A.~V. Litvinov, \emph{{Correlation functions in conformal Toda
  field theory. I.}},
  \href{http://dx.doi.org/10.1088/1126-6708/2007/11/002}{\emph{JHEP} {\bf 11}
  (2007) 002}, [\href{http://arxiv.org/abs/0709.3806}{{\tt 0709.3806}}].

\bibitem{Tseytlin:1990mz}
A.~A. Tseytlin, \emph{{On the Structure of the Renormalization Group Beta
  Functions in a Class of Two-dimensional Models}},
  \href{http://dx.doi.org/10.1016/0370-2693(90)91285-J}{\emph{Phys. Lett.} {\bf
  B241} (1990) 233--237}.

\bibitem{Grisaru:1990gf}
M.~T. Grisaru, A.~Lerda, S.~Penati and D.~Zanon, \emph{{Renormalization Group
  Flows in Generalized Toda Field Theories}},
  \href{http://dx.doi.org/10.1016/0550-3213(90)90281-H}{\emph{Nucl. Phys.} {\bf
  B346} (1990) 264--292}.

\bibitem{Zamolodchikov:1985wn}
A.~B. Zamolodchikov, \emph{{Infinite Additional Symmetries in Two-Dimensional
  Conformal Quantum Field Theory}},
  \href{http://dx.doi.org/10.1007/BF01036128}{\emph{Theor. Math. Phys.} {\bf
  65} (1985) 1205--1213}.

\bibitem{Bouwknegt:1992wg}
P.~Bouwknegt and K.~Schoutens, \emph{{W symmetry in conformal field theory}},
  \href{http://dx.doi.org/10.1016/0370-1573(93)90111-P}{\emph{Phys. Rept.} {\bf
  223} (1993) 183--276}, [\href{http://arxiv.org/abs/hep-th/9210010}{{\tt
  hep-th/9210010}}].

\bibitem{Bilal:1993rg}
A.~Bilal, \emph{{NonAbelian Toda theory: A Completely integrable model for
  strings on a black hole background}},
  \href{http://dx.doi.org/10.1016/0550-3213(94)00149-9}{\emph{Nucl. Phys.} {\bf
  B422} (1994) 258--290}, [\href{http://arxiv.org/abs/hep-th/9312108}{{\tt
  hep-th/9312108}}].

\bibitem{Bilal:1995ei}
A.~Bilal, \emph{{Consistent string backgrounds and completely integrable 2-D
  field theories}},
  \href{http://dx.doi.org/10.1016/0920-5632(95)00619-2}{\emph{Nucl. Phys. Proc.
  Suppl.} {\bf 45A} (1996) 105--111},
  [\href{http://arxiv.org/abs/hep-th/9508062}{{\tt hep-th/9508062}}].

\bibitem{Jack:1993de}
I.~Jack, D.~R.~T. Jones and J.~Panvel, \emph{{Quantum nonAbelian Toda field
  theories}}, \href{http://dx.doi.org/10.1142/S0217751X9400145X}{\emph{Int. J.
  Mod. Phys.} {\bf A9} (1994) 3631--3656},
  [\href{http://arxiv.org/abs/hep-th/9308080}{{\tt hep-th/9308080}}].

\bibitem{Bilal:1995ag}
A.~Bilal, \emph{{Nonlocal extensions of the conformal algebra: Matrix W
  algebras, matrix KdV hierarchies and nonAbelian Toda theories}},  in
  \emph{{proceedings of 59 Recontre entre Phys. Theor. et Mathem. }}, 1995.
\newblock \href{http://arxiv.org/abs/hep-th/9501033}{{\tt hep-th/9501033}}.

\bibitem{Callan:1985ia}
C.~G. Callan, Jr., E.~J. Martinec, M.~J. Perry and D.~Friedan, \emph{{Strings
  in Background Fields}},
  \href{http://dx.doi.org/10.1016/0550-3213(85)90506-1}{\emph{Nucl. Phys.} {\bf
  B262} (1985) 593--609}.

\bibitem{Tseytlin:1988rr}
A.~A. Tseytlin, \emph{{Sigma model approach to string theory}},
  \href{http://dx.doi.org/10.1142/S0217751X8900056X}{\emph{Int. J. Mod. Phys.}
  {\bf A4} (1989) 1257}.

\bibitem{Tseytlin:1991ht}
A.~A. Tseytlin, \emph{{On the form of the black hole solution in D = 2
  theory}}, \href{http://dx.doi.org/10.1016/0370-2693(91)90800-6}{\emph{Phys.
  Lett.} {\bf B268} (1991) 175--178}.

\bibitem{Jack:1992mk}
I.~Jack, D.~R.~T. Jones and J.~Panvel, \emph{{Exact bosonic and supersymmetric
  string black hole solutions}},
  \href{http://dx.doi.org/10.1016/0550-3213(93)90239-L}{\emph{Nucl. Phys.} {\bf
  B393} (1993) 95--110}, [\href{http://arxiv.org/abs/hep-th/9201039}{{\tt
  hep-th/9201039}}].

\bibitem{Tseytlin:1993df}
A.~A. Tseytlin, \emph{{On field redefinitions and exact solutions in string
  theory}}, \href{http://dx.doi.org/10.1016/0370-2693(93)91372-T}{\emph{Phys.
  Lett.} {\bf B317} (1993) 559--564},
  [\href{http://arxiv.org/abs/hep-th/9308042}{{\tt hep-th/9308042}}].

\bibitem{Dijkgraaf:1991ba}
R.~Dijkgraaf, H.~L. Verlinde and E.~P. Verlinde, \emph{{String propagation in a
  black hole geometry}},
  \href{http://dx.doi.org/10.1016/0550-3213(92)90237-6}{\emph{Nucl. Phys.} {\bf
  B371} (1992) 269--314}.

\bibitem{Tseytlin:1992ri}
A.~A. Tseytlin, \emph{{Effective action of gauged WZW model and exact string
  solutions}},
  \href{http://dx.doi.org/10.1016/0550-3213(93)90511-M}{\emph{Nucl. Phys.} {\bf
  B399} (1993) 601--622}, [\href{http://arxiv.org/abs/hep-th/9301015}{{\tt
  hep-th/9301015}}].

\bibitem{Bars:1993zf}
I.~Bars and K.~Sfetsos, \emph{{Exact effective action and space-time geometry n
  gauged WZW models}},
  \href{http://dx.doi.org/10.1103/PhysRevD.48.844}{\emph{Phys. Rev.} {\bf D48}
  (1993) 844--852}, [\href{http://arxiv.org/abs/hep-th/9301047}{{\tt
  hep-th/9301047}}].

\bibitem{Tseytlin:1993my}
A.~A. Tseytlin, \emph{{Conformal sigma models corresponding to gauged
  Wess-Zumino-Witten theories}},
  \href{http://dx.doi.org/10.1016/0550-3213(94)90461-8}{\emph{Nucl. Phys.} {\bf
  B411} (1994) 509--558}, [\href{http://arxiv.org/abs/hep-th/9302083}{{\tt
  hep-th/9302083}}].

\bibitem{Hoare:2010fb}
B.~Hoare and A.~A. Tseytlin, \emph{{On the perturbative S-matrix of generalized
  sine-Gordon models}},
  \href{http://dx.doi.org/10.1007/JHEP11(2010)111}{\emph{JHEP} {\bf 11} (2010)
  111}, [\href{http://arxiv.org/abs/1008.4914}{{\tt 1008.4914}}].

\bibitem{ORaifeartaigh:1990ved}
L.~O'Raifeartaigh and A.~Wipf, \emph{{Conformally reduced WZNW theories and
  two-dimensional gravity}},
  \href{http://dx.doi.org/10.1016/0370-2693(90)90720-Q}{\emph{Phys. Lett.} {\bf
  B251} (1990) 361--368}.

\bibitem{Balog:1990mu}
J.~Balog, L.~Feher, L.~O'Raifeartaigh, P.~Forgacs and A.~Wipf, \emph{{Toda
  Theory and $W$ Algebra From a Gauged {WZNW} Point of View}},
  \href{http://dx.doi.org/10.1016/0003-4916(90)90029-N}{\emph{Annals Phys.}
  {\bf 203} (1990) 76--136}.

\bibitem{Feher:1992yx}
L.~Feher, L.~O'Raifeartaigh, P.~Ruelle, I.~Tsutsui and A.~Wipf, \emph{{On
  Hamiltonian reductions of the Wess-Zumino-Novikov-Witten theories}},
  \href{http://dx.doi.org/10.1016/0370-1573(92)90026-V}{\emph{Phys. Rept.} {\bf
  222} (1992) 1--64}.

\bibitem{Klimcik:1994wp}
C.~Klimcik and A.~A. Tseytlin, \emph{{Exact four-dimensional string solutions
  and Toda like sigma models from 'null gauged' WZNW theories}},
  \href{http://dx.doi.org/10.1016/0550-3213(94)90089-2}{\emph{Nucl. Phys.} {\bf
  B424} (1994) 71--96}, [\href{http://arxiv.org/abs/hep-th/9402120}{{\tt
  hep-th/9402120}}].

\bibitem{Gerasimov:1990fi}
A.~Gerasimov, A.~Morozov, M.~Olshanetsky, A.~Marshakov and S.~L. Shatashvili,
  \emph{{Wess-Zumino-Witten model as a theory of free fields}},
  \href{http://dx.doi.org/10.1142/S0217751X9000115X}{\emph{Int. J. Mod. Phys.}
  {\bf A5} (1990) 2495--2589}.

\bibitem{Ford:2000eb}
C.~Ford, \emph{{Quantum parafermions in the $SL(2,\mathbb R) / U(1)$ WZNW black
  hole model}},
  \href{http://dx.doi.org/10.1016/S0034-4877(01)80065-3}{\emph{Rept. Math.
  Phys.} {\bf 48} (2001) 67--74},
  [\href{http://arxiv.org/abs/hep-th/0010123}{{\tt hep-th/0010123}}].

\bibitem{Kruger:2004jm}
C.~Kruger, \emph{{Exact operator quantization of the Euclidean black hole
  CFT}},  \href{http://arxiv.org/abs/hep-th/0411275}{{\tt hep-th/0411275}}.

\bibitem{Bakas:1991fs}
I.~Bakas and E.~Kiritsis, \emph{{Beyond the large N limit: Nonlinear
  $W_{\infty}$ as symmetry of the $SL(2,\mathbb R) / U(1)$ coset model}},
  \href{http://dx.doi.org/10.1142/S0217751X92003720}{\emph{Int. J. Mod. Phys.}
  {\bf A7S1A} (1992) 55--81}, [\href{http://arxiv.org/abs/hep-th/9109029}{{\tt
  hep-th/9109029}}].

\bibitem{Sevrin:1993si}
A.~Sevrin and W.~Troost, \emph{{Extensions of the Virasoro algebra and gauged
  WZW models}},
  \href{http://dx.doi.org/10.1016/0370-2693(93)91617-V}{\emph{Phys. Lett.} {\bf
  B315} (1993) 304--310}, [\href{http://arxiv.org/abs/hep-th/9306033}{{\tt
  hep-th/9306033}}].

\bibitem{Dolan:2011dv}
F.~A. Dolan and H.~Osborn, \emph{{Conformal Partial Waves: Further Mathematical
  Results}},  \href{http://arxiv.org/abs/1108.6194}{{\tt 1108.6194}}.

\bibitem{DHoker:1999kzh}
E.~D'Hoker, D.~Z. Freedman, S.~D. Mathur, A.~Matusis and L.~Rastelli,
  \emph{{Graviton exchange and complete four point functions in the AdS/CFT
  correspondence}},
  \href{http://dx.doi.org/10.1016/S0550-3213(99)00525-8}{\emph{Nucl. Phys.}
  {\bf B562} (1999) 353--394}, [\href{http://arxiv.org/abs/hep-th/9903196}{{\tt
  hep-th/9903196}}].

\bibitem{Dolan:2000ut}
F.~A. Dolan and H.~Osborn, \emph{{Conformal four point functions and the
  operator product expansion}},
  \href{http://dx.doi.org/10.1016/S0550-3213(01)00013-X}{\emph{Nucl. Phys.}
  {\bf B599} (2001) 459--496}, [\href{http://arxiv.org/abs/hep-th/0011040}{{\tt
  hep-th/0011040}}].

\bibitem{Dolan:2003hv}
F.~A. Dolan and H.~Osborn, \emph{{Conformal partial waves and the operator
  product expansion}},
  \href{http://dx.doi.org/10.1016/j.nuclphysb.2003.11.016}{\emph{Nucl. Phys.}
  {\bf B678} (2004) 491--507}, [\href{http://arxiv.org/abs/hep-th/0309180}{{\tt
  hep-th/0309180}}].

\bibitem{Fateev:1987zh}
V.~A. Fateev and S.~L. Lukyanov, \emph{{The Models of Two-Dimensional Conformal
  Quantum Field Theory with $Z_{n}$ Symmetry}},
  \href{http://dx.doi.org/10.1142/S0217751X88000205}{\emph{Int. J. Mod. Phys.}
  {\bf A3} (1988) 507}.

\bibitem{Braaten:1982yn}
E.~Braaten, T.~Curtright and C.~B. Thorn, \emph{{An Exact Operator Solution of
  the Quantum Liouville Field Theory}},
  \href{http://dx.doi.org/10.1016/0003-4916(83)90214-2}{\emph{Annals Phys.}
  {\bf 147} (1983) 365}.

\bibitem{Schomerus:2002dc}
V.~Schomerus, \emph{{Lectures on branes in curved backgrounds}},
  \href{http://dx.doi.org/10.1088/0264-9381/19/22/305}{\emph{Class. Quant.
  Grav.} {\bf 19} (2002) 5781--5847},
  [\href{http://arxiv.org/abs/hep-th/0209241}{{\tt hep-th/0209241}}].

\bibitem{Recknagel:2013uja}
A.~Recknagel and V.~Schomerus, \emph{{Boundary Conformal Field Theory and the
  Worldsheet Approach to D-Branes}}.
\newblock Cambridge Monographs on Mathematical Physics. Cambridge University
  Press, 2013,
  \href{http://dx.doi.org/10.1017/CBO9780511806476}{10.1017/CBO9780511806476}.

\bibitem{Menotti:2003km}
P.~Menotti and E.~Tonni, \emph{{The Tetrahedron graph in Liouville theory on
  the pseudosphere}},
  \href{http://dx.doi.org/10.1016/j.physletb.2004.01.027}{\emph{Phys. Lett.}
  {\bf B586} (2004) 425--431}, [\href{http://arxiv.org/abs/hep-th/0311234}{{\tt
  hep-th/0311234}}].

\end{thebibliography}\endgroup
\bibliographystyle{JHEP}

\end{document}